\documentclass[prd,twocolumn,nofootinbib,letter,superscriptaddress]{revtex4-1}

\usepackage{amssymb}
\usepackage{amsmath}
\usepackage{graphicx,psfrag}% Include figure files
\usepackage{bm}% bold math
\usepackage{mathbbol,verbatim}
\usepackage{slashed}
\usepackage{graphics}
\usepackage{color,ulem}
\usepackage{makecell}
\usepackage[justification=justified,singlelinecheck=false]{caption}
\usepackage{subcaption}

\usepackage{times}
\usepackage{hyperref}
\usepackage{setspace,ulem}

\usepackage{graphicx}
\usepackage{amssymb}
\usepackage{textcomp}
\usepackage{amsmath}
\usepackage{bm}
\usepackage{times}
\usepackage{epsfig}
\usepackage{color}
\usepackage{graphics}
\usepackage{hyperref}
\usepackage{setspace,ulem}

\allowdisplaybreaks
\usepackage{multirow}

\hypersetup{
    pdfnewwindow=true,      % links in new window
    colorlinks=true,       % false: boxed links; true: colored links
    linkcolor=black,          % color of internal links
    citecolor=blue,        % color of links to bibliography
    filecolor=blue,      % color of file links
    urlcolor=blue           % color of external links
}

\makeatletter
\renewcommand\@makecaption[2]{%
  \par
  \vskip\abovecaptionskip
  \begingroup
   \small\rmfamily
    \begingroup
     \samepage
     \flushing
     \let\footnote\@footnotemark@gobble
     \@make@capt@title{#1}{#2}\par
    \endgroup
  \endgroup
  \vskip\belowcaptionskip
}
\makeatother

\definecolor{greeen}{rgb}{0.03,0.84,0.13}
\definecolor{test}{rgb}{0.03,0.74,0.33}
\definecolor{viol}{rgb}{0.44,0,0.94}
\definecolor{or}{rgb}{0.95,0.65,0}

\begin{document}

\title{Probing TeV scale origin of neutrino mass at future lepton colliders \\
via neutral and doubly-charged scalars}

\author{P. S. Bhupal Dev}
\email{bdev@wustl.edu}
\affiliation{Department of Physics and McDonnell Center for the Space Sciences,  \\
Washington University, St. Louis, MO 63130, USA}

\author{Rabindra N. Mohapatra}
\email{rmohapat@umd.edu}
\affiliation{Maryland Center for Fundamental Physics, Department of Physics, \\University of Maryland, College Park, MD 20742, USA}

\author{Yongchao Zhang}
\email{yongchao.zhang@physics.wustl.edu}
\affiliation{Department of Physics and McDonnell Center for the Space Sciences,  \\
Washington University, St. Louis, MO 63130, USA}

%\date{\today}

\begin{abstract}
  We point out how future lepton colliders can provide unique insight into the scalar sector of TeV scale models for neutrino masses with local $B-L$ symmetry. Our specific focus is on the TeV scale left-right model, which naturally embeds this $B-L$ symmetry. In particular, we make a detailed study of the lepton collider implications of the neutral ($H_3$) and doubly-charged ($H^{\pm\pm}$) scalars from the right-handed triplet Higgs that is responsible for the spontaneous breaking of the $B-L$ symmetry and implementing the seesaw mechanism. Due to mixing with other scalars, the neutral scalar $H_3$ could acquire sizable flavor violating couplings to the charged leptons. Produced on-shell or off-shell at the planned $e^+e^-$ colliders, it would induce distinct lepton flavor violating signals like $e^+e^- \to \mu^\pm \tau^\mp ~ (+H_3)$, with the couplings probed up to $\sim 10^{-4}$ for a wide range of neutral scalar mass, which is well beyond the reach of current searches for charged lepton flavor violation. The Yukawa couplings of the doubly-charged scalar $H^{\pm\pm}$ to the charged leptons might also be flavor-violating, which is correlated to the heavy right-handed neutrino masses and mixings. With a combination of the pair, single and off-shell production of $H^{\pm\pm}$ like $e^+e^- \to H^{++} H^{--},\, H^{\pm\pm} e^\mp \mu^\mp,\, \mu^\pm \tau^\mp$, the Yukawa couplings can be probed up to $10^{-3}$ at future lepton colliders, which is allowed by current lepton flavor data in a large region of parameter space. For both the neutral and doubly-charged cases, the scalar masses could be probed up to the few-TeV range in the off-shell channel.
\end{abstract}


\maketitle

\tableofcontents

\section{Introduction} Exploring the scale of new physics responsible for neutrino masses is one of the major topics under intense focus in particle physics today. If the scale is near the grand unification theory (GUT) scale (or much higher than the multi-TeV range), only ways to explore this would be via rare decays of leptons, and via proton decay in GUT models, providing only a limited window for their search. The situation however changes drastically if the new physics scale is around a few TeV, which is theoretically quite plausible, thus opening up many more possible experimental probes. In particular, explorations at the high energy frontier using existing as well as planned future colliders primarily designed for other studies of beyond the Standard Model (SM) physics become viable and effective for probing the neutrino mass physics.
The lepton colliders are particularly well-suited to study this question since they provide clean signals, not ``muddied'' by the QCD jet background from the SM and other unrelated new physics scenarios. It is therefore no surprise that the literature using  lepton colliders for studying neutrino mass physics is already quite extensive; see e.g. Refs.~\cite{Gluza:1995js, Gluza:1996bz, Porod:2000hv, Deppisch:2003wt, Cakir:2004cv, Ciftci:2005gf, Atwood:2007zza, Liu:2007yka, Kadastik:2007yd, McDonald:2007ka, Sierra:2008wj, Arhrib:2009mz, Yue:2010zzb, Das:2012ze, List:2013dga, Freitas:2014fda, Blondel:2014bra, Antusch:2015mia, Banerjee:2015gca, Antusch:2015gjw, Antusch:2016vyf, Antusch:2016ejd, Chekkal:2017eka, Yue:2017mmi, Liao:2017jiz, Dev:2017ftk, Sui:2017qra, Agrawal:2018pci, delAguila:1987nn, Buchmuller:1991tu, Djouadi:1993pe, Azuelos:1993qu, Gluza:1997ts, delAguila:2005pin, Biswal:2017nfl, Bray:2005wv, Gluza:1995ix, Belanger:1995nh, Ananthanarayan:1995cn, Rodejohann:2010jh, Achard:2001qv, Asaka:2015oia, Wang:2016eln,AristizabalSierra:2003ix, Raidal:2000ru, Barry:2012ga, delAguila:2005ssc, Dittmar:1989yg, Abreu:1996pa, Cho:2018mro} for a partial list and Refs.~\cite{Deppisch:2015qwa, Cai:2017mow} for reviews.

In this paper, we add to this literature by exploring how the planned lepton colliders can provide unique insight into a specific class of TeV-scale seesaw models~\cite{seesaw1,seesaw2,seesaw3,seesaw4,seesaw5}  for neutrino masses based on left-right symmetric model (LRSM) of weak interactions~\cite{LR, Mohapatra:1974gc, Senjanovic:1975rk}. The LRSM, originally proposed as a {well-motivated} extension of SM for providing an alternative approach to parity violation in low energy processes, has since emerged as a model for neutrino masses via the seesaw mechanism. In this note we focus particularly on the scalar sector of the LRSM where the parity symmetry has been broken at high scale so that the low energy theory does not contain the left triplet with $B-L=2$. We call this LRSM (in contrast with {the version} where the discrete parity is broken at the TeV scale). This model contains two unique particles, which are not part of the SM nor many beyond SM scenarios and are particularly suited for the lepton collider searches: one is a hadrophobic neutral scalar connected to spontaneous breaking of $B-L$ symmetry that goes into the seesaw mechanism (called $H_3$, following the convention in Ref.~\cite{Dev:2016dja}) and another is the right-handed (RH) doubly-charged scalar partner of this (called $H^{++}$), that is part of the $SU(2)_R$ multiplet that contains $H_3$. The hadron collider  implications of the hadrophobic scalar was discussed in previous papers~\cite{Dev:2016dja, Dev:2016vle, Dev:2017dui} and we continue this exploration in this paper in the context of lepton colliders like the Circular Electron-Positron Collider (CEPC)~\cite{CEPC-SPPCStudyGroup:2015csa}, International Linear Collider (ILC)~\cite{Baer:2013cma}, Future Circular Collider (FCC-ee)~\cite{Gomez-Ceballos:2013zzn} and Compact Linear Collider (CLIC)~\cite{Battaglia:2004mw}.

The reason these two scalars are particularly interesting for studying the origin of neutrino masses is that in the LRSM, the couplings (denoted by $f_{\alpha\beta}$) of the hadrophobic scalar  and the doubly-charged scalar to leptons are the ones that are responsible for the seesaw masses of the RH neutrinos (RHNs) i.e. $M_{N,\alpha\beta}=2f_{\alpha\beta}v_R$. For a $B-L$ breaking scale $v_R$ in the 5-8 TeV range so that the $W_R$ is accessible at the LHC~\cite{Keung:1983uu, Chen:2013fna, Nemevsek:2018bbt, Dev:2015kca, Nemevsek:2011hz, Ferrari:2000sp, Das:2012ii, Mitra:2016kov}, the magnitudes of $f_{\alpha\beta}$ responsible to fit the neutrino oscillation data~\cite{PDG}, i.e. the mass square differences and the mixing angles, are sizable and hence accessible at the lepton colliders, unless all the entries of the Dirac mass matrix $m_D$ in the seesaw formula $m_\nu\simeq -m_DM_N^{-1}m_D^T$ are very tiny. There are of course constraints on some subsets of these couplings from rare lepton decays but they leave enough room for some of them being of order one. Furthermore, rare lepton decays generally probe products of two different $f$ couplings whereas the lepton collider probes them individually~\cite{Dev:2017ftk}. A question one can of course ask is: do we really know that all the $m_D$ entries in the seesaw formula are not very tiny making the $f$'s similarly tiny and inaccessible? Two reasons to think that this may not be the case are that
\begin{itemize}
\item [(i)] there are interesting seesaw textures for neutrino masses where some of the $m_D$ elements are sizable~\cite{Kersten:2007vk,Dev:2013oxa,Mitra:2011qr, Gluza:2002vs, Xing:2009in, He:2009ua, Adhikari:2010yt, Ibarra:2010xw} for TeV scale seesaw; and

\item [(ii)] if we make the $f$ couplings very small, the Yukawa couplings that go into $m_D$ become very small and give a feeling of being unnatural in the sense that we could as well have had just Dirac neutrinos of sub-eV mass with Yukawa couplings $\lesssim 10^{-12}$ without requiring any seesaw in the first place.
\end{itemize}
Add to this the possibility that  sizable Dirac mass terms can be measured in colliders by measuring the heavy-light neutrino mixing parameters $V_{\nu N}$  in the process $pp \to \ell^\pm \ell^\pm jj$~\cite{AguilarSaavedra:2012gf, Chen:2013fna}. Finally, it is not just that left-right models provide a venue for such sizable $f$ couplings but there can be other models (e.g. models with global $B-L$~\cite{Dev:2017xry}) with similar properties, making the probes of sizable $f$  of great theoretical as well as experimental interest. Also, there exist well-motivated $f$ matrix textures within the context of LRSM fully consistent with current neutrino mass and other observations e.g. rare lepton decays~\cite{Dev:2013oxa}. We show in this paper the interesting range of $f$ values that can be measured in the planned lepton colliders such as CEPC and ILC and can provide new ways to test these models. They will in any case provide complementary information to rare lepton decay constraints on the $f$ couplings and makes such studies interesting from the synergistic viewpoint of energy and intensity frontiers.

This paper is organized as follows: In Section~\ref{sec:model}, we describe the essentials of the LRSM relevant for our discussion.
%specially the couplings of the neutral hadro-phobic and the doubly charged scalars;
We then focus in Section~\ref{sec:H3} on the production of the neutral scalar $H_3$ in future lepton colliders like CEPC and ILC, given its couplings to the other scalars, the heavy RHNs and the heavy $W_R$ and $Z_R$ bosons, as well as other relevant couplings. Special attention is paid to the production of lepton flavor violating (LFV) signals induced by the neutral scalar $H_3$, as an explicit example of the general proposal in Ref.~\cite{Dev:2017ftk}. Prospects in all the possible on-shell and off-shell production channels of $H_3$ are given in this section, some of which are well below the current low energy LFV constraints~\cite{PDG}. The lepton collider physics for the RH doubly-charged scalar follows  in Section~\ref{sec:Hpp}, where all the flavor and collider constraints are summarized, and all the possible pair production, single production and off-shell production modes of the doubly-charged scalar are discussed, with potential LFV signals. As far as we know, this is the first complete list of the production of RH doubly-charged scalar at lepton colliders in the literature, with all the possible accompanying LFV signal taken into consideration, though some of the channels (and some of the flavor combinations) have been separately investigated before. Three particular textures of the $f_{\alpha\beta}$ matrix are also exemplified for the doubly-charged scalar in this section, as seen in Table~\ref{tab:examples}. We will summarize the main results and conclude in Section~\ref{sec:conclusion}.

\section{Seesaw scalars in the left-right model}
\label{sec:model}

%The LFV process $e^+ e^- \to e^+ \ell^-$ is discussed in Ref.~\cite{Kabachenko:1997aw, Cho:2016zqo} in the effective four-fermion coupling and in~\cite{Yue:2005yr} in technicolor model, and the effect of doubly-charged scalar on $e^+ e^- \to \ell^+ \ell^-$ is studied in~\cite{Nomura:2017abh}.

%Main backgrounds: $e^+ e^- \to W^+ W^- \to \ell^\pm_\alpha \ell^\mp_\beta \nu \bar{\nu}$~\cite{Kabachenko:1997aw}.
The LRSM is based on the gauge group ${\cal G}_{\rm LR}\equiv SU(3)_C \times SU(2)_L\times SU(2)_R\times U(1)_{B-L}$~\cite{LR, Mohapatra:1974gc, Senjanovic:1975rk}. The quarks $Q$ and leptons $\psi$ are assigned to the following irreducible representations of ${\cal G}_{\rm LR}$:
\begin{align}
& Q_{L,\alpha} \ = \ \left(\begin{array}{c}u_L\\d_L \end{array}\right)_\alpha : \: \left({ \bf 3}, {\bf 2}, {\bf 1}, \frac{1}{3}\right), \nonumber \\
& Q_{R,\alpha} \ = \ \left(\begin{array}{c}u_R\\d_R \end{array}\right)_\alpha : \: \left({ \bf 3}, {\bf 1}, {\bf 2}, \frac{1}{3}\right), \nonumber \\
& \psi_{L,\alpha} \ = \  \left(\begin{array}{c}\nu_L \\ e_L \end{array}\right)_\alpha : \: \left({ \bf 1}, {\bf 2}, {\bf 1}, -1 \right), \nonumber \\
& \psi_{R,\alpha} \ = \ \left(\begin{array}{c} N_R \\ e_R \end{array}\right)_\alpha : \: \left({ \bf 1}, {\bf 1}, {\bf 2}, -1 \right),
\label{lrSM}
\end{align}
where $\alpha=1,2,3$ represents the family indices, and the subscripts $L,R$ denote  the left and right chiral projection operators $P_{L,R} = (1\mp \gamma_5)/2$, respectively.

In the LRSM, in addition to a bidoublet Higgs field $\Phi(2,2,0)$, an RH triplet Higgs field is introduced:
\begin{eqnarray}
\Delta_R \ = \ \left(\begin{array}{cc}\Delta^+_R/\sqrt{2} & \Delta^{++}_R\\\Delta^0_R & -\Delta^+_R/\sqrt{2}\end{array}\right) : ({\bf 1}, {\bf 1}, {\bf 3}, 2) \, ,
\label{eq:scalar}
\end{eqnarray}
in order to break the $SU(2)_R\times U(1)_{B-L}$ symmetry down to $U(1)_Y$ and to give mass to the heavy RHNs from the following Yukawa coupling:
\begin{eqnarray}
\label{eqn:LYukawa}
{\cal L}_Y \  \supset \ f_{\alpha\beta} \psi^T_{R,\alpha}C^{-1}\sigma_2\Delta_R\psi_{R,\beta}~+~{\rm H.c.} \,,
\end{eqnarray}
where $C$ is the charge conjugation matrix, $\sigma_2$ is the second Pauli matrix, and $f_{\alpha\beta}$ tare he Yukawa coupling matrix elements with $\alpha,\,\beta$ the lepton flavor indices. The singly-charged member of this triplet gets absorbed as the longitudinal mode of the heavy $W_R$ boson and the imaginary part of the neutral component ${\rm Im}\,\Delta_{R}^0$ as the longitudinal mode of the $Z_R$ boson, leaving $H_3 \equiv {\rm Re}(\Delta^0)$ and $H^{\pm\pm}\equiv \Delta^{\pm\pm}_R$ as physical fields. Note that the RHN mass matrix that goes into the seesaw formula is now determined by the vacuum expectation value (VEV) $\langle \Delta^0_R\rangle= v_R$, viz. $M_{N,\alpha\beta}=2f_{\alpha\beta}v_R$. Therefore measuring the couplings of $H^{\pm\pm}$ is tantamount to measuring the RHN mass matrix. We show below how the planned lepton colliders like CEPC and ILC can probe deeper into the ranges of $f_{\alpha\beta}$ providing tests of certain LRSM-based neutrino mass models. We call $H_3$ and $H^{\pm\pm}$ the ``seesaw Higgs bosons" due to their role in implementing the seesaw mechanism. For a detailed discussion of the symmetry breaking scalar sector of the LRSM, see e.g. Refs.~\cite{Grifols:1989xe, Zhang:2007da,Dev:2016dja,Maiezza:2016ybz}.

%With the Yukawa couplings in Eq.~(\ref{eqn:LYukawa}), the doubly-charged scalar couples

\section{The neutral scalar}
\label{sec:H3}

A unique property of the $SU(2)_R$ symmetry breaking neutral scalar $H_3$ in the LRSM is that it directly couples only to the heavy doubly-charged scalar $H^{\pm\pm}$ from the RH triplet $\Delta_R$ and the heavy scalars from the bidoublet $\Phi$, the heavy gauge bosons $W_R$, $Z_R$, and the heavy RHNs $N_\alpha$, but does not interact directly with the SM quarks, i.e. it is naturally a hadrophobic scalar~\cite{Dev:2016dja}. Following Ref.~\cite{Dev:2016dja}, the interaction Lagrangian involving $H_3$ is given by
\begin{widetext}
\begin{eqnarray}
\label{eqn:Lagrangian}
{\cal L}_{H_3} & \ = \ &
\frac{1}{2\sqrt2} \alpha_1 v_R H_3 h h
- \frac{1}{\sqrt2} \alpha_1 v_{\rm EW} h H_3 H_3
+ 2\sqrt2 (\rho_1 + 2\rho_2) v_R H_3 H^{++} H^{--} \nonumber \\
&& + \bigg(\frac{1}{\sqrt2}\left[\widehat{Y}_U
  \sin\tilde\theta_1
  - \left( V_L \widehat{Y}_D V_R^\dagger \right) \sin\tilde\theta_2  \right]_{\alpha\beta} H_3 \bar{u}_\alpha u_\beta
+ \frac{1}{\sqrt2}\left[  \widehat{Y}_D
  \sin\tilde\theta_1
  -  \left( V_L \widehat{Y}_U V_R^\dagger \right) \sin\tilde\theta_2  \right]_{\alpha\beta} H_3 \bar{d}_\alpha d_\beta \nonumber \\
&&  + \frac{1}{\sqrt2} \left[\widehat{Y}_{E}  \sin\tilde\theta_1
- Y_{\nu N} \sin\tilde\theta_2\right]_{\alpha\beta}H_3\bar{\ell}_\alpha\ell_\beta +  f_{\alpha\beta} H_3 \overline{N_\alpha^C} N_\beta ~+~ {\rm H.c.} \bigg)
+ \frac{c_{H_3 \gamma\gamma} \, \alpha_{\rm EM}}{4v_R} H_3 F_{\mu\nu} F^{\mu\nu}  \nonumber \\
&& +\frac{1}{\sqrt2}g_L^2 \sin\theta_1\, v_{\rm EW} H_3 W_\mu^+ W^{- \mu}
+ \frac{g_L^2 \sin\theta_1\, v_{\rm EW}}{2\sqrt2 \cos^2\theta_w}
H_3 Z_\mu Z^{\mu}
+ \sqrt2  g_R^2 v_R H_3 W_{R \mu}^+ W_R^{- \mu} \,.
\end{eqnarray}
\end{widetext}
%with $v_{\rm EW}$ the electroweak VEV, $\widehat{Y}_{U,D, E}$ the diagonal Yukawa couplings of the SM Higgs to the up- and down-type quarks and charged leptons respectively, $V_{L,R}$ the left- and right-handed quark matrix matrices, $Y_{\nu N} = m_D/v_{\rm EW}$, $\sin\tilde\theta_{1,\,2} \equiv \sin\theta_{1,\,2} + \xi \sin\theta_{2,\,1}$ with $\xi = \kappa^\prime/\kappa$ the ratio of the VEVs from the bidoublet $\Phi$, $\alpha_1$, $\rho_{1}$, $\rho_2$ the coefficients of the quartic coupling terms ${\rm Tr}(\Phi^\dag \Phi){\rm Tr}(\Delta_R\Delta_R^\dag)$, $[{\rm Tr}(\Delta_R\Delta_R^\dag)]^2$ and ${\rm Tr}(\Delta_R\Delta_R){\rm Tr}(\Delta_R^\dag\Delta_R^\dag)$ respectively in the scalar potential, $g_L$ the gauge coupling for the $SU(2)_L$ gauge group and $\theta_w$ the weak mixing angle, and $\alpha,\, \beta$ are the quark/lepton flavor indices.
%See more details in Ref.~\cite{Dev:2016dja}.
In Eq.~(\ref{eqn:Lagrangian}), $v_{\rm EW}$ is the electroweak VEV; $\hat{Y}_{U,D,E}$ are respectively the diagonal Yukawa couplings of the SM Higgs to the up-type quarks, down-type quarks and charged leptons; $V_{L,R}$ are the left- and right-handed quark mixing matrices; $Y_{\nu N} = m_D / v_{\rm EW}$; $\sin\tilde\theta_{1,2}  \equiv \sin\theta_{1,2} + \xi \sin\theta_{2,1}$ with $\xi = \kappa^\prime / \kappa$ the ratio of the VEVs from the bidoublet $\Phi$; $\alpha_1$, $\rho_1$ and $\rho_2$ are respectively the coefficients for the quartic coupling terms ${\rm TR} (\Phi^\dagger \Phi) {\rm TR} (\Delta \Delta^\dagger)$, $[{\rm TR} (\Phi^\dagger \Phi)]^2$ and ${\rm TR} (\Delta \Delta) {\rm TR} (\Delta^\dagger \Delta^\dagger)$ in the scalar potential; $g_L$ is the gauge coupling for the $SU(2)_L$ gauge group; $\theta_w$ is the weak mixing angle; $\alpha$ and $\beta$ are the quark/lepton flavor indices.

At the tree-level, the scalar mass $m_{H_3}^2 = 4 \rho_1 v_R^2$. For sufficiently small coupling $\rho_1$, the scalar mass $m_{H_3}$ could be much smaller than the RH scale $v_R$. Then in this case the radiative corrections to $m_{H_3}$ become important, and might even dominate the $H_3$ mass. However, using the Coleman-Weinberg effective potential~\cite{Coleman:1973jx}, it was found that~\cite{Dev:2016vle, Dev:2017dui} the bosonic contributions to the $H_3$ mass from the heavy scalars and the heavy gauge bosons are partially canceled by those from fermionic heavy RHNs, which allows for $H_3$ masses much lighter than the $v_R$ scale.\footnote{As long as the scalar, Yukawa and gauge couplings for the loop contribution to $m_{H_3}$ are perturbative, the higher-loop contributions are expected to be smaller than the one-loop diagrams.} Here we treat $m_{H_3}$ as a free parameter in an effective theory approach.

The couplings of $H_3$ to the heavy $W_R$ boson and the singly and doubly-charged scalars induce an effective 1-loop couplings of $H_3$ to two photons, as shown in Eq.~(\ref{eqn:Lagrangian}), with $\alpha_{\rm EM}$ the fine-structure constant and $c_{H_3 \gamma\gamma}$ depending on the scalar and vector loop functions as a function of the mass $m_{H_3}$~\cite{Dev:2016vle, Dev:2017dui} . Note that this coupling is effectively suppressed by the RH scale $v_R$ and in the limit of $m_{H_3} \ll v_R$, the coefficient $c_{H_3 \gamma\gamma}$ approaches a numerical constant $\simeq 4\sqrt2 / 3\pi$.

The $H_3$ could of course mix with the SM Higgs $h$ as well as with the heavy CP-even scalar $H_1$ from the bidoublet $\Phi$, due to the ${\rm Tr}(\Phi^\dag \Phi){\rm Tr}(\Delta_R\Delta_R^\dag)$ type quartic interaction, with the corresponding mixing angles $\sin\theta_1$ and $\sin\theta_2$ respectively. This induces the (flavor-changing) couplings of $H_3$ to the SM fermions and to the $W$ and $Z$ bosons.\footnote{The $W-W_R$ and $Z-Z_R$ mixings contribute also to the couplings of $H_3$ to the SM $W$ and $Z$ bosons, but they are heavily suppressed respectively by the mass ratios $\sim m_W^2 / m_{W_R}^2$ and $m_Z^2 / m_{Z_R}^2$~\cite{Zhang:2007da, Dev:2016dja} and are neglected here.}
In some specific scenarios the heavy-light neutrino mixings might be sizable~\cite{Dev:2013oxa}, and generate the couplings of $H_3$ to the light active neutrinos, depending also on the magnitude of the $f$ couplings in Eq.~(\ref{eqn:LYukawa}).
The $H_3$ could mix with the heavy neutral component $H_1$ from the bidoublet $\Phi$ thereby acquiring tree-level flavor-changing neutral current (FCNC) couplings  in both the quark and lepton sectors~\cite{Dev:2016vle, Dev:2017dui}. In light of the seesaw mechanism, the flavor structure of the FCNC couplings of $H_3$ in the lepton sector is different from that in the quark sector.
In particular, the flavor-changing couplings of $H_3$ to the charged leptons are dictated by the Dirac neutrino mass matrix $m_D$. In the most general cases, $m_D$ is not diagonal, as it has to be used to generate the mixings of active neutrinos. Furthermore, in some circumstances some elements of $Y_{\nu N}$ might be sizable, even of order ${\cal O} (0.1)$~\cite{Dev:2013oxa}, which is essential to have large effective couplings of $H_3$ to the charged leptons. For a heavy $H_3$, the mixing angle with $H_1$ could even reach up to the order of 0.1~\cite{Dev:2016vle, Dev:2017dui},  then it is possible to to have an effective coupling $h_{\alpha\beta}$ of order 0.01 in some region of the parameter space. For the sake of concreteness and conciseness, we write the LFV couplings of $H_3$ in an effective form, without explicit dependence on the parameters in the charged lepton
mass matrix, the Dirac neutrino mass matrix, or the RHN mass matrix:
\begin{eqnarray}
\label{eqn:LYukawa:H3}
{\cal L}_{H_3} \ \supset \ h_{\alpha \beta} H_3 \overline{\ell}_{\alpha} \ell_{\beta,\, } ~+~ {\rm H.c.} \,,
\end{eqnarray}
with $\alpha,\, \beta = e,\, \mu,\, \tau$.
For simplicity we also assume the coupling matrix $h_{\alpha\beta}$ is real and symmetric~\cite{Dev:2017ftk}.\footnote{The symmetric properties of the leptonic Yukawa couplings and the Dirac matrix $m_D$ depend on how the left and right-handed leptons $\psi_{L,R}$ and the bidoublet $\Phi$ transform under the parity symmetry, but not on how parity is broken at the TeV or higher energy scale, e.g. whether the left-handed triplet decouples from the TeV scale physics, as the matrix $m_D$ does not come from couplings to the triplets. These fields transform as $\Phi \leftrightarrow \Phi^\dagger$ and $\psi_L \leftrightarrow \psi_R$ under parity $P$, so the Yukawa coupling matrices for the leptons are required to be Hermitian~\cite{Maiezza:2010ic}. There are of course small radiative corrections below the scale of triplet mass but they are proportional to $y^2/16\pi^2$ (with $y$ the Yukawa couplings) and are therefore very small maintaining the symmetric property for all practical purpose.  More discussion of generalized parity can be found in e.g. Ref.~\cite{Senjanovic:2016vxw}.} The LFV couplings of $H_3$ to the charged leptons could also be induced radiatively through the doubly-charged loop, which is however highly suppressed by the mass ratio $m_{\ell}^2/M_{\pm\pm}^2$ ($m_\ell$ and $M_{\pm\pm}$ being respectively the charged lepton and doubly-charged scalar mass) and the loop factor~\cite{Nemevsek:2016enw}. For few TeV RH scale $v_R$, the loop-level decay is comparable to the tree-level width only when the effective coupling $h \sim 10^{-8}$ in Eq.~(\ref{eqn:Lagrangian}), which is far below the future prospects (see Figures~\ref{fig:H3:prospect1} and \ref{fig:H3:prospect2}); thus the loop-level LFV couplings of $H_3$ can be safely neglected in this paper.  One should note the Dirac neutrino mass matrix $m_D$, and consequently the effective coupling $h_{\alpha\beta}$ in Eq.~(\ref{eqn:Lagrangian}), might also be tested by other future lepton data, e.g. the rare RHN decay $N \to W^\pm \ell^\mp$, the electron EDM, neutrinoless double beta decay and neutrino transition moments~\cite{Nemevsek:2012iq}. These experiments depend however also on the RHN masses and mixings, the CP violating phases in $m_D$, and/or even the $W- W_R$ mixing angle, and might be largely complementary to the direct searches of the ``smoking-gun'' beyond SM LFV signals at future lepton colliders.

The couplings of $H_3$ collected in  Eq.~(\ref{eqn:Lagrangian})  lead to very rich phenomenology for the production and decay of $H_3$ in future lepton colliders like CEPC, ILC, FCC-ee and CLIC, even though some of the heavy particles can not be directly produced on-shell at these colliders. We first present the decay branching ratios (BR) of $H_3$, before moving on to its production modes.

\subsection{$H_3$ decay branching ratios}

\begin{figure*}[!t]
  %\centering
  \includegraphics[width=0.42\textwidth]{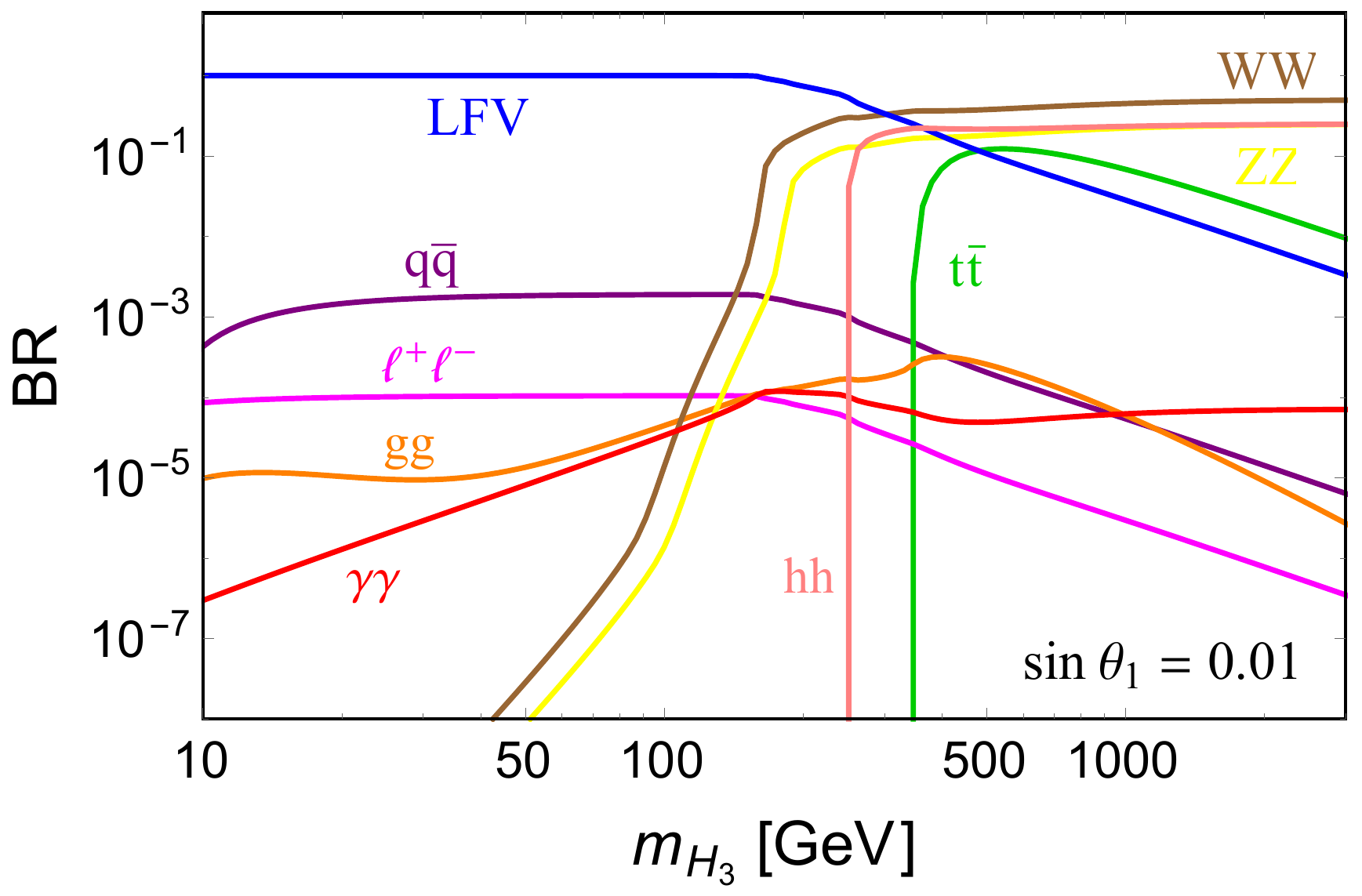}
  \includegraphics[width=0.42\textwidth]{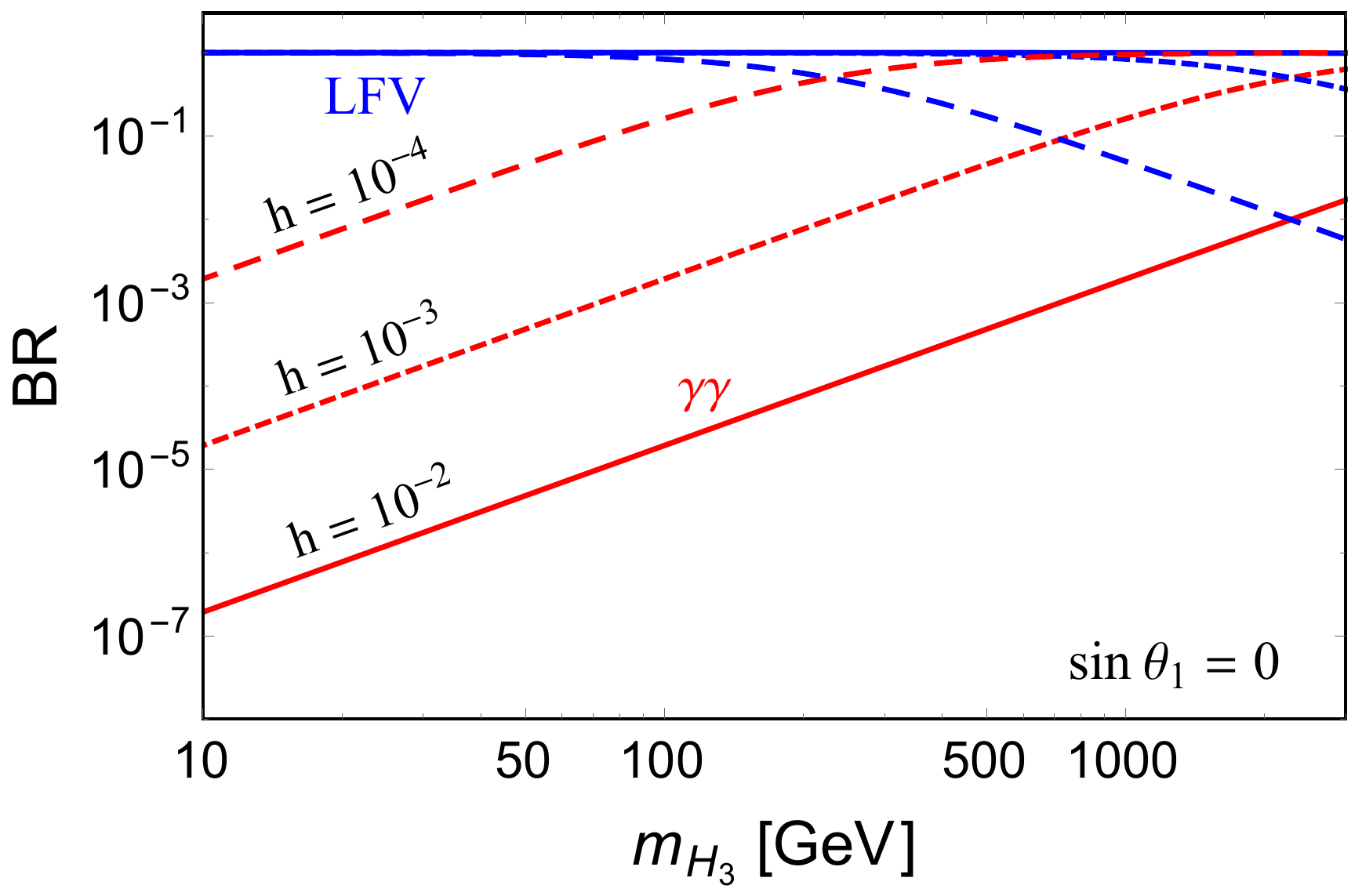}
  \caption{{\it Left}: The decay BRs of the neutral scalar $H_3$ into the $WW$, $ZZ$ bosons, the light quarks $q\bar{q}$ (with $q = u,\,d,\,s,\,c,\,b$), the top pair $t\bar{t}$, the charged leptons $\ell^+ \ell^-$, the gluon pair $gg$, the photon pair $\gamma\gamma$, and the charged lepton pairs with LFV signal (LFV).  The mixing with the SM Higgs $\sin\theta_1 = 0.01$, the RH scale $v_R = 5$ TeV and the LFV coupling $h = 0.01$. For concreteness, the bidoublet VEV ratio $\xi = \kappa'/\kappa$ has been set to $m_b/m_t$ as in Refs.~\cite{Dev:2016vle, Dev:2017dui}, and the mixing angle $\sin\theta_2$ with the heavy scalar $H_1$ is set to be zero. {\it Right}: The decay BRs of $H_3$ into $\gamma\gamma$ (red) and LFV lepton pairs (blue) without mixing with the SM Higgs ($\sin\theta_1 = 0$) and $v_R = 5$ TeV, for three benchmark values of $h = 10^{-2}$ (solid), $10^{-3}$ (short-dashed) and $10^{-4}$ (long-dashed).}
  \label{fig:BR_H3}
\end{figure*}

For illustration purpose, two examples of decay BRs of the neutral scalar $H_3$ are shown in Figure~\ref{fig:BR_H3}, with a sizable mixing with the SM Higgs, $\sin\theta_1 = 0.01$ (left) and without any mixing, $\sin\theta_1 = 0$ (right).  The $H_3$ decays into the SM particles, like the $W$, $Z$, $h$ bosons, quarks, charge leptons and gluons only due to its mixing with $h$, and therefore,  all these partial decay widths are proportional to $\sin^2\theta_1$. The decay $H_3 \to \gamma\gamma$ receives two contributions: one from the SM top quark and $W$ boson loops via mixing with the SM Higgs, and the other one from the heavy $W_R$ boson and the heavy charged scalar loops which are suppressed by the $v_R$ scale. For concreteness we have set $v_R = 5$ TeV in Figure~\ref{fig:BR_H3}. For the LFV decays $H_3 \to \ell_\alpha^\pm \ell_\beta^\mp$, we adopt a benchmark value of $h = 10^{-2}$ for the effective LFV couplings [cf.~Eq.~(\ref{eqn:Lagrangian})], which is still allowed by current lepton flavor data in a large parameter space for $m_{H_3} \gtrsim 10$ GeV (see Figures~\ref{fig:H3:prospect1} and \ref{fig:H3:prospect2} for more details). The analytic expressions for the partial decay widths of $H_3$ can be found in Refs.~\cite{Dev:2016vle, Dev:2017dui}. It is clear in the left panel of Figure~\ref{fig:BR_H3} that when the neutral scalar is light, i.e. $m_{H_3} \lesssim 2 m_{W,\,Z}$, it decays predominantly into the LFV lepton pairs, as all the other channels are either suppressed by the small couplings in the SM (like the light quarks) or loop suppressed (like gluons and photons). When the decays $h_3 \to WW,\,ZZ,\,hh$ is open, they will take over as the  dominant channels, as they grow as $\Gamma (H_3 \to WW,\,ZZ,\,hh)\propto G_F m_{H_3}^3$ ($G_F$ being the Fermi constant), much faster than the LFV channel.

In the limit of $\sin\theta_1 \to 0$, all the couplings of the SM Higgs approach to their SM values~\cite{Dev:2016dja}. In this alignment limit, the neutral scalar could only decay into two photons induced from the heavy charged particles in the $SU(2)_R$ sector, and into the LFV charged leptons from mixing with the heavy scalar $H_1$. The fractions ${\rm BR} (H_3 \to \ell_\alpha \ell_\beta) = 1 - {\rm BR} (H_3 \to \gamma\gamma)$ depends largely on the $v_R$ scale and the effective LFV coupling $h$ [cf.~Eq.~(\ref{eqn:Lagrangian})]. Given $v_R$ fixed at 5 TeV, three benchmark values of $h = 10^{-2}$, $10^{-3}$ and $10^{-4}$ are shown in  the right panel of Figure~\ref{fig:BR_H3}, depicted respectively as solid, short-dashed and long-dashed contours. As seen in this plot, for the LFV coupling $h \gtrsim 10^{-3}$, the BR of diphoton channel is very small for $m_{H_3} \lesssim {\rm TeV}$, as it is suppressed both by the loop factors and by the $v_R$ scale. As $\Gamma (H_3 \to \gamma\gamma) \propto m_{H_3}^3/v_R^2$ and $\Gamma (H_3 \to \ell\ell) \propto m_{H_3}$, the diphoton width grows much faster than the dilepton channel when $m_{H_3}$ gets heavier and becomes important for $h \sim 10^{-4}$ and $m_{H_3} \gtrsim $ few hundred GeV.

%%%%%%%%%%%%%%%%%%%%%%%%%%%%%%%%%

\subsection{Production of $H_3$}
Based on the scalar, Yukawa and gauge interactions in  Eq.~(\ref{eqn:Lagrangian}), the production of $H_3$ at an $e^+e^-$ collider can be categorized into five groups:
\begin{itemize}
  \item {\it Doubly-charged scalar portal:} Through the fusion of the doubly-charged scalar pair: $H^{++\, \ast} H^{--\, \ast} \to H_3$ with the trilinear scalar interaction $H_3 H^{++} H^{--}$ related to their masses~\cite{Dev:2016dja}. The virtual doubly-charged scalars are emitted from the initial electron/positron via the Yukawa interaction $f_{\alpha\beta}$, inducing potentially LFV signal if $\alpha \neq \beta$; see   Figure~\ref{fig:diagram1}.
  \item {\it Gauge portal:} Through the effective 1-loop coupling to diphoton $\gamma\gamma \to H_3$, with subleading contribution from $e^+ e^- \to \gamma^\ast \to \gamma H_3$ and $\gamma^\ast \gamma^\ast \to H_3$; see Figure~\ref{fig:diagram2}.
  \item {\it SM Higgs portal:} Through mixing with SM Higgs $e^+ e^- \to Z^\ast \to Z H_3$ (and other subleading production modes like through the fusion of SM $W$ and $Z$ bosons); see Figure~\ref{fig:diagram3}.
  \item {\it Neutrino portal:} Through the fusion of (heavy) RHNs, which couples to the initial electron/positron via the heavy-light neutrino mixing $N_\alpha^\ast N_\beta^\ast \to H_3$; see Figure~\ref{fig:diagram4}.
  \item {\it Heavy scalar portal:} The LFV couplings of $H_3$ [cf. Eq.~\eqref{eqn:Lagrangian}] will induce various production modes, like the on-shell production $e^+ e^- \to (\gamma/Z) H_3$ and $e^+ e^- \to (\ell_\alpha^+ \ell_\beta^- / \nu_\alpha \bar\nu_\beta) H_3$ and the $H_3$-mediated processes $e^+ e^- \to \ell_\alpha^+ \ell_\beta^-$~\cite{Dev:2017ftk}. See Figures~\ref{fig:diagram5}, \ref{fig:diagram6}, \ref{fig:diagram7} and \ref{fig:diagram8}.
\end{itemize}
We  discuss below each of these production modes in some details.
To be specific, we focus in this paper only on the CEPC and ILC as two benchmark machines for future lepton colliders. However, for comparison, we present in Table~\ref{tab:colliders} the planned final center-of-mass energy $\sqrt{s}$, the expected integrated luminosity and the required running time to achieve this luminosity for the four lepton colliders under discussion, viz. CEPC, ILC, FCC-ee and CLIC~\cite{colliders}. The center-of-mass energies and luminosities at FCC-ee and CLIC are expected to be comparable or higher than those at CEPC and ILC and could therefore improve to some extent the prospects at CEPC and ILC discussed in this paper. In this sense, the results in this paper can be considered to be conservative, at least in some of the channels.

\begin{table}[!t]
  %\centering
  \caption[]{The planned final center-of-mass energy $\sqrt{s}$, the expected integrated luminosity and the required running time to achieve this luminosity for the four future lepton colliders~\cite{colliders}. }
  \label{tab:colliders}
  \begin{tabular}[t]{cccc}
  \hline\hline
  collider & \makecell{$\sqrt{s}$ \\ (GeV)} &
  \makecell{luminosity \\ (ab$^{-1}$)} & \makecell{running time \\ (year)} \\ \hline
  CEPC & 240 & 5 & 7 \\
FCC-ee & 250 & 5 & 3 \\
  ILC & 1000 & 1 & 8 - 10 \\
  CLIC & 3000 & 3 & 6 - 8 \\
  \hline\hline
  \end{tabular}
\end{table}

\subsection{Doubly-charged scalar portal}

\begin{figure}[t!]
  \centering
  \includegraphics[width=0.28\textwidth]{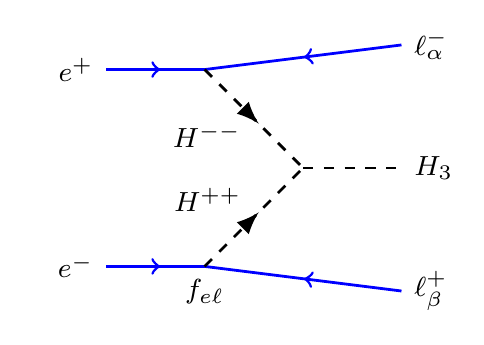} \vspace{-10pt}
  \caption{Feynman diagram for the production of $H_3$ at lepton colliders from the doubly-charged scalar fusion.}
  \label{fig:diagram1}
\end{figure}

\begin{figure}[t!]
  %\centering
  \includegraphics[width=0.4\textwidth]{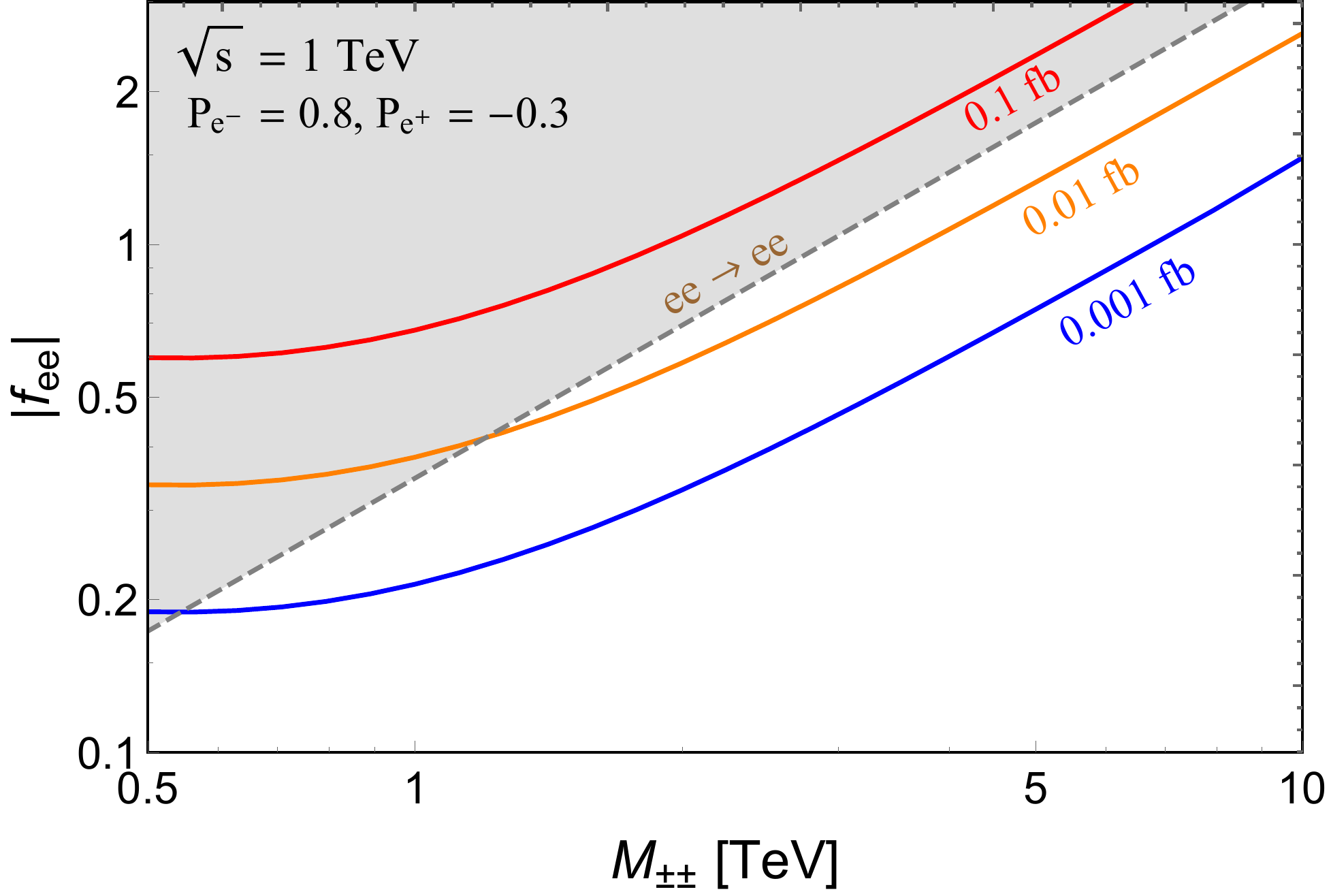}
  \caption{Production cross section contours of $H_3$ at ILC 1 TeV with polarized beams ($P_{e^-} = 0.8$, $P_{e^+} = -0.3$) through the doubly-charged scalar fusion in the channel $e^+ e^- \to e^+ e^- H_3$ (see Figure~\ref{fig:diagram1}), as functions of $M_{\pm\pm}$ and $|f_{ee}|$, in the limit of $m_{H_3} \ll M_{\pm\pm}$. The shaded region is excluded by the LEP $ee \to ee$ data~\cite{Abdallah:2005ph}. See text for more details.  }
  \label{fig:production1}
\end{figure}

For the coupling of $H_3$ to the doubly-charged scalar $H^{\pm\pm}$ in  Eq.~(\ref{eqn:Lagrangian}),
%in the limit of $M_{\pm\pm} \equiv M_{H_2^{\pm\pm}} \gg m_{H_3}$,
\begin{eqnarray}
\label{eqn:triple}
(\rho_1 + 2 \rho_2) v_R \ = \
\frac{m^2_{H_3}}{4 v_R} +
\frac{M^2_{\pm\pm}}{2 v_R} \,,
\end{eqnarray}
where we have used the fact that for the neutral and doubly-charged scalar masses are respectively $m_{H_3}^2 = 4\rho_1v_R^2$ and $M_{\pm\pm}^2 = 4 \rho_2 v_R^2$~\cite{Dev:2016dja}. In the limit of $m_{H_3} \ll v_R$ and $M_{\pm\pm} \gg \sqrt{s}$, the dependence of the trilinear scalar coupling on the doubly-charged scalar mass cancels partially the suppression of the heavy $H^{\pm\pm}$ propagator $(1/M_{\pm\pm}^2)^2$ in Figure~\ref{fig:diagram1}, leaving only a $\sim 1/M_{\pm\pm}^2$ dependence on the doubly-charged scalar mass. The couplings of $H^{\pm\pm}$ to the SM charged leptons stem from the $f_{\alpha\beta}$ term in Eq.~(\ref{eqn:LYukawa}) and might be flavor-changing, i.e.
\begin{eqnarray}
\label{eqn:H3production1}
e^+ e^- \ \to \ \ell_\alpha^+ \ell_\beta^- H_3 \,
\end{eqnarray}
with $\alpha,\beta\neq e$. In light of the clean environment at lepton colliders, this would definitely point to  new physics beyond the SM. The production cross section contours at $\sqrt{s} = 1$ TeV are shown in Figure~\ref{fig:production1}, as functions of the doubly-charged scalar mass $M_{\pm\pm}$ and $|f_{ee}|$, for the benchmark values of $v_R = 5$ TeV and in the limit of small $m_{H_3}$. The initial state radiation (ISR) and beamstrahlung have been taken into consideration for the colliding beams and $p_T > 10$ GeV imposed for the $e^\pm$ in the final state, implemented by using {\tt CalcHEP}~\cite{Belyaev:2012qa}. We have assumed the initial beams are polarized, i.e. $P_{e^-} = 0.8$ and $P_{e^+} = -0.3$, which enhance the couplings to the RH doubly-charged scalar by a factor of $(1+P_{e^-}) (1-P_{e^+}) = 2.34$ compared to the unpolarized beams. Limited by a smaller $\sqrt s=240$ GeV, the cross sections at CEPC are expected to be much smaller. %To be specific, we have taken the RH scale $v_R = 5$ TeV and $f_{ee} = 1$.
In the limit of small mixing with other scalar particles, the neutral scalar $H_3$ decays predominantly into two photons $H_3 \to \gamma\gamma$ through the $W_R$ and charged scalar loops, or into a pair of charged leptons, as detailed in Section~\ref{sec:heavyscalar} with the signal of $e^+ e^- \to \ell_\alpha \ell_\beta (H_3 \to \gamma\gamma,\, \ell_\gamma \ell_\delta)$.

%with an integrated luminosity of 5 ab$^{-1}$, we could expect up to hundreds of $e^+ e^- \gamma\gamma$ signal with almost free (or well-understood) background.

%In the light $H_3$ limit, the production cross section at $\sqrt{s} = 240$ GeV and 1 TeV are respectively 0.032 fb and 0.34 fb, doubly-charged scalar mass $M_{\pm\pm} = 500$ GeV and .

%which generate the RHN masses and the seesaw mechanism, with $C$ standing for charge conjugation, $\psi_R \equiv (N,\, \ell_R)^{\sf T}$ the right-handed lepton doublets and $\alpha,\,\beta$ the flavor indices.

%Note that the Yukawa matrix $f_{\alpha\beta}$ could be non-diagonal in the flavor space, which then renders LFV couplings of $H_2^{\pm\pm}$ to the charged leptons, i.e. the vertices $H_2^{\pm\pm} e^{\mp} \mu^{\mp}$ and  $H_2^{\pm\pm} e^{\mp} \tau^{\mp}$ shown in Fig.~\ref{fig:diagram1}, which produce the LFV signals like %With sizable couplings $f_{\alpha\beta}$, it is rather promising to see the LFV signals at future lepton colliders, like
%\begin{eqnarray}
%e^+ e^- \to e^+ \mu^- H_3, \, \mu^+ \tau^- H_3 \,,
%\end{eqnarray}
%which is suppressed only by the doubly-charged scalar mass $1/M_{\pm\pm}^2$.

With the lepton flavor conserving and violating couplings $f_{\alpha\beta}$, the doubly-charged scalar contributes to the electron and muon $g-2$, rare charged lepton decays $\ell_\alpha \to \ell_\beta \gamma$, $\ell_\alpha \to \ell_\beta \ell_\gamma \ell_\delta$, muonium-anti-muonium oscillation and $ee \to \ell\ell$. All the flavor limits on the Yukawa couplings are collected in Table~\ref{tab:limits} in Section~\ref{sec:constraints}. The LFV signals $\ell_\alpha \ell_\beta = e \mu$, $e\tau$ and $\mu\tau$ in Eq.~(\ref{eqn:H3production1}) depend on the combinations of Yukawa couplings $|f_{ee}^\dagger f_{e\mu}|$, $|f_{ee}^\dagger f_{e\tau}|$ and $|f_{e\mu}^\dagger f_{e\tau}|$, which are tightly constrained respectively by the rare decays $\mu \to eee$, $\tau \to eee$ and $\tau^- \to \mu^- e^+ e^-$. For instance, for $M_{\pm\pm}=500$ GeV, the limits are respectively
\begin{eqnarray}
|f_{ee}^\dagger f_{e\mu}|^2 & \ < \ & 3.3 \times 10^{-11} \,, \nonumber \\
|f_{ee}^\dagger f_{e\tau}|^2 & \ < \ & 5.2 \times 10^{-6} \,, \nonumber \\
|f_{e\mu}^\dagger f_{e\tau}|^2 & \ < \ & 2.6 \times 10^{-6} \,,
\end{eqnarray}
which make the corresponding LFV cross section too small to be observable in the doubly-charged scalar fusion portal.\footnote{Note that the low-energy rare decays $\mu \to eee$ and the production $ee \to e\mu H_3$ at high-energy lepton colliders have the same dependence on the combination of $|f_{ee} f_{e\mu}|/M_{\pm\pm}^2$ in the limit of heavy doubly-charged scalar mass, and thus, a lighter $H^{\pm\pm}$ pushes the limits on the $f_{\alpha\beta}$ couplings to more stringent values (note that the triple scalar coupling in Eq.~(\ref{eqn:triple}) becomes smaller when the doubly-charged scalar gets lighter), and does not help to alleviate the limits from the LFV decays. } The flavor-conserving process $e^+ e^- \to e^+ e^- H_3$ in Figure~\ref{fig:diagram1} depends only on the coupling $f_{ee}$ in the Yukawa sector and is thus free of the LFV decay limits. It also contributes to Bhabha scattering and is constrained by the LEP $ee \to ee$ data~\cite{Abdallah:2005ph}, which is shown in Figure~\ref{fig:production1} by the shaded region. The production cross section contours of $\sigma (ee \to ee H_3)$ are also presented in Figure~\ref{fig:production1}, as functions of the doubly-charged scalar mass $M_{\pm\pm}$ and $|f_{ee}|$ in the limit of small $m_{H_3}$. It is clear that even if the doubly-charged scalar is heavier than 1 TeV, there is still ample parameter region to have an observable production cross section (up to 0.03 fb), that is allowed by the existing limits from LEP.

\subsection{Gauge portal}
\label{sec:gaugeportal}

\begin{figure}
  \centering
  \hspace{-30pt}
  \begin{subfigure}[b]{0.15\textwidth}
  \includegraphics[height=\textwidth]{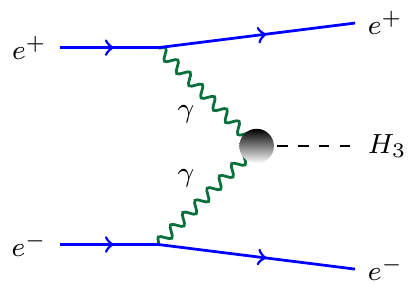}
  \caption{}
  \label{fig:4a}
  \end{subfigure}

  \hspace{-30pt}
  \begin{subfigure}[b]{0.15\textwidth}
  \includegraphics[height=\textwidth]{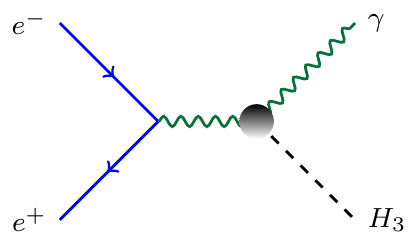}
  \caption{}
  \label{fig:4b}
  \end{subfigure} \hspace{50pt}
  \begin{subfigure}[b]{0.15\textwidth}
  \includegraphics[height=\textwidth]{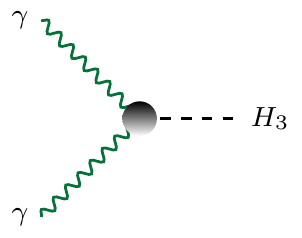}
  \caption{}
  \label{fig:4c}
  \end{subfigure}
  \caption{Feynman diagrams for the production of $H_3$ at lepton colliders from the effective radiative coupling to photons (denoted by the black blob) in the process (a) $e^+ e^- \to e^+ e^- H_3$, (b) $e^+ e^- \to \gamma H_3$ and (c) $\gamma \gamma \to H_3$. }
  \label{fig:diagram2}
\end{figure}

\begin{figure}[t!]
  \centering
  \includegraphics[width=0.4\textwidth]{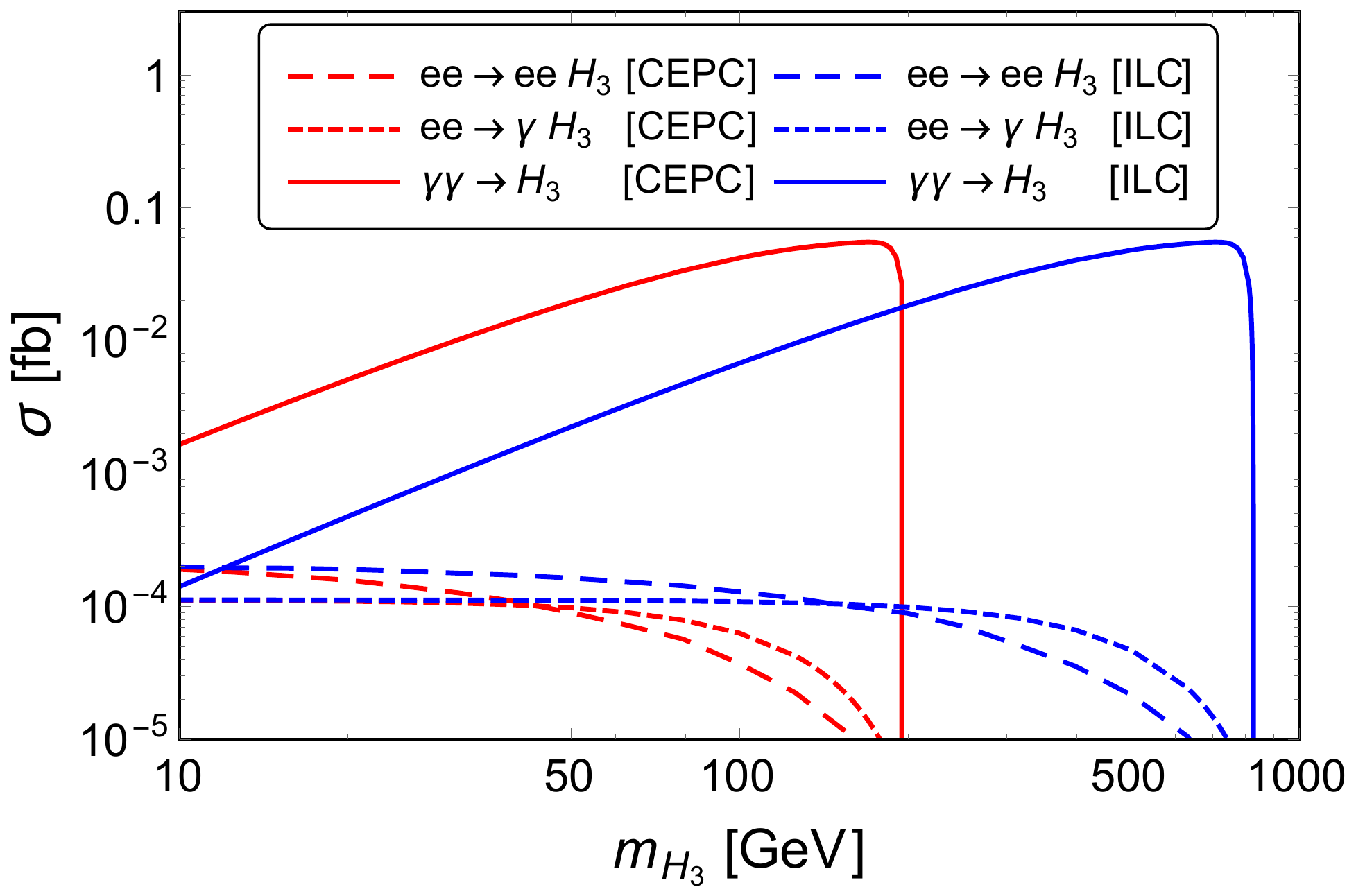}
  \caption{Production cross section contours of $H_3$ at CEPC 240 GeV (red) and ILC 1 TeV (blue) from the radiative couplings to photons (see Figure~\ref{fig:diagram2}), as functions of the scalar mass $m_{H_3}$. The $p_T$ cuts on the electrons and photon are all taken to be 5 GeV and the RH scale $v_R = 5$ TeV.}
  \label{fig:production2}
\end{figure}

The scalar $H_3$ can also be produced from the radiative couplings to photons, mediated by the heavy scalar and gauge bosons, as shown in Figure~\ref{fig:diagram2}. All the diagrams in Figure~\ref{fig:diagram2} are effectively suppressed by the RH scale $v_R$, which is typically (much) higher than the center-of-mass energy.
The cross section of photon fusion channel $\gamma^\ast \gamma^\ast \to H_3$ (Figure~\ref{fig:4a}) depends largely on the momentum of the leptons in the final state: if the transverse momenta of the outgoing electron/positron are small, then the cross section would rise significantly like in Compton scattering. With an aggressive cut of $p_T (e^\pm) = 1$ GeV, the photon fusion cross section is respectively $2.3 \times 10^{-4}$ fb and $5.1 \times 10^{-4}$ fb at CEPC $240$ GeV and ILC 1 TeV, in the light $H_3$ limit for the RH scale $v_R = 5$ TeV. If the cuts are more realistic, like $p_T > 5$ GeV, the cross section is smaller, for instance $7.0 \times 10^{-5} \, (2.1 \times 10^{-4})$ fb at $\sqrt{s} = 240$ GeV (1 TeV), as shown in Figure~\ref{fig:production2} (the long-dashed curves).
The production cross section $\sigma (e^+ e^- \to \gamma H_3)$ is also very small, roughly $1.1 \times 10^{-4} \, {\rm fb}$ in the light $H_3$ limit at both CEPC 240 GeV and ILC 1 TeV, with a $p_T > 5$ GeV cut on the photon, which are also presented in Figure~\ref{fig:production2}.  Even though the production cross sections would go larger as $\propto 1/v_R^2$ when the RH scale is to some extent lower, they are still too small.
%photon fusion
%pt = 1 GeV:	0.00023149
%pt = 5 GeV:     6.9786 x10-05	
%1 TeV, pt = 1 GeV: 0.00050762
%3 TeV, pt = 1 GeV: 0.00080703

In future lepton colliders, high luminosity photon beams can be obtained by Compton backscattering of low energy, high intensity laser beam off the high energy electron beam, and then $H_3$ can be produced from the laser ``photon fusion'' processes as shown in Figure~\ref{fig:4c}. The effective photon luminosity distribution reads~\cite{Ginzburg:1981vm, Ginzburg:1982yr, Telnov:1989sd}
\begin{eqnarray}
\label{eqn:photon_pdf}
f_{\gamma/e} (x) & \ = \ &
\frac{1}{D(\xi)} \left[ (1-x) + \frac{1}{(1-x)} \nonumber \right. \\
&& \left. - \frac{4x}{\xi (1-x)} + \frac{4x^2}{\xi^2 (1-x)^2} \right] \,, \\
{\rm with} \quad
D (\xi) & \ = \ &
\left( 1 - \frac{4}{\xi} - \frac{8}{\xi^2} \right) \log (1+\xi)
+ \frac12 \nonumber \\
&& + \frac{8}{\xi} - \frac{1}{2(1+\xi)^2} \,,
\end{eqnarray}
where $x =\omega/E_e$ is the fraction of electron energy carried away by the scattered photon, with $\omega$ and $E_e$ respectively the energies of scattered photon and initial electron. The parameter $\xi = 4\omega_0 E_e / m_e^2$ depends on the energy $\omega_0$ of initial laser photon. When $\xi \gtrsim 4.8$ the photon conversion efficiency drops drastically, as a consequence of the $e^+ e^-$ pair production from the laser photons and the photon backscattering, which sets an upper bound on the energy fraction $x < x_{\rm max} = \xi / (1+\xi) \simeq 0.83$. The production cross sections $\sigma (\gamma\gamma \to H_3)$ at $\sqrt{s} = 240$ GeV and 1 TeV are shown in Figure~\ref{fig:production2}. The cross section could reach up to 0.05 fb for the RH scale $v_R = 5$ TeV.

\subsection{SM Higgs portal}
\label{sec:SMHiggsportal}

\begin{figure}[t!]
  \centering
  \includegraphics[width=0.28\textwidth]{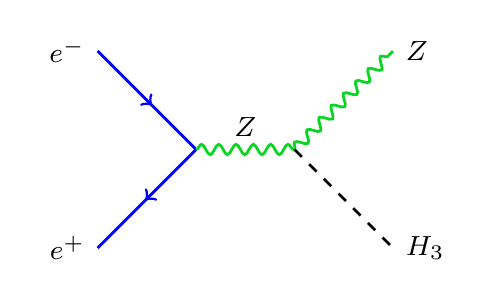}
  \caption{Feynman diagram for the production of $H_3$ at lepton colliders from mixing with the SM Higgs.}
  \label{fig:diagram3}
\end{figure}

\begin{figure}[t!]
  \centering
  \includegraphics[width=0.4\textwidth]{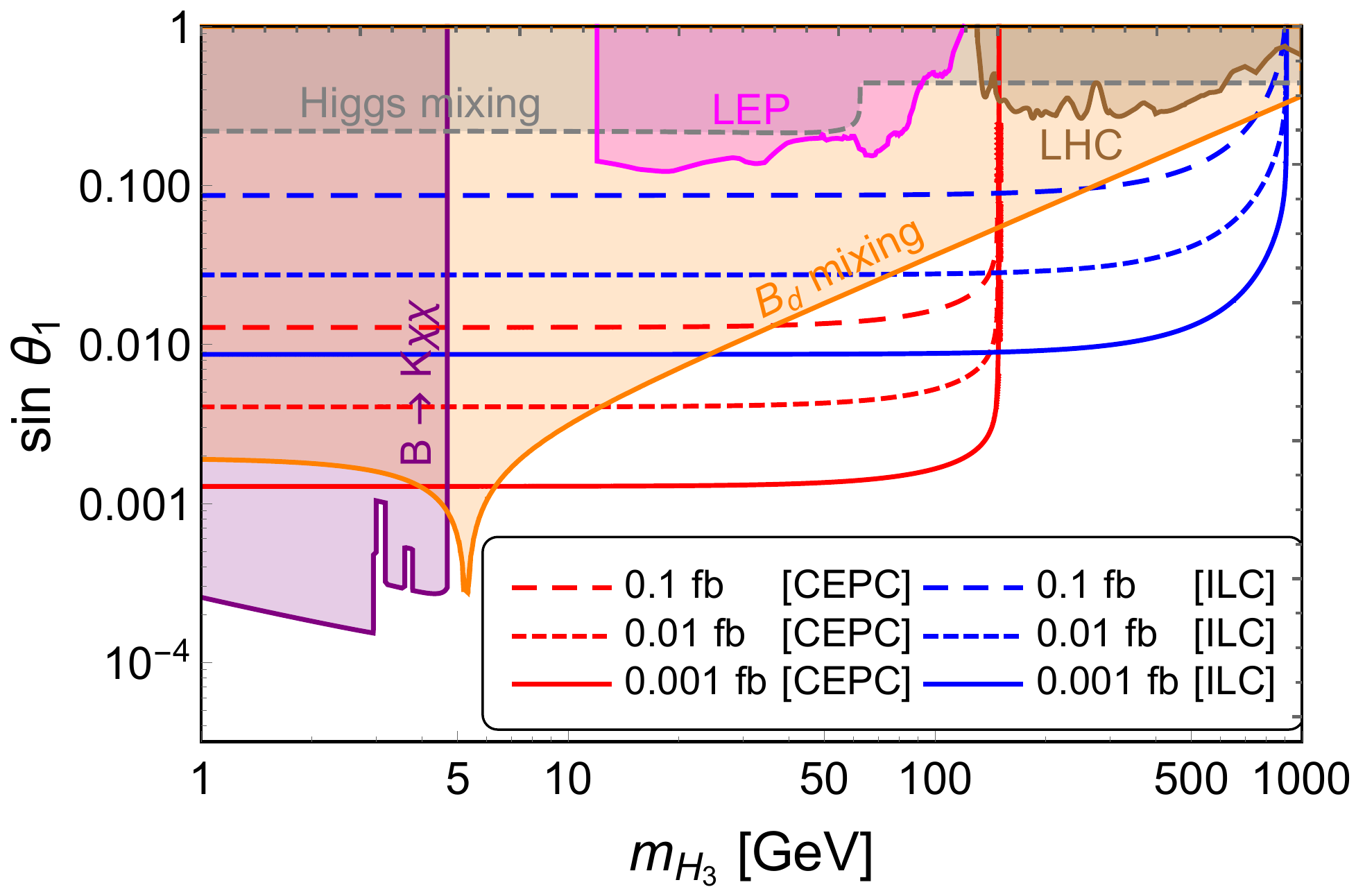}
  \caption{Experimental limits on the mixing angle $\sin\theta_1$ of $H_3$ with the SM Higgs {(with the mixing angle $\sin\theta_2$ set to zero)}, as function of the scalar mass $m_{H_3}$: the flavor-changing decays $B \to K \chi\chi$ (purple), the $B_d - \bar{B}_d$ oscillation data (orange)~\cite{Dev:2016vle, Dev:2017dui}, the Higgs precision measurements (gray)~\cite{Falkowski:2015iwa}, the direct searches of scalars at LEP (pink) and LHC (brown)~\cite{Dev:2017xry}. For the $B$ meson limits, the bidoublet VEV ratio $\xi  = \kappa'/\kappa$ is set to be $m_b/m_t$ as in Refs.~\cite{Dev:2016vle, Dev:2017dui}. All the shaded regions are excluded. The red and blue contours are respectively the production cross sections of 0.1 fb (long dashed), 0.01 fb (short-dashed) and 0.001 fb (solid) at CEPC 240 GeV and ILC 1 TeV, as functions of the scalar mass $m_{H_3}$ and mixing angle $\sin\theta_{1}$.}
  \label{fig:scalarmixing}
\end{figure}

At lepton colliders, the most important channel to produce the SM Higgs is through the so-called Higgstrahlung process, i.e. in association with a SM $Z$ boson. If $H_3$ mixes with the SM Higgs, then it could be produced from the diagram in Figure~\ref{fig:diagram3}, %\footnote{The Diagram in Fig.~\ref{fig:diagram3} could also induced from the $Z - Z_R$  mixing, which is however highly suppressed by the small $Z- Z_R$ mixing angle which is roughly $m_Z^2 / m_{Z_R}^2$, and is not considered here.}
with the amplitude proportional to the mixing angle $\sin\theta_1$. All the existing limits on the scalar mixing $\sin\theta_1$ are collected in Figure~\ref{fig:scalarmixing}. When $H_3$ is light, say lighter than the $B$ meson masses, the mixing angle $\sin\theta_1$ is tightly constrained by the rare flavor-changing $B$ meson decays like $B \to K \ell^+ \ell^-$. For heavier $H_3$, the most stringent limit is from the $B_d - \bar{B}_{d}$ oscillation, which originates from the FCNC couplings of the heavy doublet scalar $H_1$ to the SM quarks and its mixing with the SM Higgs, which is governed by the VEV ratio $\kappa^\prime/\kappa$ in the bidoublet sector. See more details in Refs.~\cite{Dev:2016vle, Dev:2017dui}. In the limit of $m_B \ll m_{H_3}$ the $B$ meson oscillation limit scales like $1/m_{H_3}^2$, as shown in Figure~\ref{fig:scalarmixing}. There are also constraints on the mixing angle from the precision Higgs data~\cite{Falkowski:2015iwa} and the direct searches of scalars at LEP and LHC~\cite{Dev:2017xry}, which are however weaker than that from $B_d$ oscillation data for the mass range $m_{H_3} < 1$ TeV.\footnote{In models where the FCNC couplings to the SM quarks are loop-induced, like a $U(1)_{B-L}$ gauge extension of the SM, the constraints from $B$ oscillation data would be much weaker and the limits from Higgs precision data and direct searches at LEP and LHC take over as the primary constraints for a heavy scalar~\cite{Dev:2017dui}.}

The total production cross section at the leading order can be easily obtained from that of the SM Higgs,
\begin{eqnarray}
\sigma (e^+ e^- \to Z H_3) & \ = \ &
\frac{G_F m_Z^4 \sin^2\theta_1}{96\pi s} \left( v_e^2 + a_e^2 \right) \nonumber \\
&& \times \lambda^{1/2} \frac{\lambda + 12 m_Z^2 / s}{(1- m_Z^2/s)^2} \,,
\end{eqnarray}
with $v_e = -1$, $a_e = -1 + 4 \sin\theta_w$, and
\begin{eqnarray}
\lambda \ = \ \left( 1 - \frac{m_{H_3}^2}{s} - \frac{m_Z^2}{s} \right)^2
- \frac{4 m_{H_3}^2 m_Z^2}{s^2} \,.
\end{eqnarray}
The production cross sections at CEPC 240 GeV and ILC 1 TeV are shown in Figure~\ref{fig:scalarmixing} respectively as the red and blue contours, with the values of 0.1 fb, 0.01 fb and 0.001 fb. Even if the flavor and direct search constraints are taken into consideration, for a wide range of $H_3$ mass, from 6 GeV to 149 GeV (75 GeV to 900 GeV), the production cross section is still larger than 0.01 fb at CEPC (ILC), excluding the vicinity of SM Higgs from roughly 120 GeV to 130 GeV, with a clear signal of $e^+ e^- \to Z H_3$ with $H_3 \to \gamma \gamma$ or $H_3$ decaying into a pair of charged leptons with different flavors, as implied by Figure~\ref{fig:BR_H3}.

%In the case of $U(1)_{B-L}$ model, the flavor-changing couplings arise at 1-loop level, and thus the limits from rare $B$ decays and oscillations are much weaker. For $m_{H_3} \gtrsim 10$ GeV, the most stringent limits are from the precision Higgs data and the direct searches at LEP and LHC, which allows a mixing angle $\sin\theta_1 > 0.1$. For a scalar $H_3$ heavier than 5 GeV, the cross section is always larger than 1 fb, as shown in Fig.~\ref{fig:production2}. In the $U(1)$ model, without the charged $W_R$ and scalar bosons, the decay of $H_3$ is SM-like, i.e. decays mostly into the bottom quark pairs if kinematically allowed, i.e. $e^+ e^- \to Z H_3 \to Z b \bar{b}$. The branching fraction into two photons ${\rm BR} (h \to b \bar{b})$ is loop suppressed.

\subsection{Neutrino portal}

\begin{figure}[t!]
  \centering
  \includegraphics[width=0.28\textwidth]{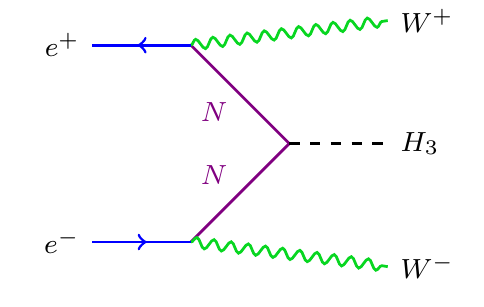}
  \caption{Feynman diagram for the production of $H_3$ at lepton colliders from heavy-light neutrino mixing.}
  \label{fig:diagram4}
\end{figure}

In the LRSM, the tiny neutrino masses are generated via the type-I seesaw mechanism (and/or type-II seesaw). With the Dirac and Majorana mass matrices $m_D$ and $M_N$, the heavy and light neutrinos could mix with each other, and the Feynman diagram in Figure~\ref{fig:diagram4} can lead to production of an $H_3$. For ${\cal O} (1)$ Yukawa couplings $f_{\alpha\beta}$, the heavy neutrino masses $M_N \simeq 2 f v_R$ are expected to be at the TeV scale (for few-TeV scale $v_R$), and their mixing with the active neutrino $V_{\nu N}\sim 10^{-6}$. However, with specific textures of the matrices $m_D$ and $M_N$, (some of) the heavy-light neutrino mixing angles might be sizable~\cite{Kersten:2007vk,Dev:2013oxa}, or one or two of the RHNs masses could be much smaller than the RH scale, say at the GeV scale~\cite{Antusch:2016vyf, Helo:2013esa, Cottin:2018kmq}. The production cross section via the neutrino portal is, however, suppressed by both the heavy-light neutrino mixing $V^4_{\nu \, N}$ and the small Yukawa couplings $f^2 = (M_N / 2 v_R)^2$ in the LRSM when the RHN is light; therefore there is no hope to see the neutral scalar $H_3$ or to set any limits on the neutrino mixing by producing off-shell RHNs as in Figure~\ref{fig:diagram4}.

%If $H_3$ are not responsible for the RHN mass generation, or in other words the couplings of $H_3$ to the RHNs are not directly related to the RHN masses, then we could

%which is not in conflict with any basic principles in the LR framework. In addition, in the LRSMs, the RHNs decays mostly into a charged lepton and an (off-shell) $W_R$ boson which decays further into hadron jets. When the RHNs are sufficiently light, the three-body decay length $b c \tau_N^0$ might be large enough to produce displaced signals.
%For the sake of simplicity, we assume one of the RHN is light, and its mixing with the three active neutrinos are free parameters.

\subsection{Heavy scalar portal}
\label{sec:heavyscalar}

The $h_{\alpha\beta}$ couplings in Eq.~\eqref{eqn:Lagrangian} will induce the on-shell production processes $e^+ e^- \to (\gamma/Z) H_3$, $e\gamma \to \ell H_3$ and $e^+ e^- \to (\ell_\alpha^+ \ell_\beta^- / \nu_\alpha \bar\nu_\beta) H_3$ as well as the the off-shell processes $e^+ e^- \to \ell_\alpha^+ \ell_\beta^-$~\cite{Dev:2017ftk}, as aforementioned and detailed below.

\subsubsection{On-shell production}

\begin{figure}[t!]
  \centering
  \hspace{-50pt}
  \begin{subfigure}[b]{0.14\textwidth}
  \includegraphics[height=\textwidth]{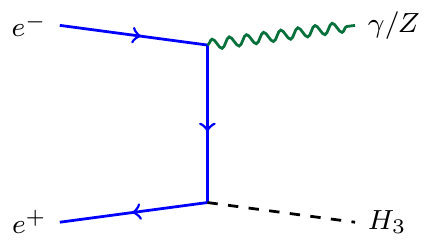}
  \caption{}
  \label{fig:9a}
  \end{subfigure} \hspace{50pt}
  \begin{subfigure}[b]{0.14\textwidth}
  \includegraphics[height=\textwidth]{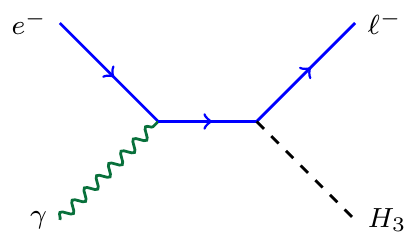}
  \caption{}
  \label{fig:9b}
  \end{subfigure}
  \caption{Feynman diagrams for the on-shell production $e^+ e^- \to (\gamma/Z)H_3$ (a) and $e^\pm \gamma \to \ell^\pm H_3$ (b) from the (LFV) couplings $h_{\alpha\beta}$ in Eq.~(\ref{eqn:Lagrangian}).}
  \label{fig:diagram5}
\end{figure}

\begin{figure*}[t!]
  \centering
  \includegraphics[width=0.4\textwidth]{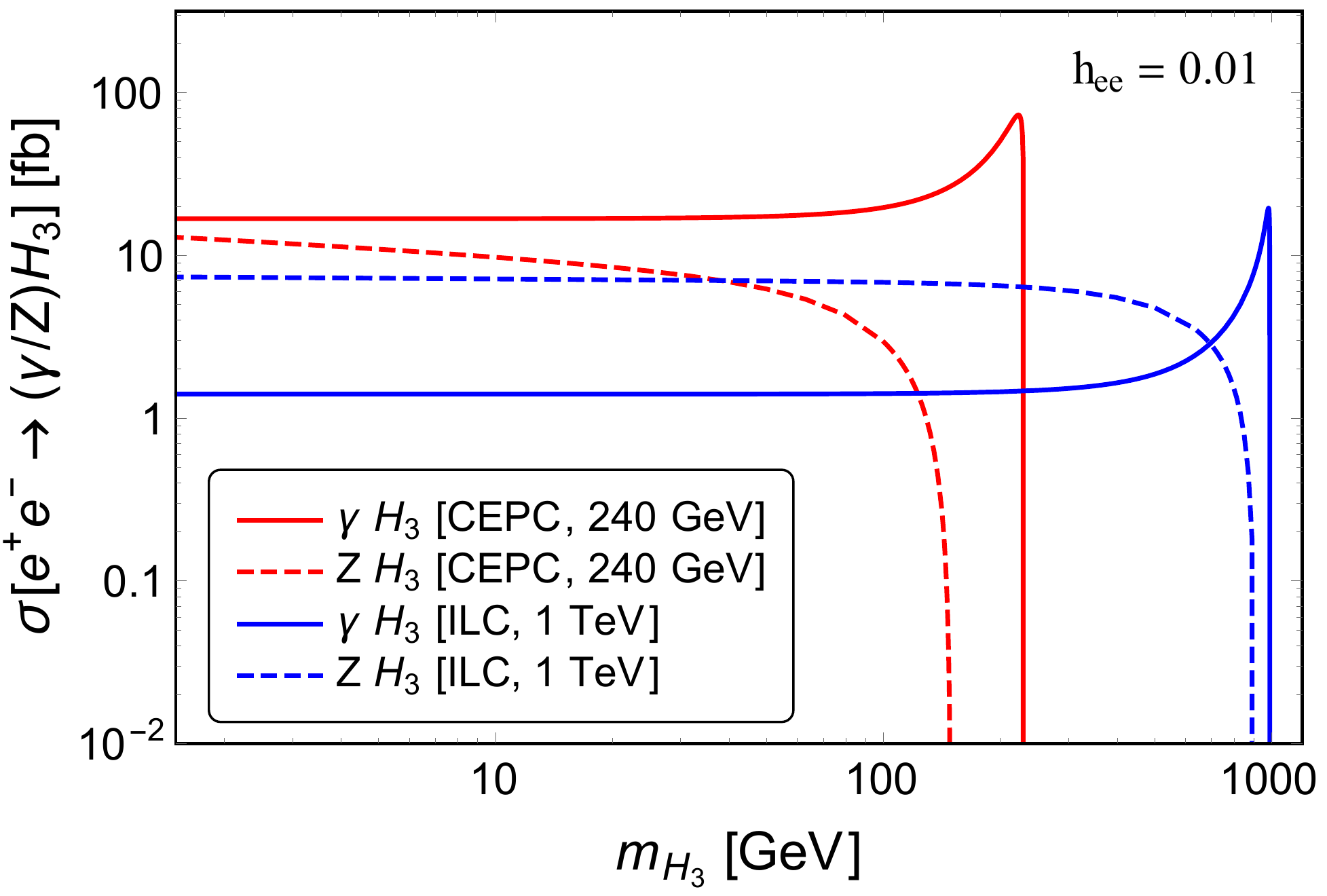}
  \includegraphics[width=0.4\textwidth]{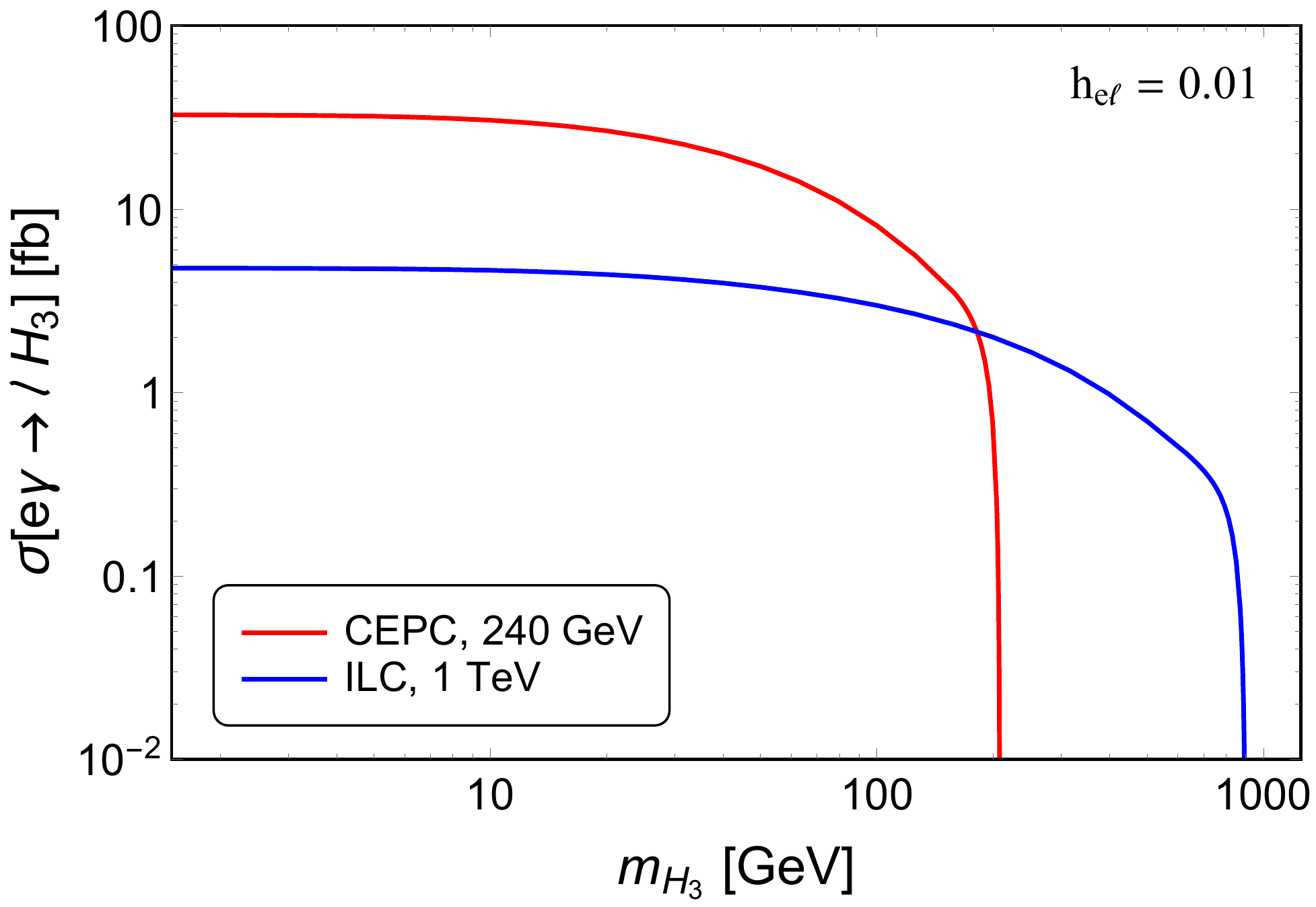}
  \caption{Production cross sections of $H_3$ in the processes $e^+ e^- \to (\gamma/Z) H_3$ (left) and $e^\pm \gamma \to \ell^\pm H_3$ (right) at CEPC $240$ GeV (red) and ILC 1 TeV (blue), as functions of $m_{H_3}$, with the Yukawa coupling $h_{ee} = 0.01$ (left) or $h_{e\ell} = 0.01$ (right). {For other values of $h_{ee}$ or $h_{e\ell}$, the cross section can be simply rescaled by a factor of $(h_{ee}/0.01)^2$ (left) or $(h_{e\ell}/0.01)^2$ (right), though $h_{e\ell} = 0.01$ has been excluded for some ranges of $m_{H_3}$ (see Figure~\ref{fig:H3:prospect1}).} See Figure~\ref{fig:diagram5} for the diagrams.}
  \label{fig:production4}
\end{figure*}

%\begin{figure}[t!]
%  \centering
%  \includegraphics[width=0.55\textwidth]{H3_heavy_scalar_1b.pdf}
%  \caption{Production cross section of $H_3$ in the process $e^\pm \gamma \to e^\pm H_3$ in Fig.~\ref{fig:diagram5b} at CEPC with $\sqrt{s} = 240$ GeV (red) and ILC with 1 TeV (blue), as functions of $m_{H_3}$, with the Yukawa coupling $h_{ee} = 1$. }
%  \label{fig:production4b}
%\end{figure}

The diagram for $e^+ e^- \to (\gamma/Z) H_3$ is shown in Figure~\ref{fig:9a}, where we have neglected the mixing of $H_3$ with the SM Higgs and the loop-induced coupling of $H_3 \gamma\gamma$. The amplitude is proportional to the effective coupling $h_{ee}$ in Eq.~(\ref{eqn:Lagrangian}). For the illustration purpose, we show in the left panel of Figure~\ref{fig:production4} the cross sections of $\sigma (e^+ e^- \to (\gamma/Z) H_3)$ at the CEPC 240 GeV and ILC 1 TeV, as function of the scalar mass $m_{H_3}$, for the value of $h_{ee} = 0.01$. Though the production of $H_3$ proceeds in this channel via the lepton flavor conserving coupling $h_{ee}$, the scalar $H_3$ could decay into two charged leptons with different flavors, i.e. $H_3 \to \ell_\alpha^\pm \ell_\beta^\mp$ ($\alpha \neq \beta$), thus generating clear LFV signals.

The neutral scalar $H_3$ could also be produced in the process
\begin{eqnarray}
e^\pm \gamma \ \to \ \ell^\pm H_3 \;\; (\ell = e,\, \mu,\, \tau) \,,
\end{eqnarray}
with the initial photon being the laser photon as in Section~\ref{sec:gaugeportal}, and the flavor conversion ``$e \to \ell$'' induced by the coupling $h_{e\ell}$. The Feynman diagrams are shown in Figure~\ref{fig:9b}, and the corresponding cross sections at CEPC and ILC are presented in the right panel of Figure~\ref{fig:production4}, again with the value of $h_{e\ell} = 0.01$. Obviously the cross sections in the $e^+ e^-$ and $e\gamma$ channels are comparable to each other in a larger region of parameter space, and largely complementary in probing the Yukawa couplings of $H_3$ at lepton colliders.

%\subsubsection{On-shell production $e^+ e^-,\,\gamma\gamma \to (\ell_\alpha^+ \ell_\beta^- / \nu_\alpha \bar\nu_\beta) H_3$}

\begin{figure*}[!t]
  \centering
  \includegraphics[width=0.75\textwidth]{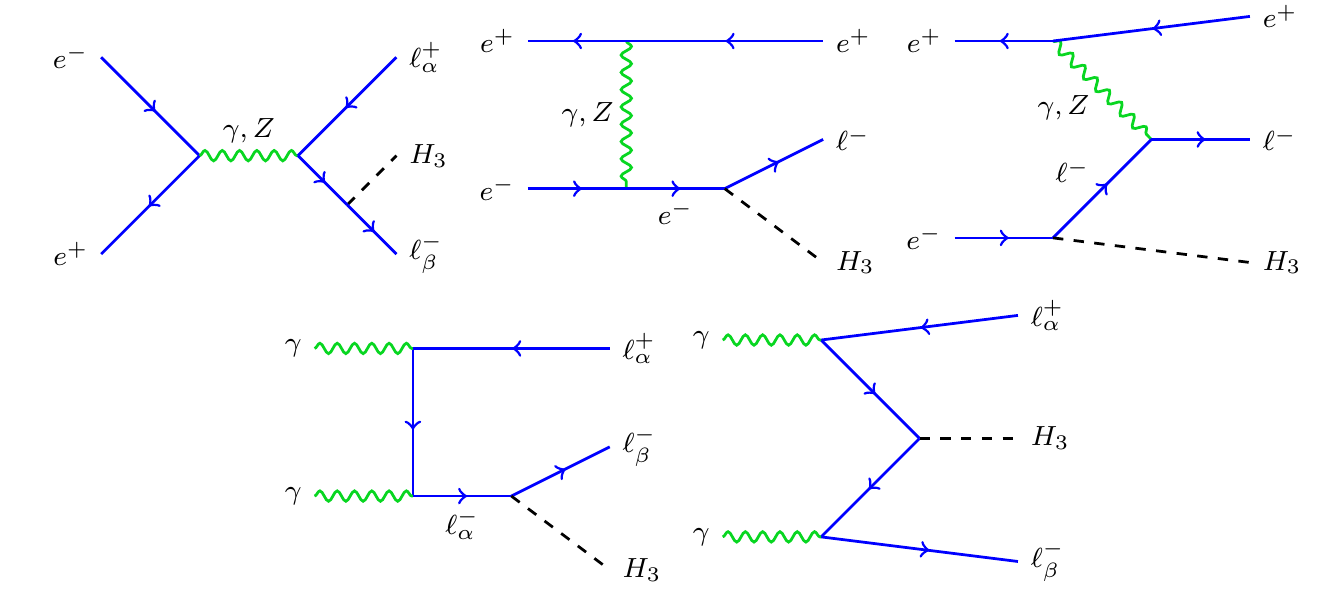}
  \caption{Feynman diagrams for the on-shell production of $H_3$ via $e^+ e^- \to \ell_\alpha^+ \ell_\beta^- H_3$ (upper) and $\gamma\gamma \to \ell_\alpha^\pm \ell_\beta^\mp H_3$ (lower) at lepton colliders from the (LFV) couplings $h_{\alpha\beta}$ in Eq.~(\ref{eqn:Lagrangian}).}
  \label{fig:diagram6}
\end{figure*}

\begin{figure*}[t!]
  \centering
  \includegraphics[width=0.4\textwidth]{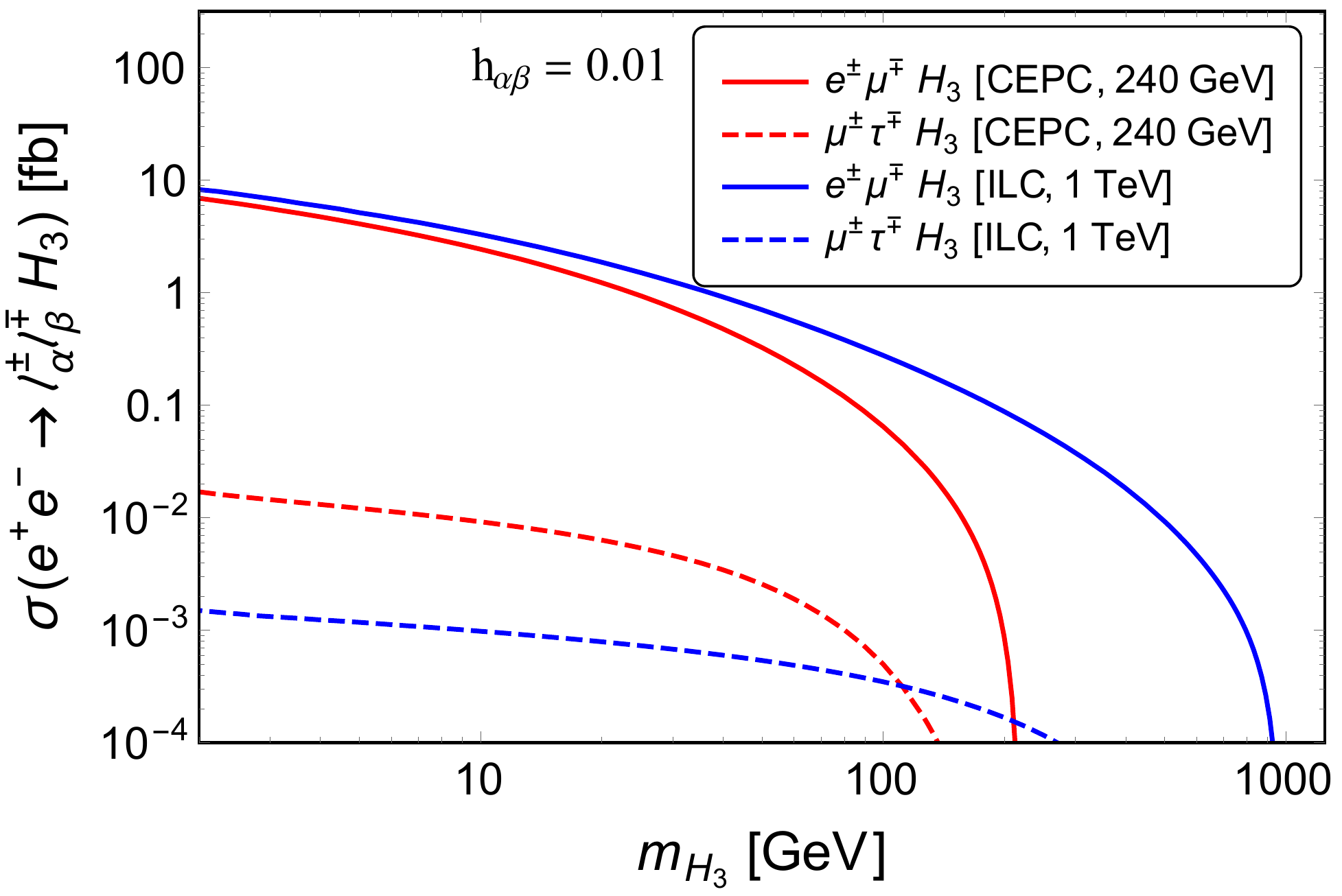}
  \includegraphics[width=0.4\textwidth]{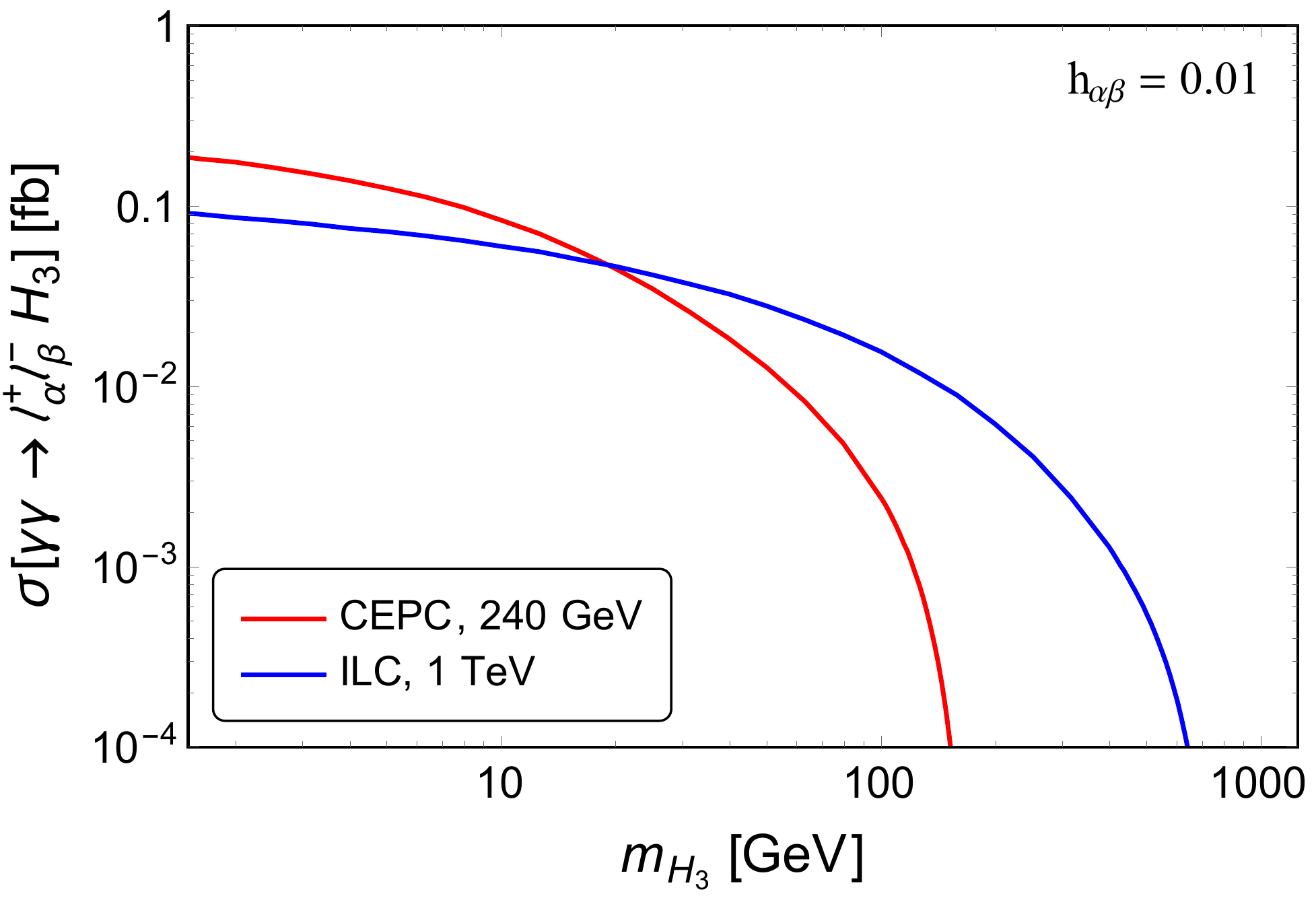}
  \caption{Production cross sections of $H_3$ in the processes $e^+ e^- \to \ell_\alpha^\pm \ell_\beta^\mp H_3$ (with $\ell_\alpha \ell_\beta = e\mu$, $\mu\tau$, left) and $\gamma\gamma \to \ell_\alpha^+ \ell_\beta^- H_3$ (right) at CEPC $240$ GeV (red) and ILC 1 TeV (blue), as functions of $m_{H_3}$. The Yukawa coupling involved $h_{\alpha\beta}$ are set to be 0.01 and all others irrelevant are zero.  {For other values of $h_{\alpha\beta}$, the cross section can be simply rescaled by a factor of $(h_{\alpha\beta}/0.01)^2$, though $h_{\alpha\beta} = 0.01$ has been excluded for some ranges of $m_{H_3}$ (see Figure~\ref{fig:H3:prospect1}).} See Figure~\ref{fig:diagram6} for the diagrams. }
  \label{fig:production5}
\end{figure*}

With the Yukawa couplings $h_{\alpha\beta}$ in Eq.~(\ref{eqn:Lagrangian}), we can have the on-shell LFV processes of the form~\cite{Dev:2017ftk}
\begin{eqnarray}
\label{eqn:LFV:H3}
e^+ e^- \ \to \ \ell^\pm_\alpha \ell^\mp_\beta H_3 \,, \quad
\alpha \neq \beta \,,
\end{eqnarray}
with the representative Feynman diagrams shown in the upper panels of Figure~\ref{fig:diagram6}. The neutral scalar $H_3$ can be emitted from any of the initial or final fermion lines in the Drell-Yan process $e^+ e^- \to \ell^+ \ell^-$ and hence potentially change the lepton flavors. The total production cross sections for $\ell_\alpha \ell_\beta = e\mu$ and $\mu\tau$ at CEPC 240 GeV and ILC 1 TeV are collected in the left panel of Figure~\ref{fig:production5}, with the nominal cuts of $p_T > 10$ GeV on the leptons. In a large parameter space the process $ee \to ee H_3$ is dominated by $ee \to Z H_3$ with the subsequent decay $Z \to ee$.
The cross sections for $e\tau$ are almost the same as that for the $e\mu$, with subleading corrections from the muon and tauon mass difference, and are thus not shown in the plot. Only some of the diagrams in Figure~\ref{fig:diagram6}, e.g. the first one, apply to the process $ee \to \mu\tau + H_3$, and therefore the cross sections for $\mu\tau$ channel is much smaller than those for $e\mu$ and $e\tau$, as clearly seen in Figure~\ref{fig:production5}. The cross sections with the final states $\mu\mu$ and $\tau\tau$ are approximately the same as that for $\mu\tau$.

$H_3$ could also be emitted from the $\gamma\gamma$ processes at lepton colliders, i.e.
\begin{eqnarray}
\gamma\gamma \ \to \ \ell_\alpha^\pm \ell_\beta^\mp H_3 \,,
\end{eqnarray}
as shown in the lower panels of Figure~\ref{fig:diagram6}, and induce LFV signals if $\alpha \neq \beta$. The cross sections at CEPC and ILC are given in the right panel of Figure~\ref{fig:production5}, with the same cuts as above on the leptons. Note that for the final state with different flavors, e.g. $e^\pm \mu^\mp$, the cross section in Figure~\ref{fig:production5} have to be multiplied by a factor of 2, to account for the two different flavor and charge combinations of $e^+ \mu^-$ and $e^- \mu^+$. The $\gamma\gamma$ processes could provide complementary prospects to those in Eq.~(\ref{eqn:LFV:H3}) in searching for the LFV signals at future lepton colliders.

\begin{figure}[!t]
  \centering
  \includegraphics[width=0.28\textwidth]{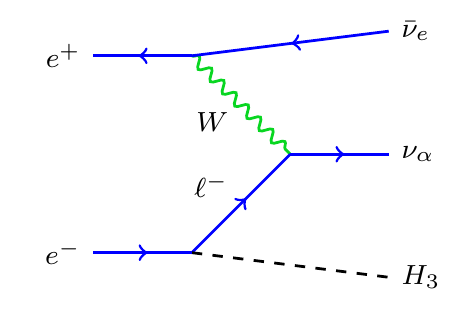} \vspace{-10pt}
  \caption{Feynman diagram for the on-shell production of $H_3$ via $e^+ e^- \to \nu_\alpha \bar\nu_e H_3$ at lepton colliders from the (LFV) couplings $h_{\alpha\beta}$ in Eq.~(\ref{eqn:Lagrangian}).}
  \label{fig:diagram7}
\end{figure}

\begin{figure}[t!]
  \centering
  \includegraphics[width=0.4\textwidth]{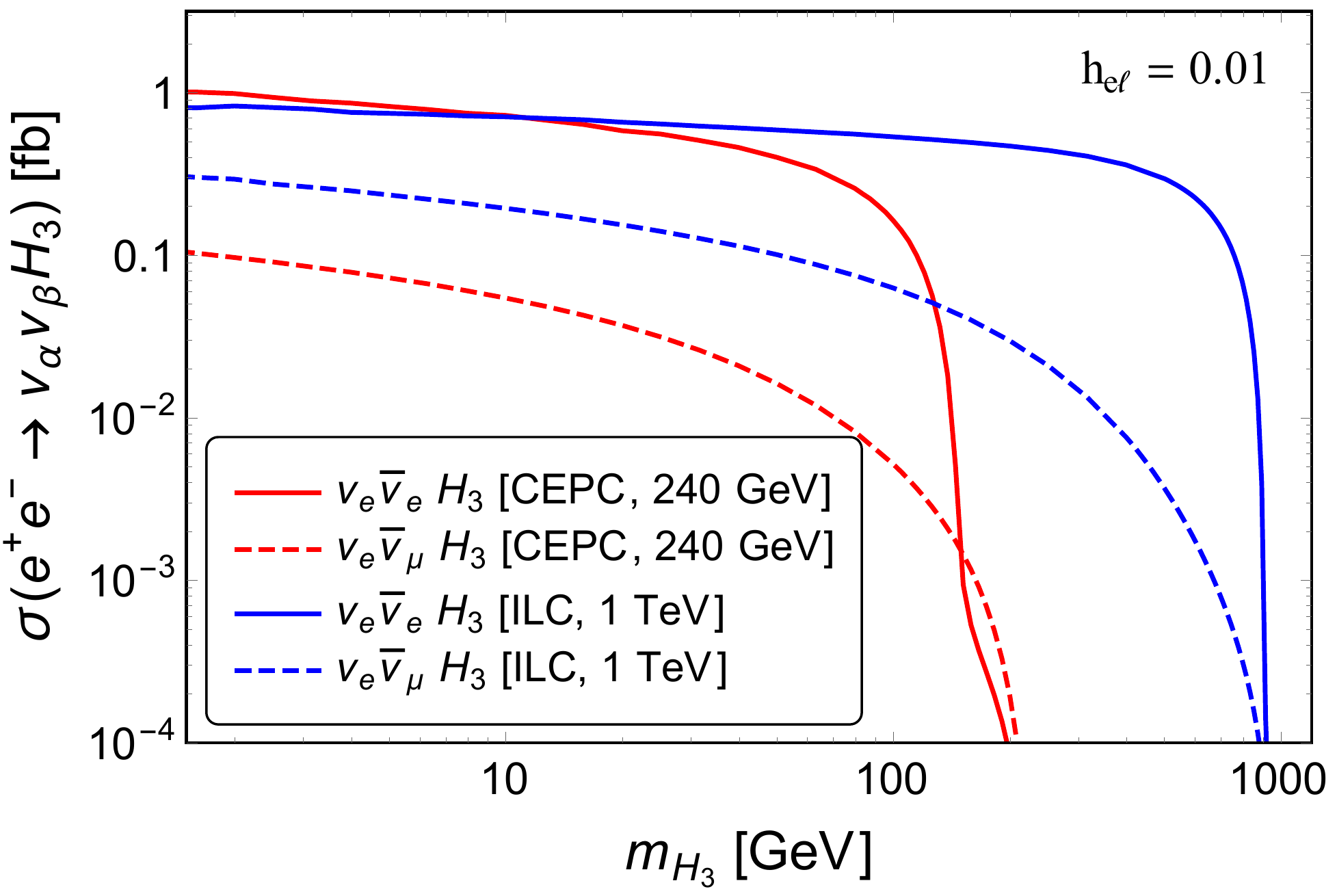}
  \caption{Production cross sections of $H_3$ in the process of $e^+ e^- \to \nu_\alpha \bar\nu_\beta H_3$ with $\nu_\alpha \bar\nu_\beta = \nu_e \bar\nu_e$ (solid) $\nu_e \bar\nu_\mu$ (dashed) at CEPC 240 GeV (red) and ILC 1 TeV (blue), as functions of $m_{H_3}$.  The Yukawa coupling involved ($h_{ee}$ or $h_{e\mu}$) are set to be 0.01 and all others irrelevant are zero.  {For other values of $h_{e\ell}$, the cross section can be simply rescaled by a factor of  $(h_{e\ell}/0.01)^2$, though $h_{e\ell} = 0.01$ has been excluded for some ranges of $m_{H_3}$ (see Figure~\ref{fig:H3:prospect1}).} See Figure~\ref{fig:diagram7} for the diagram. }
  \label{fig:production6}
\end{figure}

The neutral scalar $H_3$ could also be produced in the processes
\begin{eqnarray}
e^+ e^- \ \to \ \nu_\alpha \bar\nu_\beta + H_3 \,,
\end{eqnarray}
as shown in Figure~\ref{fig:diagram7}, with the flavors $\alpha = e$ or $\beta = e$ being induced from the LFV couplings $h_{\alpha\beta}$ in Eq.~(\ref{eqn:Lagrangian}) if $\alpha \neq \beta$. The production cross section for the cases of $\nu_e \bar\nu_e$ and $\nu_e \bar\nu_\mu$ (including also the contribution of $\nu_\mu \bar\nu_e$) at CEPC 240 GeV and ILC 1 TeV are presented in Figure~\ref{fig:production6}. It should be noted that for the case of $\nu_e \bar\nu_e$, there is extra contribution from the process $e^+ e^- \to Z H_3$ with $Z \to \nu_e \bar\nu_e$, which can not be distinguished from the $W$-mediated diagram in Figure~\ref{fig:diagram7}, as a result of the invisible nature of neutrinos. The solid red and blue lines in Figure~\ref{fig:production6} combine both the contributions. If the vertex $H_3 \bar{e} \ell$ is flavor-violating, i.e. $\ell = \mu,\, \tau$, then we have only the diagram in Figure~\ref{fig:diagram7}, and the cross sections are comparatively smaller, as indicated by the dashed lines in Figure~\ref{fig:production6}. Neglecting the small charged lepton masses, the cross sections for $\nu_e \bar\nu_\tau$ are almost the same as that for $\nu_e \bar\nu_\mu$, and are thus not shown in Figure~\ref{fig:production6}.

\subsubsection{Off-shell production}

\begin{figure}[t!]
  \centering
  \includegraphics[width=0.45\textwidth]{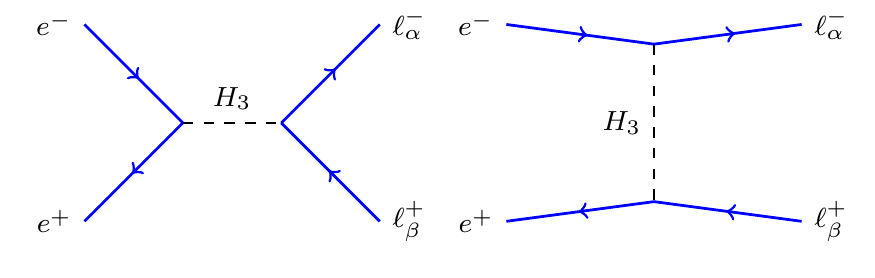}
  \caption{Feynman diagrams for the off-shell production of $H_3$ via the process $e^+ e^- \to \ell_\alpha^\pm \ell_\beta^\mp$ at lepton colliders from the (LFV) couplings $h_{\alpha\beta}$ in Eq.~(\ref{eqn:Lagrangian}).}
  \label{fig:diagram8}
\end{figure}

\begin{figure*}[t!]
  \centering
  \includegraphics[width=0.4\textwidth]{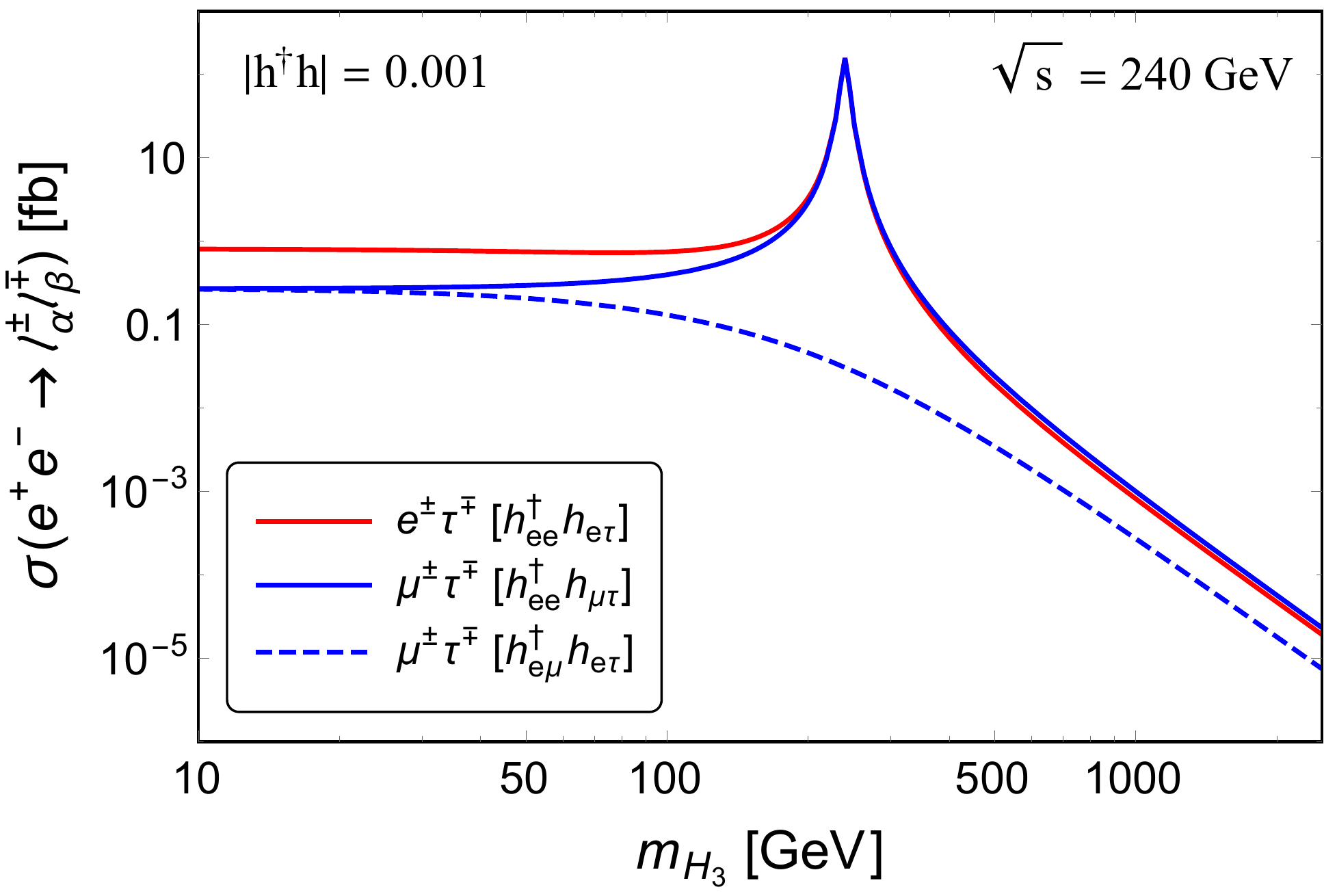}
  \includegraphics[width=0.4\textwidth]{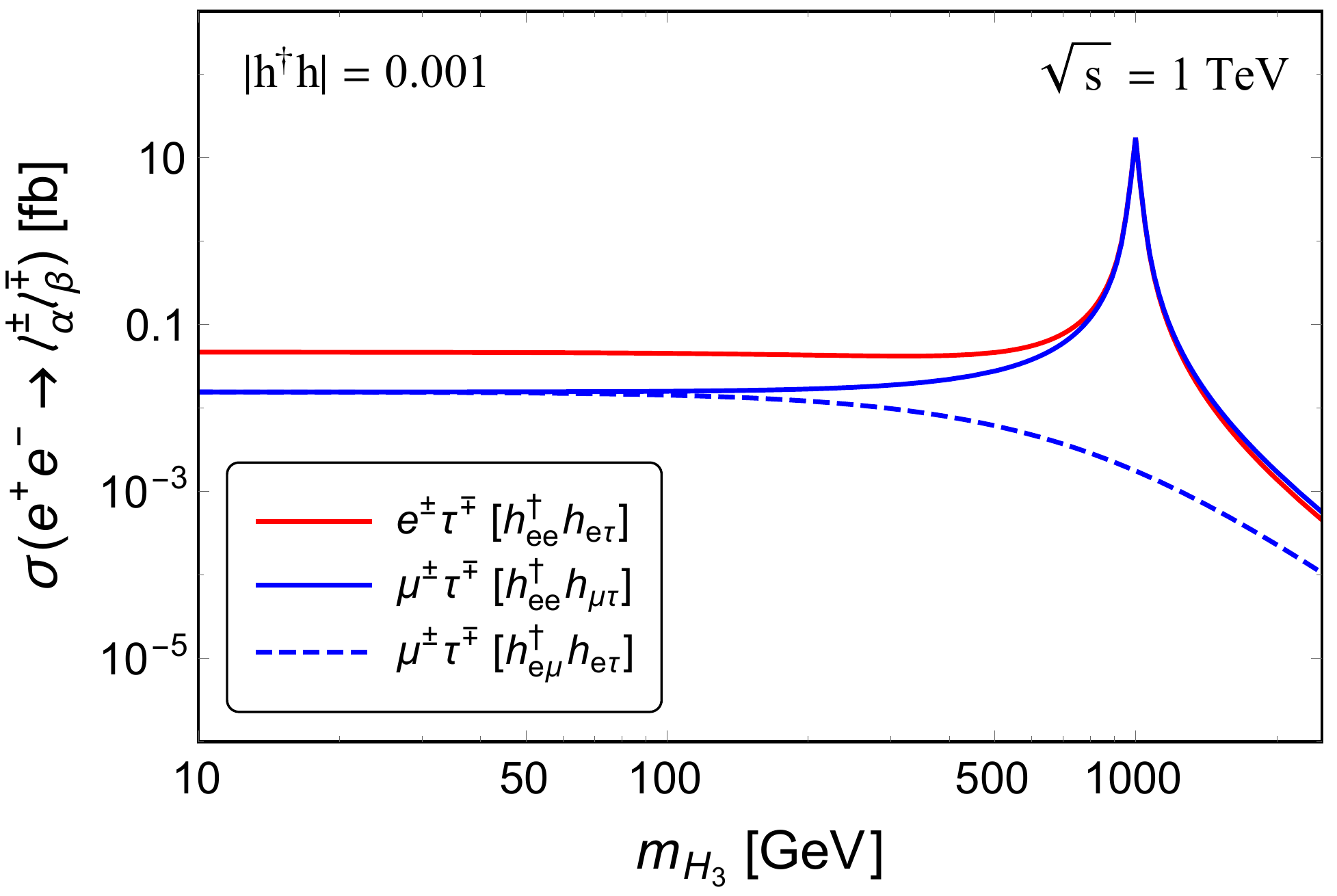}
  \caption{Cross sections for the off-shell production of $H_3$ in the process $e^+ e^- \to e^\pm \tau^\mp$, $\mu^\pm \tau^\mp$ from respectively the couplings $h_{ee}^\dagger h_{e\tau}$ (red), $h_{ee}^\dagger h_{\mu\tau}$ (blue, solid) and $h_{e\mu}^\dagger h_{e\tau}$ (blue, dashed) at CEPC 240 GeV (left) and ILC 1 TeV (right), as functions of $m_{H_3}$. The Yukawa coupling involved $|h^\dagger h|$ are set to be $10^{-3}$ and all others irrelevant are zero.  {For other values of $|h^\dagger h|$, the cross section can be simply rescaled by a factor of $(|h^\dagger h|/10^{-3})^2$, though $|h^\dagger h| = 10^{-3}$ has been excluded for some ranges of $m_{H_3}$ (see Figure~\ref{fig:H3:prospect2}).} See Figure~\ref{fig:diagram8} for the diagrams. }
  \label{fig:production7}
\end{figure*}

As studied in Ref.~\cite{Dev:2017ftk}, even if the neutral scalar $H_3$ is heavier than the center-of-mass energy, the LFV signals could still be produced from an off-shell $H_3$, i.e.
\begin{eqnarray}
e^+ e^- \ \to \ \ell^\pm_\alpha \ell^\mp_\beta \,,
\end{eqnarray}
%and induce the LFV signals if $\alpha \neq \beta$,
with the diagram presented in Figure~\ref{fig:diagram8}. This could occur in both the $s$ and $t$ channels, depending on different combinations of the couplings $h_{\alpha\beta}$. For instance, the process $e^+ e^- \to \mu^\pm \tau^\mp$ are respectively proportional to the couplings $|h_{ee}^\dagger h_{\mu\tau}|$ and $|h_{e\mu}^\dagger h_{e\tau}|$ in the $s$ and $t$ channels. The production cross sections for $e\tau$ and $\mu\tau$ (in both the $s$ and $t$-channels) are presented in Figure~\ref{fig:production7}, for the center-of-mass energies of 240 GeV and 1 TeV at CEPC and ILC, with the relevant couplings $|h^\dagger h|=0.001$. In the $s$-channel $H_3$ could be produced on-shell if the colliding energy $\sqrt{s} \simeq m_{H_3}$, as clearly shown in the plots, where we have set explicitly the widths to be 10 GeV and 30 GeV for CEPC and ILC, for the sake of concreteness.
%which are constrained respectively by the decays $\tau^- \to \mu^- e^+ e^-$ and $\tau^- \to \mu^+ e^- e^-$.

\subsection{Prospects at lepton colliders}
In this section, we compare the future prospects of the neutral scalar search at lepton colliders in light of the existing constraints from both low and high-energy sectors.

\subsubsection{Current flavor constraints}

The direct searches of neutral scalars have been performed at LEP~\cite{Barate:2003sz} and LHC, e.g. when they decay into two SM fermions~\cite{ATLAS:2016pyq}, gauge bosons~\cite{Chatrchyan:2013mxa, Khachatryan:2015cwa, Aad:2015kna, Aaboud:2016okv, Aad:2014ioa, ATLAS:2016eeo, Khachatryan:2016yec}, or di-Higgs~\cite{Khachatryan:2015yea, Khachatryan:2016sey, ATLAS:2016ixk}, as already mentioned in Section~\ref{sec:SMHiggsportal}. Here we summarize all the available flavor constraints on the Yukawa couplings $h_{\alpha\beta}$ in Eq.~(\ref{eqn:Lagrangian})~\cite{Dev:2017xry, Dev:2017ftk}, including those from the rare decays $\ell_\alpha \to \ell_\beta \ell_\gamma \ell_\delta$, $\ell_\alpha \to \ell_\beta \gamma$, the electron and muon $g-2$, the muonium-antimuonium oscillation, and the LEP $e^+ e^- \to \ell^+ \ell^-$ data~\cite{Abdallah:2005ph}. %\footnote{If the couplings $h_{\alpha\beta}$ in Eq.~(\ref{eqn:LYukawa:H3}) are complex, then we have also the 2-loop contribution to the electron electric dipole moment (EDM), mediated by the neutral scalar $H_3$ and the $W_R$ boson. For simplicity, we have assumed that these couplings are real, so that we can ignore the EDM constraints.}
All the current experimental data and constraints on the couplings $h_{\alpha\beta}$ are collected in Table~\ref{tab:limits:H3}.

In the limit of $m_{H_3} \gg m_\mu \gg m_e$, the partial decay width of $\mu \to eee$ is given by~\cite{Sher:1991km}
\begin{eqnarray}
\label{eqn:mudecay}
\Gamma (\mu^- \to e^+ e^- e^-) \ \simeq \
\frac{1}{\delta}
\frac{|h_{ee}^\dagger h_{e\mu}|^2 \, m_\mu^5}{3072 \pi^3 m_{H_3}^4} \,,
\end{eqnarray}
with the symmetry factor $\delta = 2$ accounting for the identical particles in the final state. The calculations of the widths for the LFV three-body tau decays such as $\Gamma (\tau^- \to e^+ e^- e^-)$ and $\Gamma (\tau^- \to \mu^- e^+ e^-)$ are quite similar, with the couplings and symmetry factor in Eq.~(\ref{eqn:mudecay}) appropriately replaced.

The presence of couplings $h_{\alpha\beta}$ could induce the effective Wilson operators that couple the photon to electron and muon at 1-loop level
\begin{eqnarray}
{\cal O}_{\mu e \gamma,\, (L,R)} \ = \
\frac{e}{8\pi^2} m_\mu
\left( \bar{e} \sigma^{\alpha\beta} P_{L,R} \mu \right)
F_{\alpha\beta}
\end{eqnarray}
with the effective coefficients, in the limit of $m_{H_3} \gg m_\tau \gg m_\mu \gg m_e$~\cite{Harnik:2012pb,Blankenburg:2012ex},
\begin{eqnarray}
\label{eqn:coeff}
&& c_{L}^{(\mu e \gamma)} \ = \
c_{R}^{(\mu e \gamma)} \nonumber \\
&&  \ \simeq \ \frac{1}{8 m_{H_3}^2}
\left[ \frac13 h_{ee}^\dagger h_{e\mu} +
h_{e\mu}^\dagger h_{\mu\mu}
\left( 2 \log \left( \frac{m_{H_3}^2}{m_\mu^2} \right) - 3 \right)
\right. \nonumber \\
&& \qquad\quad \left.  + \frac{m_\tau}{m_\mu} h_{e\tau}^\dagger h_{\tau\mu}
\left( 2 \log \left( \frac{m_{H_3}^2}{m_\tau^2} \right) - 3 \right) \right] \,.
\end{eqnarray}
As expected, in the large $m_{H_3}$ limit, the effective coefficients are suppressed by $m_{H_3}^2$. Then the partial decay width
\begin{eqnarray}
\Gamma (\mu \to e \gamma)  \ = \
\frac{\alpha_{\rm EM} m_\mu^5}{64 \pi^4}
\left( |c_L|^2 + |c_R|^2 \right) \,.
\label{eqn:mu2decay}
\end{eqnarray}
The calculations of $\Gamma (\tau \to e \gamma)$ and $\Gamma (\tau \to \mu \gamma)$ are quite similar, with $m_\mu^5$ in Eq.~(\ref{eqn:mu2decay}) replaced by $m_\tau^5$ and the corresponding Wilson coefficients
\begin{eqnarray}
\label{eqn:coeff2}
c_{L}^{(\tau e \gamma)} \ = \
c_{R}^{(\tau e \gamma)} \ &\simeq& \
\frac{1}{8 m_{H_3}^2}
\left[ \frac13 h_{ee}^\dagger h_{e\tau} +
\frac13 h_{e\mu}^\dagger h_{\mu\tau} \right. \nonumber \\
&& \left. + h_{e\tau}^\dagger h_{\tau\tau}
\left( 2 \log \left( \frac{m_{H_3}^2}{m_\tau^2} \right) - 3 \right) \right] \,, \\
\label{eqn:coeff3}
c_{L}^{(\tau \mu \gamma)} \ = \
c_{R}^{(\tau \mu \gamma)} \ &\simeq& \
\frac{1}{8 m_{H}^2}
\left[ \frac13 h_{\mu e}^\dagger h_{e\tau} +
\frac13 h_{\mu\mu}^\dagger h_{\mu\tau} \right. \nonumber \\
&& \left. + h_{\mu\tau}^\dagger h_{\tau\tau}
\left( 2 \log \left( \frac{m_{H_3}^2}{m_\tau^2} \right) - 3 \right) \right] \,.
\end{eqnarray}
%in the limit of $m_{H} \gg m_\tau \gg m_{e,\,\mu}$ one has only to replace the couplings in Eq.~(\ref{eqn:coeff}) by $h_{ee}^\dagger h_{e\tau}$ and $h_{e\mu}^\dagger h_{e\tau}$.

%The Feynman diagram for the magnetic dipole moment of electron is shown in Fig.~\ref{fig:diagram3}, in presence of the coupling $h_{e\mu}$.
The $H_3 - \mu$ and $H_3 - \tau$ loops contribute to the electron magnetic dipole moment, given by~\cite{Lindner:2016bgg}
\begin{eqnarray}
\label{eqn:g-2:H3}
\Delta a_e \ \simeq \ \sum_{\ell = \mu,\tau}
\frac{h_{e\ell}^2 m_e m_\ell}{16\pi^2 m_{H_3}^2}
\left[ 2 \log \left( \frac{m_{H_3}^2}{m_\ell^2} \right) - 3 \right] \,.
\end{eqnarray}
It is simple to get the extra contribution for the muon $g-2$, which is dominated by the coupling $h_{\mu\tau}$.

%by changing accordingly the flavors of the fermion lines and the couplings for the vertices.

%If the incoming electron  is replaced by $\tau$ and the internal $\mu$ line is replaced by $e$, then we have the LFV decay $\tau \to e\gamma$, which depends on the couplings $|h_{ee}^\dagger h_{e\tau}|$ (cf. Table I). In an analogous way we can have the diagram for $\tau \to \mu \gamma$ which has an electron mediator and depends on $|h_{e\mu}^\dagger h_{e\tau}|$.

%The Feynman diagrams for the $H$-induced muonium oscillation are presented in Fig.~\ref{fig:diagram2}.
The muonium-anti-muonium oscillation probability is given by~\cite{Clark:2003tv}
\begin{eqnarray}
{\cal P} \ = \
\frac{2 |\Delta M|^2}{\Gamma_\mu^2 + 4|\Delta M|^2 }
\label{eq:prob}
\end{eqnarray}
with the $H_3$-induced mass splitting
\begin{eqnarray}
|\Delta M|  \ = \
\frac{2 \alpha_{\rm EM}^3 |h_{e\mu}|^2 \mu^3}{\pi m_{H_3}^2} \,,
\end{eqnarray}
with $\mu = m_e m_\mu / (m_e + m_\mu)$ the effective mass. In the denominator of Eq.~\eqref{eq:prob}, the muon decay width $\Gamma_\mu$ is typically much larger than $2|\Delta M|$ for the range of mass and couplings we are interested in; therefore, ${\cal P} \propto |h_{e\mu}|^4/m_{H_3}^4$.

The di-muonium (not yet found experimentally), i.e. the bound state ${\rm Dm} \equiv (\mu^+ \mu^-)$, could also be affected by the LFV couplings $h_{\alpha\beta}$ in Eq.~(\ref{eqn:Lagrangian}), e.g. a $t$-channel $H_3$ could contribute to the decay ${\rm Dm} \to e^+ e^-$.\footnote{If $m_{H_3} \lesssim m_\mu$, then Dm could also decay via ${\rm Dm} \to H_3 H_3$, which would turn out to be invisible if $|h_{ee}|$ is sufficiently small such that the decay length of $H_3 \to e^+ e^-$ is long enough at the detector level.} Even if Dm is found in the low-energy experiments and its decay branching ratio (BR) into $e^+ e^-$ is consistent with the SM prediction, the theoretical uncertainties would imply a bound $|h_{e\mu}|^2 / m_{H_3}^2 \lesssim {\rm GeV}^{-2}$~\cite{Jentschura:1997tv, Karshenboim:1998we, Ginzburg:1998df}, which is much weaker than the muonium oscillation constraint.

The neutral scalar $H_3$ could mediate the scattering $e^+ e^- \to \ell^+ \ell^-$ ($\ell = e,\, \mu,\, \tau$), in the $s$ and/or $t$-channel, depending on the couplings involved, e.g. the $ee \to \mu\mu$ could be induced in the $s$-channel by the coupling $h_{ee}^\dagger h_{\mu\mu}$ and in the $t$-channel by $h_{e\mu}$. The $H_3$-mediated diagrams interfere with the pure SM process, thus altering both the total cross section and differential distributions~\cite{Hou:1995dg}. The LEP data exclude an effective cutoff scale $\Lambda_{\rm eff}$ of few TeV, by analyzing the $ee \to \ell\ell$ data~\cite{Abdallah:2005ph}, which could be used to set limits on the couplings on the couplings $\Lambda_{\rm eff} \sim m_{H_3}/\sqrt{|h^\dagger h|}$, as collected in Table~\ref{tab:limits:H3}. When $H_3$ is lighter than the center-of-mass energy $\sqrt{s}$, the limits on the effective cutoff scale $\Lambda_{\rm eff}$ do not apply, and we have to consider the $H_3$ propagator:
\begin{eqnarray}
\label{eqn:propagator}
\frac{1}{q^2 - m_{H_3}^2 } \ = \
\frac{1}{-s \cos\theta/2 - m^2_{H_3}} \,.
\end{eqnarray}
For simplicity we take an average over the scattering angle $\langle \cos\theta \rangle \simeq 1/2$ to interpret the LEP constraints. Then in the limit of $m_{H_3} \ll \sqrt{s}$, the propagator is dominated by the $q^2$ term, and the $ee \to \ell\ell$ limit in Figures~\ref{fig:H3:prospect1} and \ref{fig:H3:prospect2} approaches to be a constant, as expected.

\begin{table*}[!t]
  \centering
  \caption[]{Current experimental data of the rare LFV decays $\ell_\alpha \to \ell_\beta \ell_\gamma \ell_\delta$, $\ell_\alpha \to \ell_\beta \gamma$~\cite{PDG, Amhis:2016xyh}, the electron~\cite{Hanneke:2008tm} and muon~\cite{Bennett:2006fi} $g-2$, muonium oscillation~\cite{Willmann:1998gd} and the LEP $e^+ e^- \to \ell^+ \ell^-$ data~\cite{Abdallah:2005ph}, and the resultant constraints on the couplings $|h^\dagger h|/m_{H_3}^2$. Note that the $\ell_\alpha \to \ell_\beta \gamma$ and the electron and muon $g-2$ constraints on $|h^\dagger h|/m_{H_3}^2$ have weak dependence on the scalar mass $m_{H_3}$, due to the extra logarithm terms in Eqs.~(\ref{eqn:coeff}), (\ref{eqn:coeff2}), (\ref{eqn:coeff3}) and (\ref{eqn:g-2:H3}). The $ee \to \ell\ell$ limits do not apply when $m_{H_3} \lesssim \sqrt{s}$. See text for more details.  }
  \label{tab:limits:H3}
  \begin{tabular}[t]{ccc}
  \hline\hline
  process & current data & constraints [${\rm GeV}^{-2}$] \\ \hline
  $\mu^- \to e^- e^+ e^-$ & $< 1.0 \times 10^{-12}$ & $|h_{ee}^\dagger h_{e\mu}|/m_{H_3}^2 < 6.6 \times 10^{-11}$ \\ \hline

  $\tau^- \to e^- e^+ e^-$ & $< 1.4 \times 10^{-8}$ &
  $|h_{ee}^\dagger h_{e\tau}|/m_{H_3}^2 < 1.9 \times 10^{-8}$ \\
  $\tau^- \to e^- \mu^+ \mu^-$ & $< 1.6 \times 10^{-8}$ & $|h_{\mu\mu}^\dagger h_{e\tau}|/m_{H_3}^2 < 1.4 \times 10^{-9}$ \\
  $\tau^- \to \mu^- e^+ \mu^-$ & $< 9.8 \times 10^{-9}$ & $|h_{e\mu}^\dagger h_{\mu\tau}|/m_{H_3}^2 < 1.5 \times 10^{-9}$ \\

  $\tau^- \to \mu^- e^+ e^-$ & $< 1.1 \times 10^{-8}$ &
  $|h_{ee}^\dagger h_{\mu\tau}|/m_{H_3}^2 < 1.2 \times 10^{-8}$ \\
  $\tau^- \to e^- \mu^+ e^-$ & $< 8.4 \times 10^{-9}$ &
  $|h_{e\mu}^\dagger h_{e\tau}|/m_{H_3}^2 < 1.4 \times 10^{-8}$ \\
  $\tau^- \to \mu^- \mu^+ \mu^-$ & $< 1.2 \times 10^{-8}$ & $|h_{\mu\mu}^\dagger h_{\mu\tau}|/m_{H_3}^2 < 1.7 \times 10^{-8}$ \\ \hline

  $\mu^- \to e^- \gamma$ & $< 4.2 \times 10^{-13}$ &
  \makecell{$|h_{ee}^\dagger h_{e\mu}|/m_{H_3}^2 < 1.5 \times 10^{-9}$ \\
  $|h_{e\mu}^\dagger h_{\mu\mu}|/m_{H_3}^2 < 2.1 \times 10^{-11}$ \\
  $|h_{e\tau}^\dagger h_{\tau\mu}|/m_{H_3}^2 < 2.3 \times 10^{-12}$ }\\ \hline

  $\tau^- \to e^- \gamma$ & $< 3.3 \times 10^{-8}$ &
  \makecell{ $|h_{ee}^\dagger h_{e\tau}|/m_{H_3}^2 < 1.0 \times 10^{-6}$ \\
  $|h_{e\mu}^\dagger h_{\mu\tau}|/m_{H_3}^2 < 1.0 \times 10^{-6}$ \\
  $|h_{e\tau}^\dagger h_{\tau\tau}|/m_{H_3}^2 < 2.6 \times 10^{-8}$} \\ \hline

  $\tau^- \to \mu^- \gamma$ & $< 4.4 \times 10^{-8}$ &
  \makecell{ $|h_{\mu e}^\dagger h_{e\tau}|/m_{H_3}^2 < 1.2 \times 10^{-6}$ \\
  $|h_{\mu\mu}^\dagger h_{\mu\tau}|/m_{H_3}^2 < 1.2 \times 10^{-6}$ \\
  $|h_{\mu\tau}^\dagger h_{\tau\tau}|/m_{H_3}^2 < 3.0 \times 10^{-8}$ } \\ \hline

  %$e^+ e^- \to ee ,\, \tau\tau$ & $< 3.3 \times 10^{-8}$ &
  %$|h_{ee}^\dagger h_{e\tau}|/m_{H}^2 < 1.0 \times 10^{-6}$ \\

  electron $g-2$ & $< 5.2 \times 10^{-13}$ &
  \makecell{$|h_{e\mu}|^2/m_{H_3}^2 < 6.2 \times 10^{-8}$  \\
  $|h_{e\tau}|^2/m_{H_3}^2 < 6.9 \times 10^{-9}$ } \\ \hline
  muon $g-2$ & $< 4.0 \times 10^{-9}$ & $|h_{\mu\tau}|^2/m_{H_3}^2 < 4.4 \times 10^{-7}$  \\ \hline

  muonium oscillation & $<8.2 \times 10^{-11}$ &  $|h_{e\mu}^2|/m_{H_3}^2 < 1.0 \times 10^{-7}$ \\ \hline

  $ee \to ee$ (LEP) & $\Lambda_{\rm eff} > 5.7$ TeV &
  $|h^\dagger h|^2/m_{H_3}^2 < 1.9 \times 10^{-7}$ \\
  $ee \to \mu\mu$ (LEP) & $\Lambda_{\rm eff} > 6.3$ TeV &
  $|h^\dagger h|^2/m_{H_3}^2 < 1.6 \times 10^{-7}$ \\
  $ee \to \tau\tau$ (LEP) & $\Lambda_{\rm eff} > 7.9$ TeV &
  $|h^\dagger h|^2/m_{H_3}^2 < 1.0 \times 10^{-7}$ \\
  \hline\hline
  \end{tabular}
\end{table*}
\subsubsection{Comparison of different production portals}

\begin{table}[!t]
  %\centering
  \caption[]{Production channels of $H_3$ at future $e^+ e^-$ colliders, and the correponding Feynman diagrams. See text for more details.}
  \label{tab:production}
  \begin{tabular}[t]{lll}
  \hline\hline
  channel & diagram(s) & comment \\ \hline
  $H_3$ & Figure~\ref{fig:4c} & loop coupling to photons \\ \hline
  $H_3 \ell_\alpha^\pm$ & Figure~\ref{fig:9b} & (LFV) couplings to $\ell^\pm$ \\ \hline

  \multirow{2}{*}{$H_3 + \slashed{E}_T$} &
  Figure~\ref{fig:diagram7} & (LFV) couplings to $\ell^\pm$ \\
  & Figures~\ref{fig:diagram3}, \ref{fig:9a} &
  from $H_3 Z$, $Z \to \nu \bar\nu$ \\  \hline

  \multirow{2}{*}{$H_3 \gamma$} &
  Figure~\ref{fig:4b} & loop coupling to photons \\
  & Figure~\ref{fig:9a} & coupling to $e^\pm$ \\ \hline

  \multirow{2}{*}{{$H_3 Z$  ($Z \to$ visible)}} &
  Figure~\ref{fig:diagram3} & mixing with SM Higgs  \\
  & Figure~\ref{fig:9a} & coupling to $e^\pm$ \\ \hline

  $H_3 W^+ W^-$ & Figure~\ref{fig:diagram4} & couplings to RHNs \\ \hline

  \multirow{6}{*}{$H_3 \ell_\alpha^\pm \ell_\beta^\mp$} &
  Figure~\ref{fig:diagram1} & coupling to $H^{\pm\pm}$ \\
  & \multirow{2}{*}{Figure~\ref{fig:4a}} & loop coupling to photons \\
  && ($\ell_\alpha^\pm \ell_\beta^\mp = e^+ e^-$) \\
  & Figure~\ref{fig:diagram6} & (LFV) couplings to $\ell^\pm$ \\
  & \multirow{2}{*}{Figures~\ref{fig:diagram3}, \ref{fig:9a}} &
  from $H_3 Z$, $Z \to \ell^+ \ell^-$ \\
  && ($\alpha = \beta$) \\ \hline

  $\ell_\alpha^\pm \ell_\beta^\mp$ & Figure~\ref{fig:diagram8} & off-shell production \\
  \hline\hline
  \end{tabular}
\end{table}

All the Feynman diagrams of $H_3$ production through its couplings to leptons, scalars and gauge bosons (including also the loop-level coupling to photons) at future $e^+ e^-$ colliders can be found in Figures~\ref{fig:diagram1}, \ref{fig:diagram2}, \ref{fig:diagram3}, \ref{fig:diagram4}, \ref{fig:diagram5}, \ref{fig:diagram6}, \ref{fig:diagram7} and \ref{fig:diagram8}. Categorizing by the particles in the final state,  all these production channels are collected in Table~\ref{tab:production}. For the associated production  of $H_3 Z$, we include only the visible decays of the $Z$ boson, while the invisible decays of $Z \to \nu\bar\nu$ are categorized into the channel with large missing energy $H_3 + \slashed{E}_T$.

\begin{figure}[t!]
  \centering
  \includegraphics[width=0.4\textwidth]{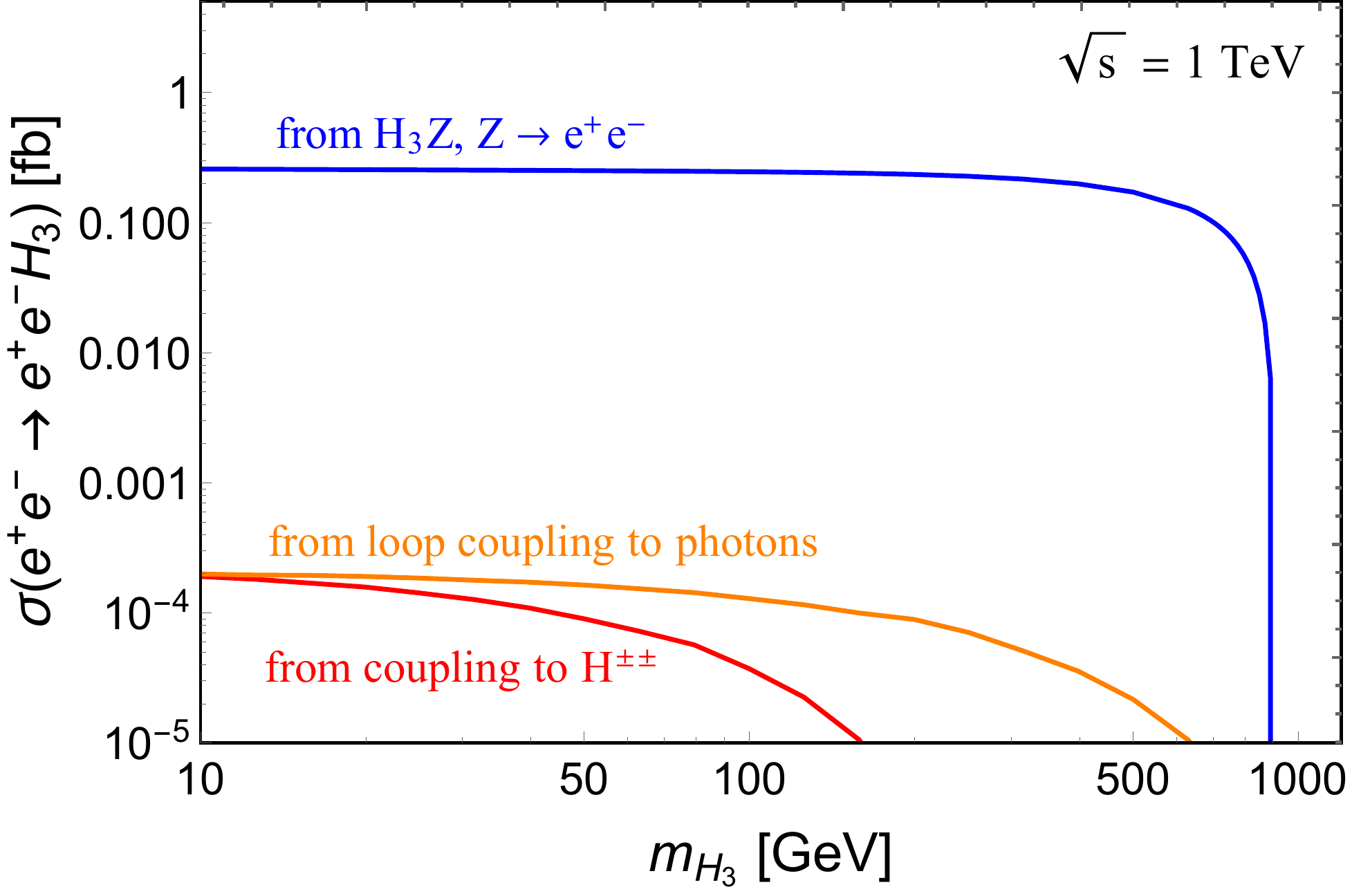}
  \caption{Production cross sections of $H_3$ in the process of $e^+ e^- \to e^+ e^- H_3$ at ILC 1 TeV from coupling to the doubly-charged scalar $H^{\pm\pm}$ (red), from loop-level coupling to photons (orange) and from the process $H_3 Z$ in Figure~\ref{fig:9a} with the subsequent decay $Z \to e^+ e^-$ (blue), as functions of $m_{H_3}$, with the benchmark parameters in Eq.~(\ref{eqn:benchmark}). }
  \label{fig:benchmark}
\end{figure}

It should be noted that for some of the production channels, we can have more than one portals depending on  the different couplings in the Lagrangian~\eqref{eqn:Lagrangian}. For instance, the associated production with a pair of leptons $H_3 \ell_\alpha^\pm \ell_\beta^\mp$ can be induced from the trilinear scalar coupling to the doubly-charged scalars $H^{\pm\pm}$ (Figure~\ref{fig:diagram1}), from the fusion of photons if $\ell_\alpha^\pm \ell_\beta^\mp = e^+ e^-$ via the loop-level effective coupling $H_3 \gamma\gamma$  (Figure~\ref{fig:4a}), from the (LFV) couplings $h_{\alpha\beta}$ to the charged leptons  (Figure~\ref{fig:diagram6}), or from the process $H_3 Z$ with the subsequent decay $Z \to \ell^+ \ell^-$ if $\alpha =\beta$ (Figure~\ref{fig:9a}).
In most of the parameter space, it is likely that one (or more) of the portals will dominate, depending largely on the relevant parameters. In some cases, the kinematic distributions of the final states might also be used to distinguish different production portals. A thorough comparison of all the portals for the processes in Table~\ref{tab:production} goes beyond the main scope of this work. However, for the illustration purpose we present the production cross sections $\sigma (e^+ e^- \to e^+ e^- H_3)$ in Figure~\ref{fig:benchmark} from all the scalar, gauge and the $H_3 Z$ ($Z \to e^+ e^-$) portals collected in Table~\ref{tab:production}, with the benchmark values of the parameters
\begin{eqnarray}
\label{eqn:benchmark}
\sin\theta_1 = 0 , \quad
M_{\pm\pm} = 1 \, {\rm TeV},\quad
f_{ee} = 0.2 ,\quad
h_{ee} = 0.01. \nonumber \\
\end{eqnarray}
Note that in the $H_3 Z$ process we have set the $h - H_3$ mixing to be zero such that it receives only the contribution from the coupling $H_3 e^+ e^-$ as shown in Figure~\ref{fig:9a}, with subleading contribution from the diagrams $e^+ e^- \to e^+ e^- H_3$ in Figure~\ref{fig:diagram6} which are induced by the same coupling $H_3 e^+ e^-$. For the parameters chosen in Eq.~(\ref{eqn:benchmark}), the doubly-charged scalar portal is highly suppressed by the mass $M_{\pm\pm}$ (red line in Figure~\ref{fig:benchmark}), while the gauge portal is highly suppressed by the loop-induced effective $H_3 \gamma\gamma$ coupling (orange line in Figure~\ref{fig:benchmark}), and the process $e^+ e^- \to e^+ e^- H_3$ is dominated by the associated $H_3 Z$ production with the subsequent decay $Z \to e^+ e^-$ (blue line in Figure~\ref{fig:benchmark}).

\subsubsection{Prospects and LFV signals}
\label{sec:prospect:H3}

\begin{figure*}[!t]
  \centering
  \includegraphics[width=0.4\textwidth]{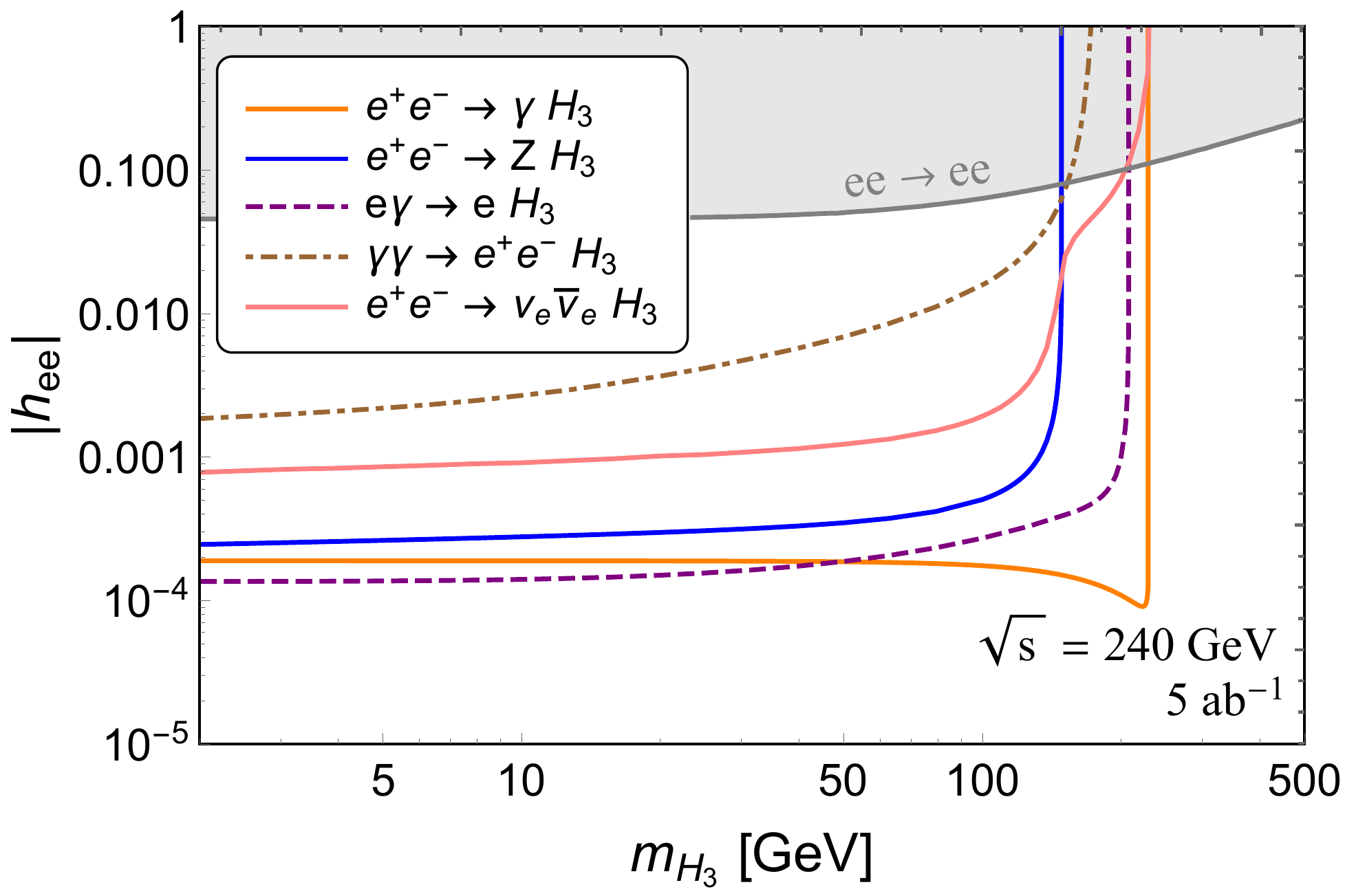}
  \includegraphics[width=0.4\textwidth]{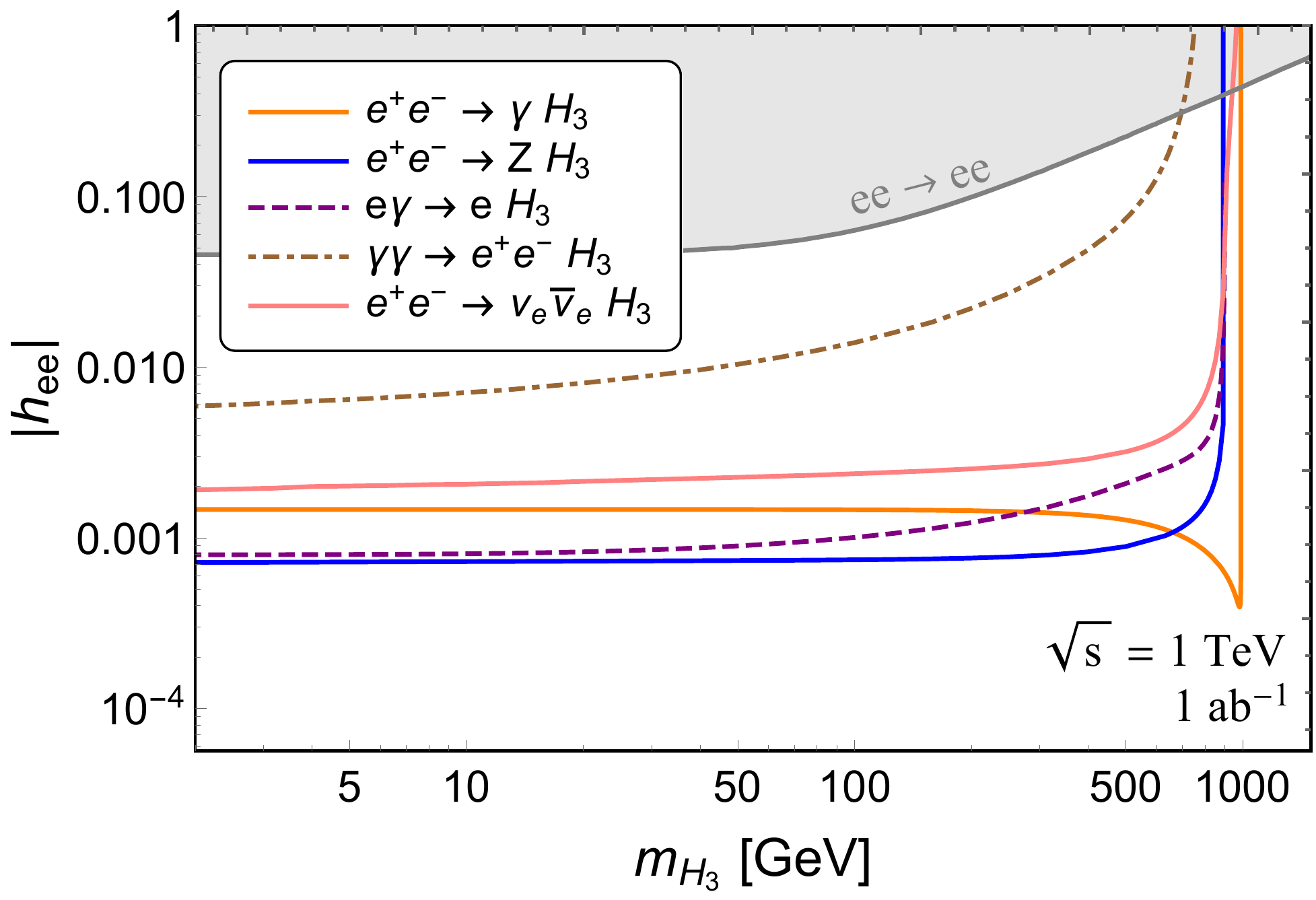} \vspace{-2pt} \\
  \includegraphics[width=0.4\textwidth]{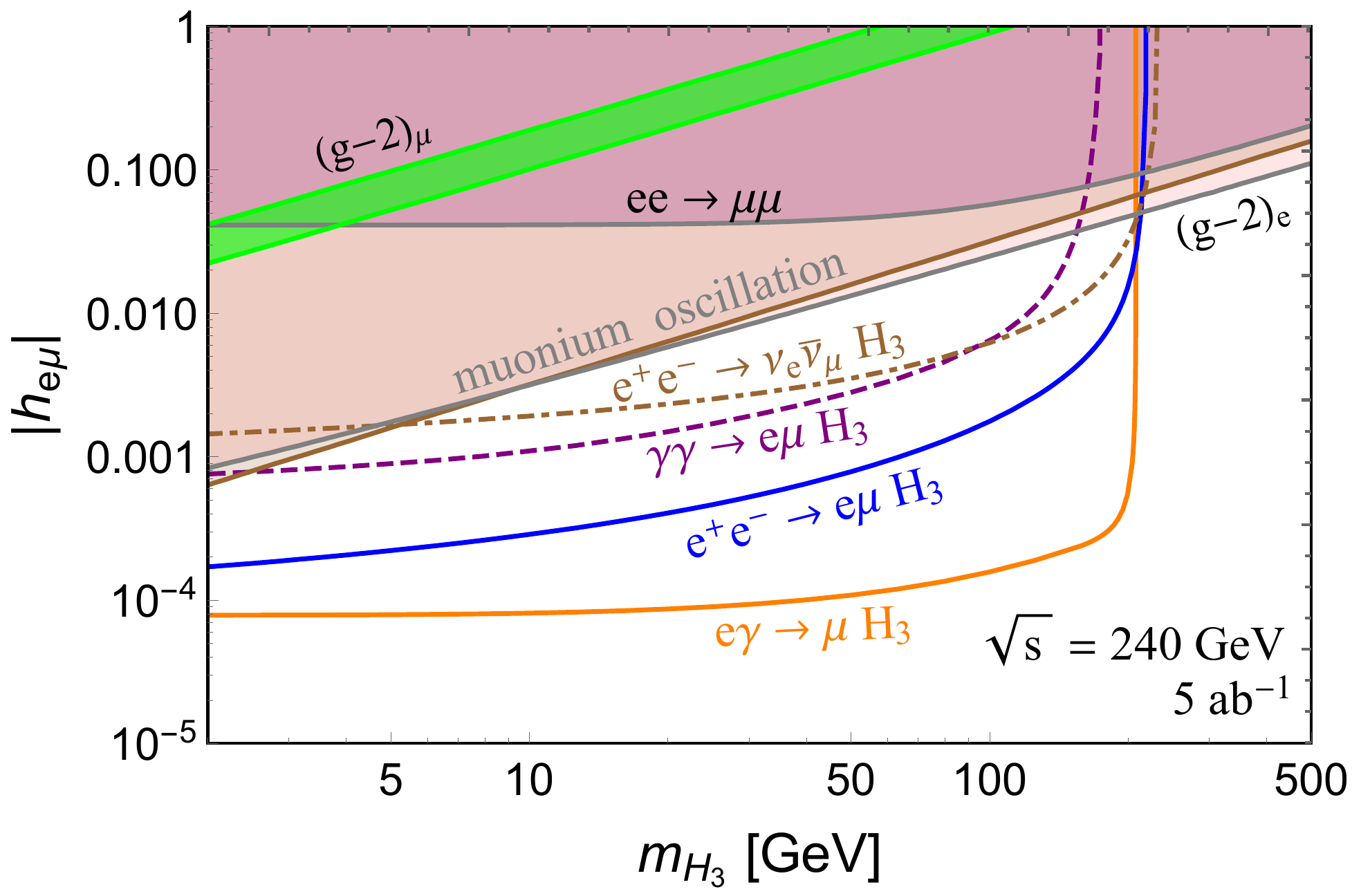}
  \includegraphics[width=0.4\textwidth]{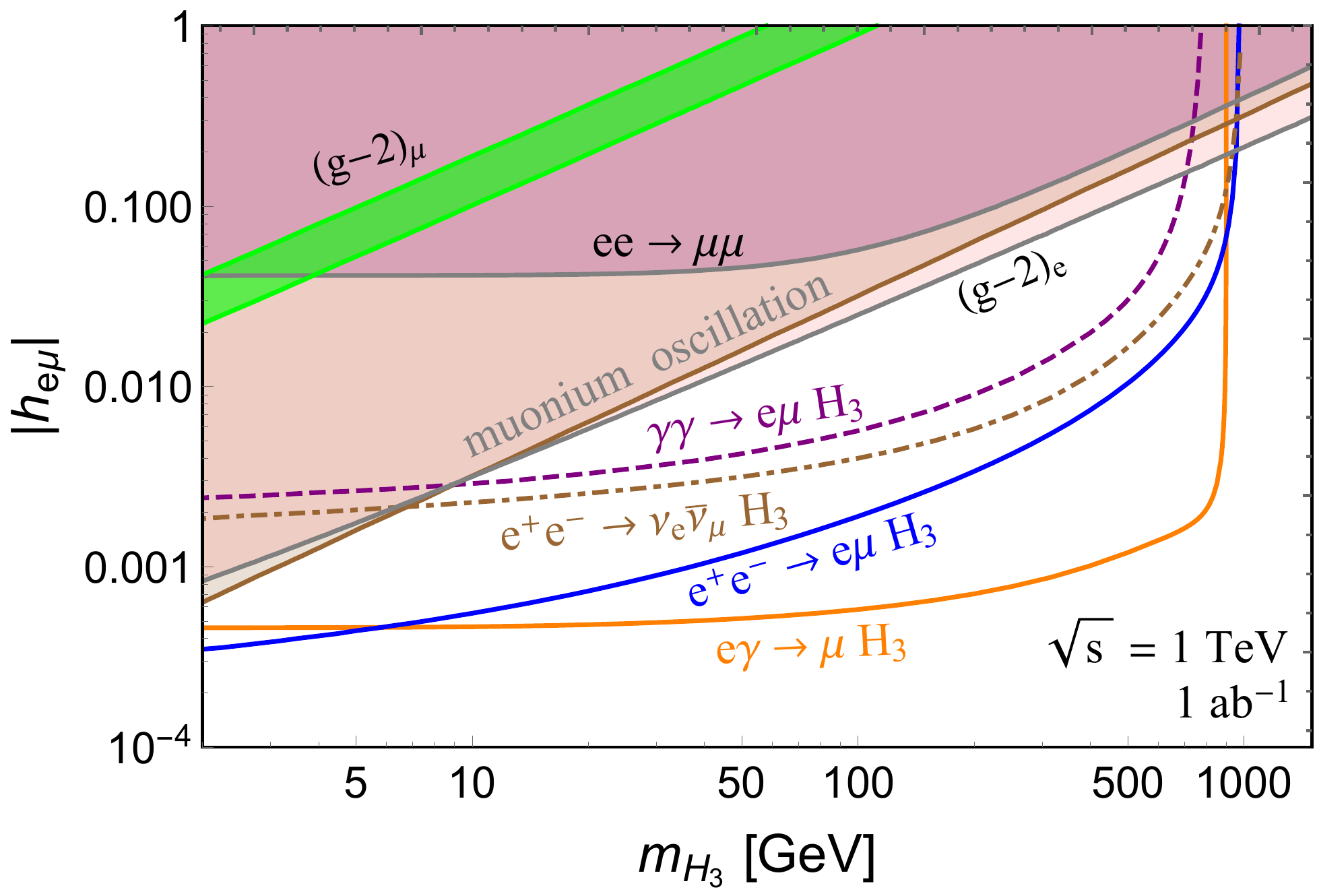} \vspace{-2pt} \\
  \includegraphics[width=0.4\textwidth]{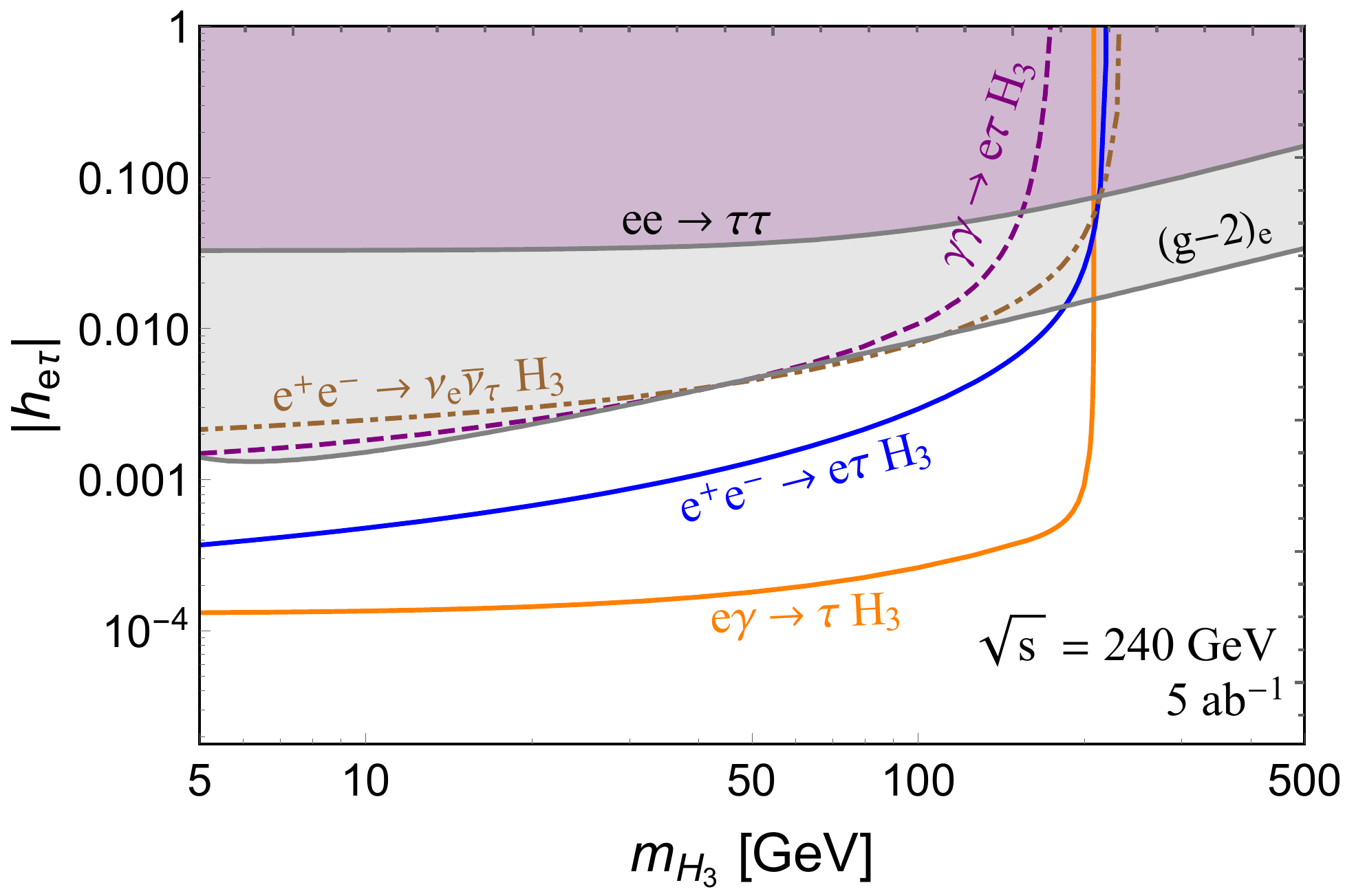}
  \includegraphics[width=0.4\textwidth]{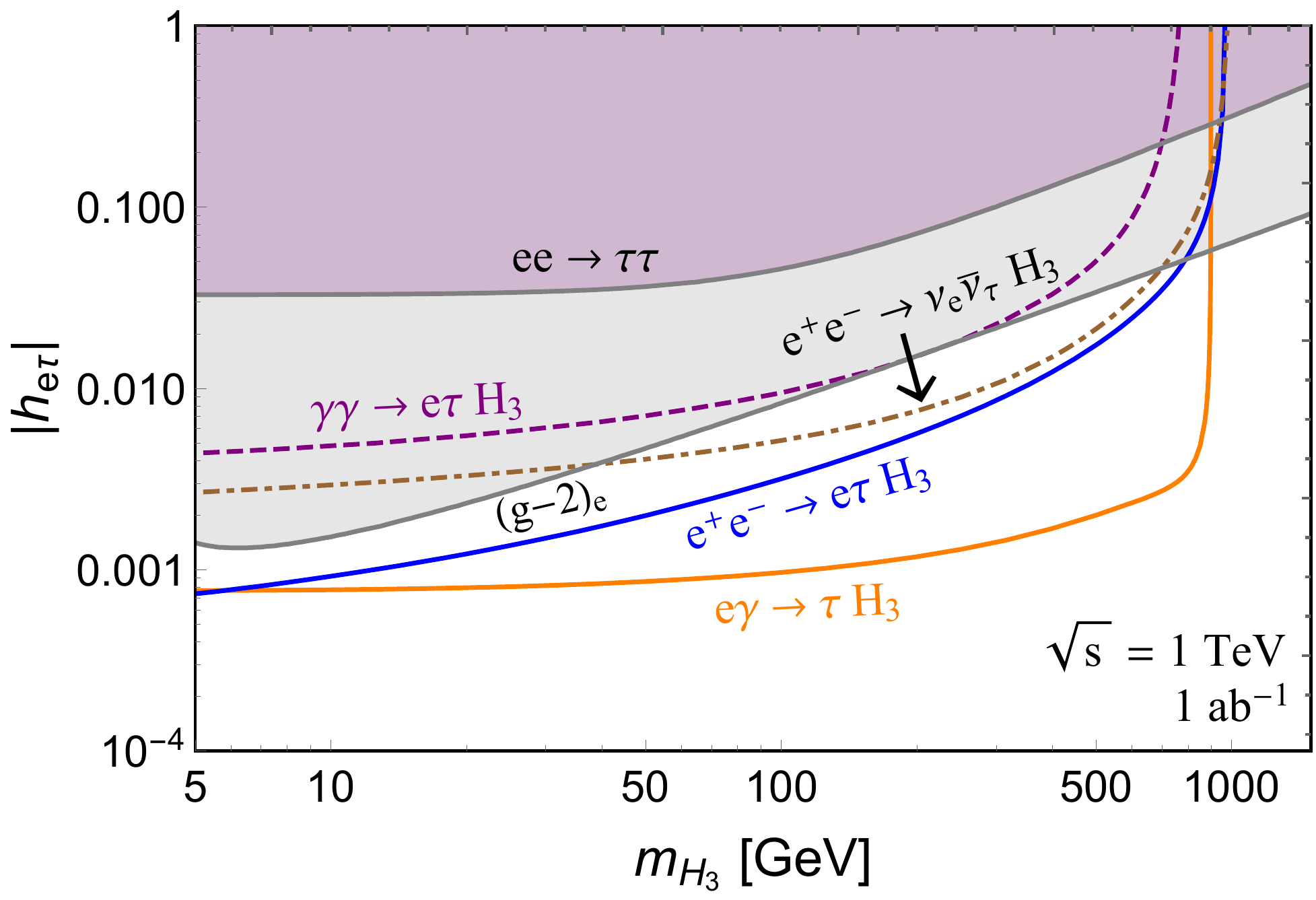} \vspace{-2pt} \\
  \includegraphics[width=0.4\textwidth]{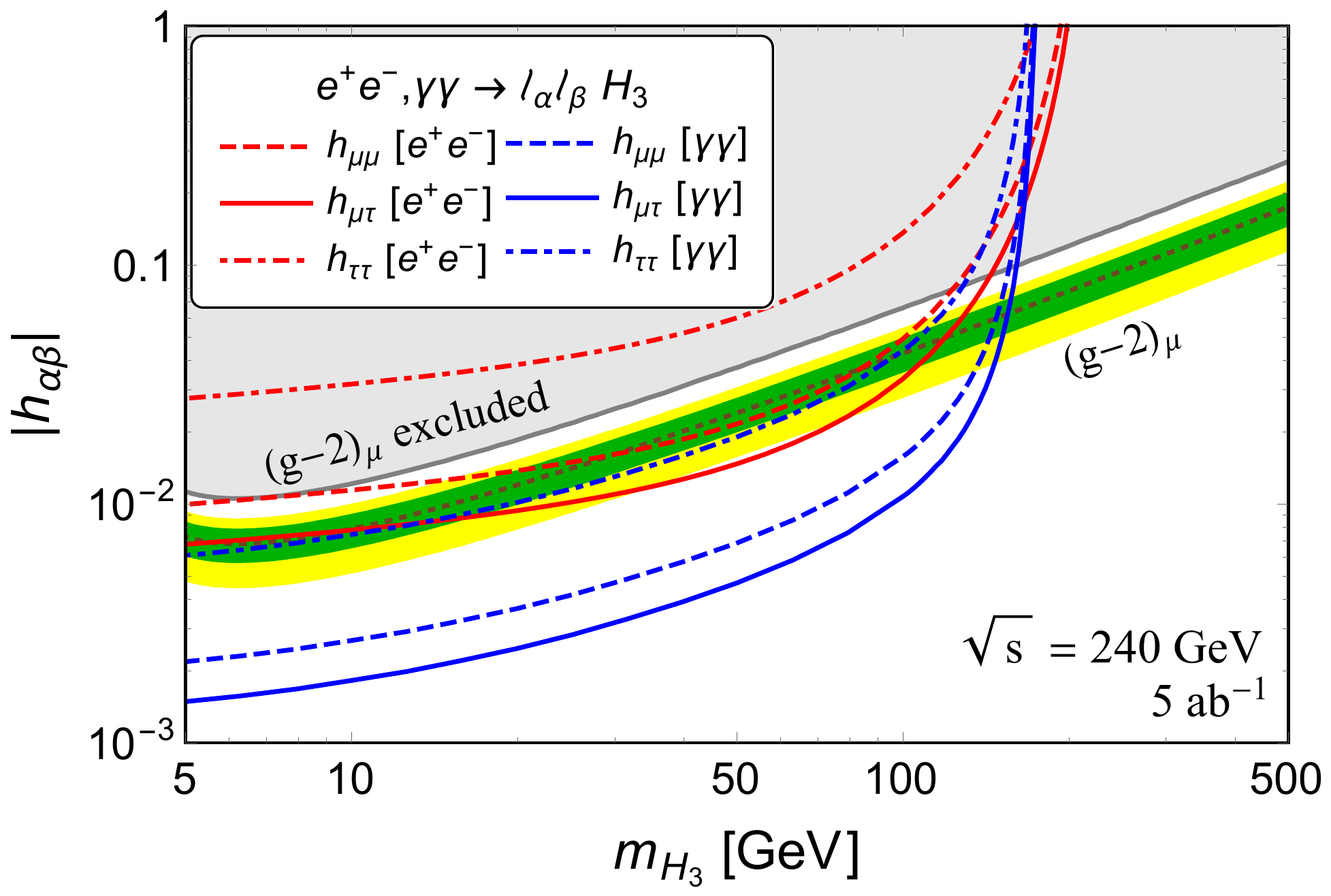}
  \includegraphics[width=0.4\textwidth]{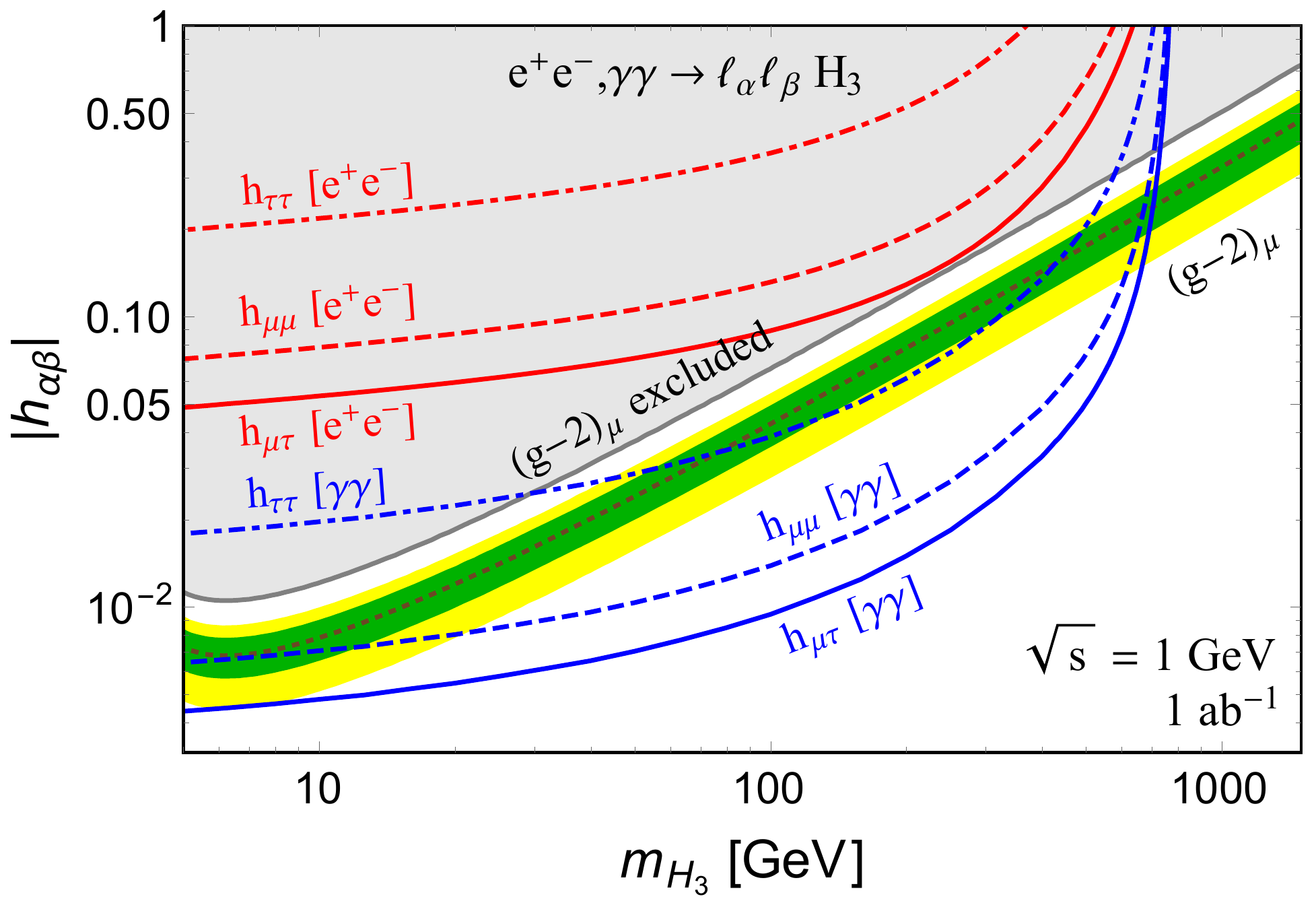}
  \caption{Prospects of the couplings $h_{\alpha\beta}$ from the on-shell production of $H_3$ at CEPC (240 GeV and 5 ab$^{-1}$, left) and ILC (1 TeV and 1 ab$^{-1}$, right), in the channels of $e^+ e^- \to (\gamma/Z) H_3$, $e\gamma \to \ell H_3$, $e^+ e^-,\, \gamma\gamma \to \ell_\alpha^\pm \ell_\beta^\mp H_3$ and $e^+ e^- \to \nu \bar\nu H_3$. The shaded regions are excluded by the muonium oscillation, electron $g-2$, muon $g-2$ (excluded by the theoretical-experimental discrepancy at the $5\sigma$ CL)~\cite{PDG} and the LEP $e^+e^- \to \ell^+ \ell^-$ data~\cite{Abdallah:2005ph}, as indicated in Table~\ref{tab:limits:H3}. The green and yellow bands in the second and fourth rows can explain the muon $g-2$ anomaly at the $1\sigma$ and $2\sigma$ CL, respectively, while the dotted line at the center of the $1\sigma$ band corresponds to the central value. }
  \label{fig:H3:prospect1}
\end{figure*}

All the amplitudes for the on-shell production of $H_3$ depend linearly on the couplings $h_{\alpha\beta}$, as shown in Figures~\ref{fig:diagram5}, \ref{fig:diagram6} and \ref{fig:diagram7}, thus free of the constraints from the rare LFV decays such as $\mu \to eee$ and $\tau \to e \gamma$ which depend quadratically on the Yukawa couplings $|h^\dagger h|$, as clearly presented in Table~\ref{tab:limits:H3}. With the production cross sections in Figures~\ref{fig:production4}, \ref{fig:production5} and \ref{fig:production6}, one can readily estimate the prospects of all the independent couplings $h_{\alpha\beta}$ at future lepton colliders, which are collected in Figure~\ref{fig:H3:prospect1}. As stated in Ref.~\cite{Dev:2017ftk}, the SM backgrounds are expected to be small, in particular for the LFV processes.\footnote{A detailed analysis of the SM backgrounds for the LFV processes is outside the scope of this paper. As the neutral scalar $H_3$ is hadrophobic and its couplings to the charged leptons are relatively small (compared to the SM gauge couplings), the decay width of $H_3$ is expected to be very small. Therefore the decay products of $H_3$ should form a sharp peak over the continuum SM backgrond, mostly due to the gauge boson decays, which  improve further the distinguishability of the signals and backgrounds, as illustrated in Ref.~\cite{Dev:2017ftk}. For LFV processes such as $e^+e^- \rightarrow \mu^+ e^- H_3$
with $H_3 \rightarrow \mu^-e^+$, we expect a peak in the $\mu^-e^+$ invariant mass distribution for the signal, which is absent in the SM, although one could get the same $e^+e^-\mu^+\mu^-$ final state from $ZZ$ decay. Similarly, for lepton flavor conserving processes, e.g. $e^+e^- \rightarrow e^+ e^- H_3$
with $H_3 \rightarrow e^-e^+$, the $e^+e^-$ invariant mass peak due to $H_3$ can in principle be distinguished from $Z\to e^+e^-$ SM background, by putting an invariant mass cut to exclude the $Z$-pole region.} For simplicity, we have turned on only one of the couplings $h_{\alpha\beta}$ and set all others irrelevant to be zero. Neglecting the mixing of $H_3$ with the SM Higgs, the $v_R^2$-suppressed loop-decay $H_3 \to \gamma\gamma$ and the decay $H_3 \to \nu\bar\nu$ suppressed by the heavy-light neutrino mixing $V_{\nu N}^4$, the neutral scalar $H_3$ decays predominantly into a pair of leptons, i.e. $H_3 \to \ell_\alpha^\pm \ell_\beta^\mp$. To be concrete, we assume a minimum number of 10 (30) for the signals with (without) LFV, and adopt an efficiency factor of $60\%$ for the tau lepton~\cite{Baer:2013cma}. In the process $e^+ e^- \to Z H_3$, only the visible decay products of $Z$ are taken into account for the prospects of $h_{ee}$ in the first row panels of Figure~\ref{fig:H3:prospect1}, with roughly a ${\rm BR} (Z \to {\rm visible}) \simeq 80\%$.

Regarding the flavor-conserving coupling $h_{ee}$, there exist only constraint from the LEP data $ee \to ee$~\cite{Abdallah:2005ph} (the limit from electron $g-2$~\cite{Hanneke:2008tm} is highly suppressed by the electron mass and thus not considered). Given an integrated luminosity of 5 ab$^{-1}$ (1 ab$^{-1}$) at CEPC (ILC), the coupling $h_{ee}$ could be probed up to the order of $10^{-4}$ ($10^{-3}$), orders of magnitude lower than the current LEP constraints, as seen in the first row panels of Figure~\ref{fig:H3:prospect1}. With three particles in the final state, the cross sections (and future prospects) for $\gamma\gamma \to ee H_3$ and $e^+ e^- \to \nu\bar\nu H_3$ are comparatively weaker than those with only two particles such as $e^+ e^- \to \gamma H_3$ and $e\gamma \to e H_3$.

As in Ref.~\cite{Dev:2017ftk}, the most stringent constraints on $h_{e\mu}$ come from the muonium oscillation~\cite{Willmann:1998gd}, the electron $g-2$~\cite{Hanneke:2008tm} and the LEP $ee \to \mu\mu$ data~\cite{Abdallah:2005ph}. A broad range of the mass $m_{H_3}$ and $h_{e\mu}$ can be probed in the channels of $e\gamma \to \mu H_3$, $e^+ e^- ,\, \gamma\gamma \to e\mu H_3$ and $ee \to \nu\bar\nu H_3$ at future lepton colliders, as shown in the second row panels of Figure~\ref{fig:H3:prospect1}. The case for $h_{e\tau}$ are quite similar, with the existing limits primarily from the electron $g-2$ and $ee \to \tau\tau$ data, as presented in the third row panels of Figure~\ref{fig:H3:prospect1}.

The processes involving only the muon and tauon flavors in the final state are limited, i.e. $e^+ e^- \to (\mu\mu,\, \mu\tau,\,\tau\tau) H_3$ and $\gamma\gamma \to (\mu\mu,\, \mu\tau,\, \tau\tau) H_3$, as shown in the last row panels of Figure~\ref{fig:H3:prospect1}. If $h_{ee} \neq 0$, we also have the contribution from $e^+ e^- \to Z H_3$, $Z \to \mu\mu,\, \tau\tau$, which are not included in the last two panels, as they do not depend on the couplings $h_{\mu\mu,\,\tau\tau}$. As a result of the smaller cross sections, the prospects for the couplings $h_{\mu\mu,\, \mu\tau,\, \tau\tau}$ are comparatively weaker than those for $h_{e\ell}$. However, the coupling $h_{\mu\tau}$ could provide a natural explanation for the muon $g-2$ anomaly~\cite{PDG}
\begin{eqnarray}
\Delta a_\mu \ = \
(2.87 \pm 0.80) \times 10^{-9}
\end{eqnarray}
in presence of the neutral scalar $H_3$, as indicated by  the green and yellow bands in the last two panels of Figure~\ref{fig:H3:prospect1} covering the $1\sigma$ and $2\sigma$ ranges respectively around the central value (dotted line). More importantly, it could be directly tested at the future lepton colliders like CEPC and ILC (and FCC-ee and CLIC), by searching for the LFV signals
\begin{eqnarray}
e^+ e^- ,\, \gamma\gamma \ \to \ \mu^\pm \tau^\mp + H_3 \,,
\end{eqnarray}
with $H_3$ decaying back into $\mu - \tau$ pairs or other particles such as two photons (cf. the solid blue and red lines in the last two panels of Figure~\ref{fig:H3:prospect1}). Almost the whole (kinematically allowed) parameter space could be covered, if a sizable BR of the $H_3$ decay products, say $\gtrsim10\%$, could be reconstructed at lepton colliders.

\begin{figure*}[!t]
  \centering
  \includegraphics[width=0.4\textwidth]{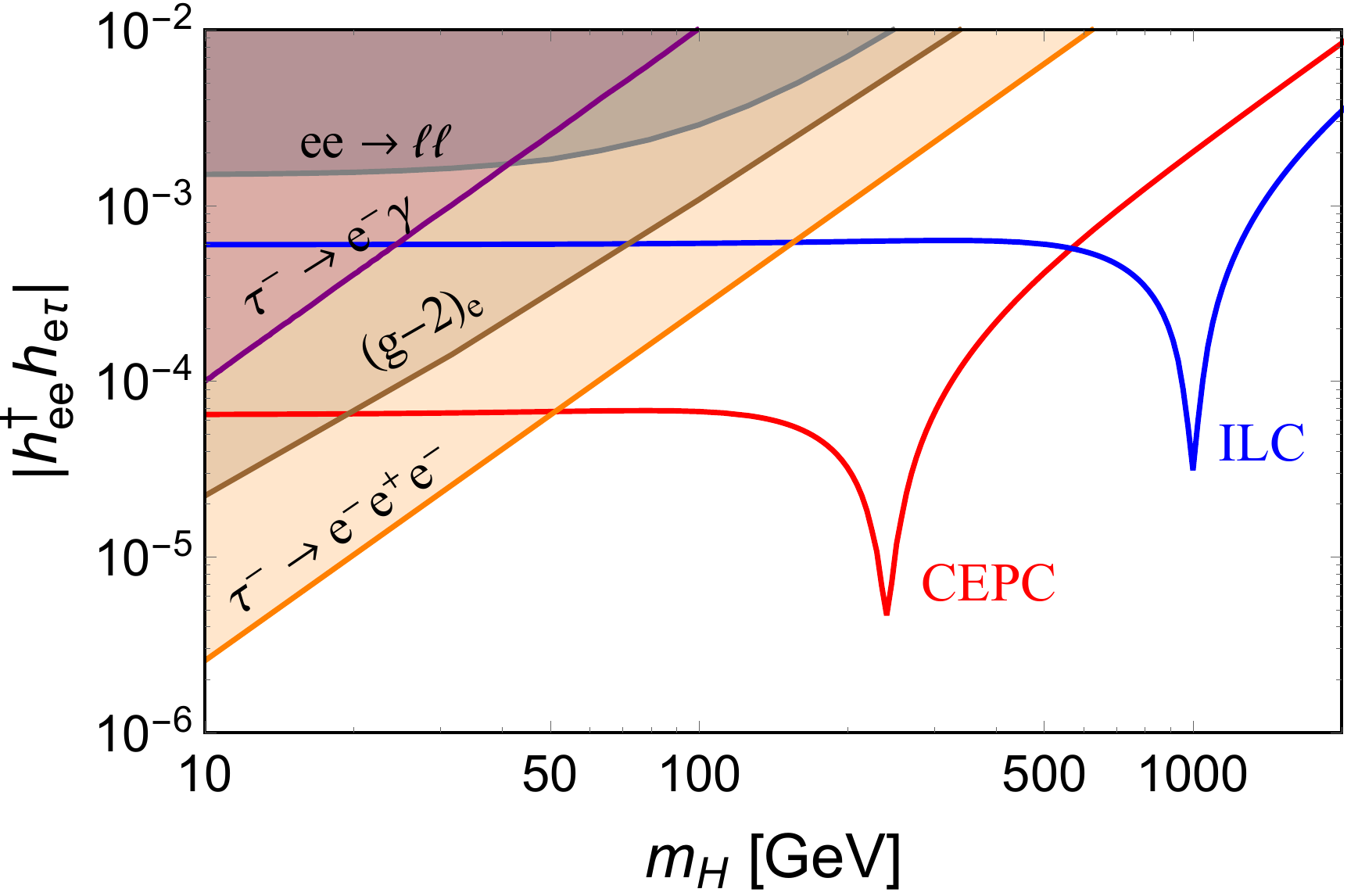} \\
  \includegraphics[width=0.4\textwidth]{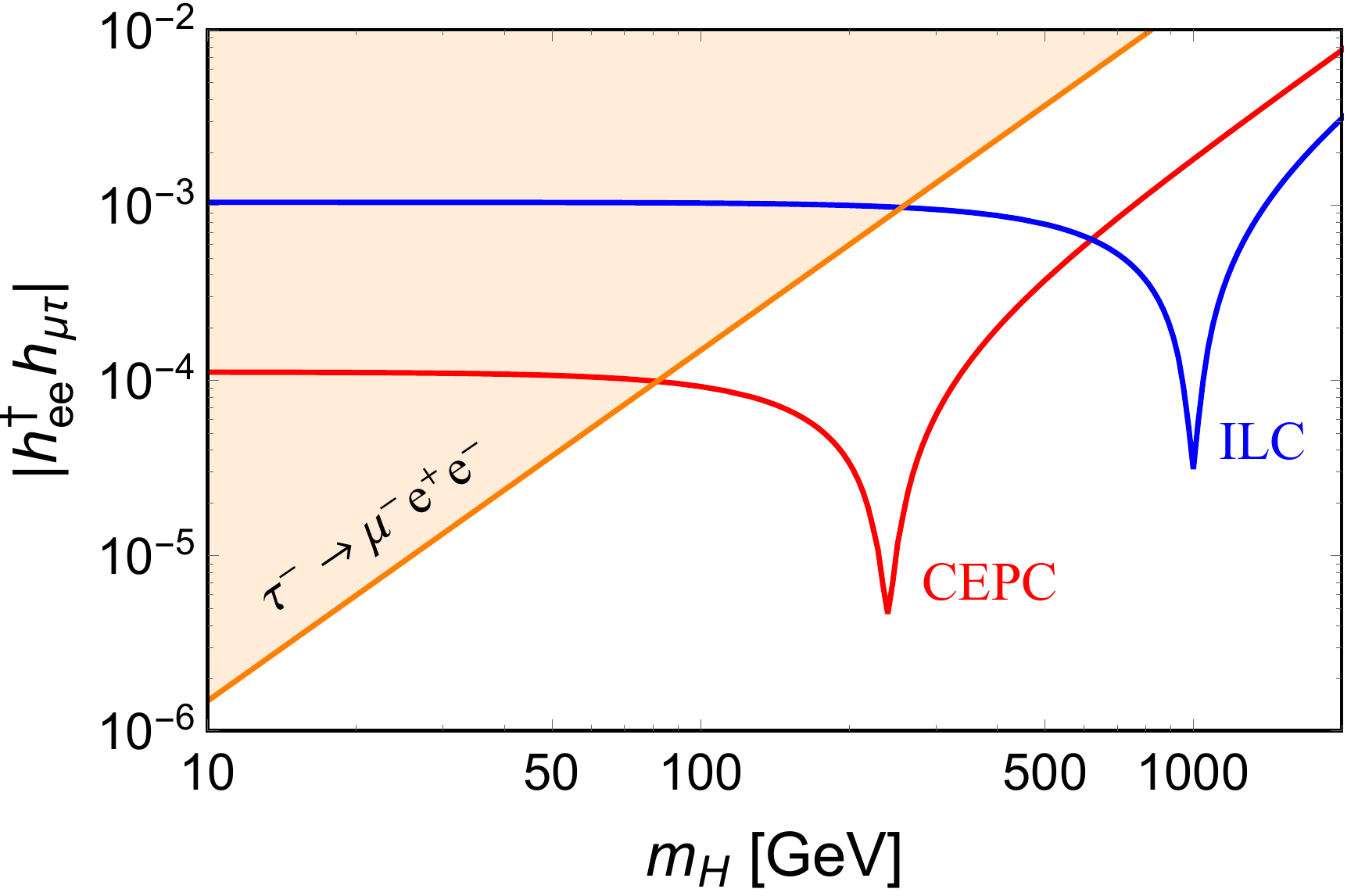}
  \includegraphics[width=0.4\textwidth]{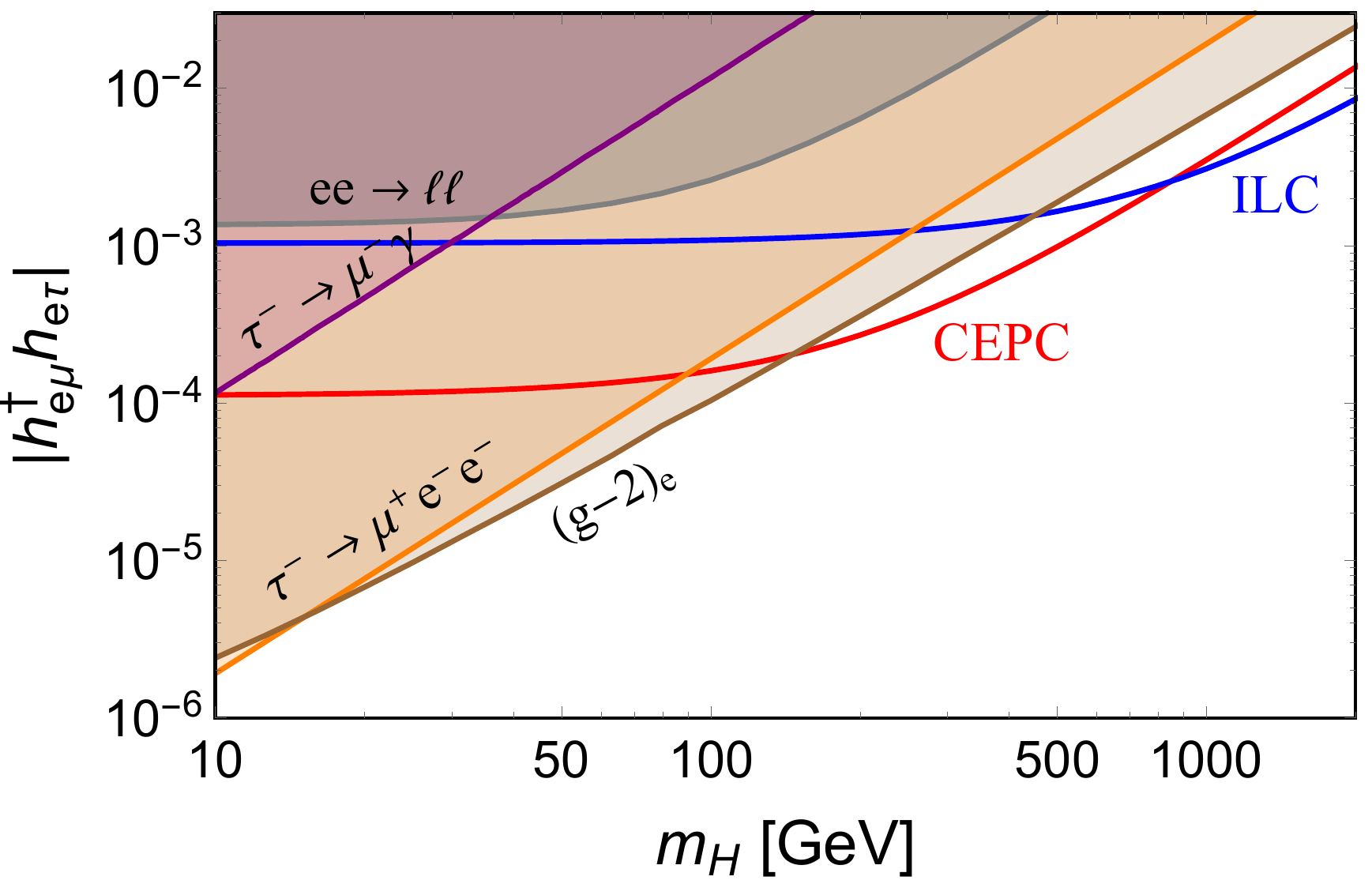}
  \caption{Prospects of $|h^\dagger_{ee} h_{e\tau}|$ (upper), $|h^\dagger_{ee} h_{\mu\tau}|$ (lower left) and $|h^\dagger_{e\mu} h_{e\tau}|$ (lower right) from searches of $e^+ e^- \to e^\pm \tau^\mp,\, \mu^\pm \tau^\mp$ at CEPC (red, $\sqrt{s} = 240$ GeV, ${\cal L} = 5$ ab$^{-1}$) and ILC (blue, 1 TeV and 1 ab$^{-1}$).  Also shown are the constraints from the rare lepton decays $\ell_\alpha \to \ell_\beta \ell_\gamma \ell_\delta$, $\ell_\alpha \to \ell_\beta \gamma$, electron $g-2$~\cite{PDG}, and the LEP $e^+ e^- \to \ell^+ \ell^-$ data~\cite{Abdallah:2005ph}, as indicated [cf.~Table~\ref{tab:limits:H3}]. Figure from Ref.~\cite{Dev:2017ftk}. }
  \label{fig:H3:prospect2}
\end{figure*}

As seen in Table~\ref{tab:limits:H3}, the decay $\mu \to eee$ sets very stringent limit on the coupling $|h_{ee}^\dagger h_{e\mu}|$; thus there is no hope to see any signal of $e^+ e^- \to e^\pm \mu^\mp$. In the $\tau$ lepton sector, the constraints from the LFV decays $\tau \to \ell_\alpha \ell_\beta \ell_\gamma$ are much weaker, which leaves us large parameter space to probe by direct searches at future lepton colliders. In light of the large production cross sections in Figure~\ref{fig:production7} and the small SM background~\cite{Kabachenko:1997aw,
Cho:2016zqo}, the couplings $|h^\dagger h|$ can be measured down to $10^{-4}$ ($10^{-3}$) at CEPC (ILC), as shown in Figure~\ref{fig:H3:prospect2}~\cite{Dev:2017ftk}, well beyond the existing flavor limits in Table~\ref{tab:limits:H3}. In vicinity of $m_{H_3} \simeq \sqrt{s}$, the production cross sections are largely enhanced by the resonance effect, if $H_3$ could be produced in the $s$-channel, as seen in the upper and lower left panel of Figure~\ref{fig:H3:prospect2}. For the sake of concreteness, we have set the width of $H_3$ to be 10 GeV (30 GeV) at 240 GeV (1 TeV). Even if $m_{H_3}$ is larger than the center-of-mass energy, the LFV signals would still reveal new physics beyond the SM, if they were found at the future lepton colliders.

%All the relevant constraints from LFV $\tau$ decay are collected in Table~\ref{tab:limits}, and depicted as the shaded regions in the plots of Fig.~\ref{fig:production5}.

\section{The doubly-charged scalar}
\label{sec:Hpp}
In this section, we study the production and detection prospects of the doubly-charged scalar $H^{\pm\pm}$ in the LRSM at future lepton colliders.

\subsection{Current experimental constraints}
\label{sec:constraints}
We first examine the current experimental constraints from the direct same-sign dilepton searches at LHC~\cite{ATLAS:2017iqw,CMS:2017pet}, the low-energy flavor constraints from rare LFV decays of charged leptons like $\ell_\alpha \to \ell_\beta \gamma$ and $\ell_\alpha \to \ell_\beta \ell_\gamma \ell_\delta$~\cite{PDG}, the anomalous magnetic moments of electron and muon, the LEP $e^+ e^- \to \ell^+ \ell^-$ data (with $\ell = e,\, \mu,\, \tau$, with the diagram shown in the left panel of Figure~\ref{fig:diagram:Hpp:3})~\cite{Abdallah:2005ph}, and muonium-anti-muonium oscillation~\cite{Willmann:1998gd}, as in the neutral scalar case in Section~\ref{sec:H3}.

Following the Yukawa Lagrangian in Eq.~(\ref{eqn:LYukawa}), the doubly-charged scalar $H^{\pm\pm}$ is RH, i.e. coupling only to the RH charged leptons, with the decay width
\begin{eqnarray}
\label{eqn:Hppdecay1}
\Gamma (H^{\pm\pm} \to \ell_\alpha^\pm \ell_\beta^\pm) \ \simeq \
\frac{S_{\alpha\beta} M_{\pm\pm} |f_{\alpha\beta}|^2}{8\pi} \,,
\end{eqnarray}
where $\alpha,\,\beta = e,\, \mu,\, \tau$ run over all the three flavors, and $S_{\alpha\beta} = 1 \, (2)$ for $\alpha=\beta$ ($\alpha \neq \beta$) is the symmetry factor. In the case where parity and $SU(2)_R$ breaking scales are decoupled~\cite{Chang:1983fu}\footnote{Such models allow $SO(10)$ embedding of TeV scale $W_R$ left-right models and are compatible with coupling unification. Parity becomes a symmetry called $D$-parity in $SO(10)$, a discrete symmetry that transforms $f\to f^c$, and is broken close to the GUT scale (by the VEV of a parity-odd singlet scalar field) well before the $SU(2)_R$ gauge symmetry breaks. The $D$-parity is different from the Lorentz parity, in the sense that Lorentz parity only interchanges the left-handed fermions with the right-handed ones, whereas the $D$-parity also interchanges the $SU(2)_L$ Higgs fields with the $SU(2)_R$ Higgs fields.}, the left-handed triplet $\Delta_L$ decouples in such models from the TeV scale physics, and the Yukawa couplings $f_{\alpha\beta}$ in Eq.~(\ref{eqn:LYukawa}) are not directly connected to the active neutrino masses and mixings like in type-II seesaw~\cite{type2a,type2b,type2c,type2d,type2e}, and all the entries can in principle be totally free parameters.

Regarding the decay of $H^{\pm\pm}$, in the LRSM, it couples also to the heavy $W_R$ boson, dictated by the gauge coupling $g_R$. The current $K$ and $B$ meson oscillation data require that the $W_R$ boson is beyond roughly 3 TeV~\cite{Zhang:2007da, Bertolini:2014sua}; thus a TeV-scale doubly-charged scalar could decay only into two off-shell heavy $W_R$ bosons, which decay further into the SM quarks (plus charged leptons and heavy RHNs if kinematically allowed), with the partial width~\cite{Djouadi:2005gi}%\footnote{Even if the doubly-charged scalar could be probed up to 10 TeV or even beyond in the off-shell mode, the production cross sections for the LFV processes like $e^+ e^- \to e^\pm \tau^\mp$ depend only on the Yukawa couplings in Eq.~(\ref{eqn:LYukawa}) but not on the gauge couplings of $H^{\pm\pm}$, nor on the decay width of $H^{\pm\pm}$ in this case, thus the prospects in this paper are not affected by the couplings to the $W_R$ boson, as long as $W_R$ can not be pair produced on-shell at future lepton colliders. %{What if the Yukawa couplings $f_{\alpha\beta}$ are such small that the BR into off-shell $W_R$ is comparable to those to the charged leptons?}}
\begin{eqnarray}
&& \Gamma (H^{\pm\pm} \to W_R^{\pm\ast} W_R^{\pm\ast})  \ \simeq \
\frac{1}{\pi^2}
\int_{0}^{M_{\pm\pm}^2} {\rm d} p
\int_{0}^{(M_{\pm\pm} -\sqrt{p})^2} {\rm d} q  \nonumber \\
&& \qquad \qquad \qquad  \times \frac{M_{W_R} \Gamma_{W_R}}{(p-M_{W_R}^2)^2 + M_{W_R}^2 \Gamma_{W_R}^2} \nonumber \\
&& \qquad \qquad \qquad \times \frac{M_{W_R} \Gamma_{W_R}}{(q-M_{W_R}^2)^2 + M_{W_R}^2 \Gamma_{W_R}^2}
\Gamma_0 \,, \\
&& {\rm where} \quad
\label{eqn:Hppdecay2}
\Gamma_0  \ = \
\frac{M_{\pm\pm}^3}{16\pi v_R^2}
\lambda^{1/2} (p,q,M_{\pm\pm}^2) \nonumber \\
&& \qquad \qquad \qquad \times \left( \lambda (p,q,M_{\pm\pm}^2) + \frac{12pq}{M_{\pm\pm}^4} \right) \,, \\
&&{\rm and} \quad
\lambda (a,b,c) \ \equiv \
\left( 1 - \frac{a}{c} - \frac{b}{c} \right)^2 - \frac{4ab}{c^2} \,.
\end{eqnarray}
For sufficiently small $f_{\alpha\beta}$ couplings which is of great interest for the prospects of the $f_{\alpha\beta}$ couplings at future lepton colliders, the RHNs are expected to be lighter than the $W_R$ boson. Then taking into account all the decays $W_R \to q\bar{q}$ and $W_R \to \ell N$ leads to the width $\Gamma_{W_R} \simeq g_R^2 M_{W_R}/4\pi$. %with $g_R$ the gauge coupling for the $SU(2)_R$ group.

The SM $W$ boson mixes with the heavy $W_R$ boson, but the mixing angle is highly suppressed by the mass ratio via~\cite{Zhang:2007da, Dev:2016dja}
\begin{eqnarray}
\tan\zeta_W & \ \simeq \ &
- \frac{g_R}{g_L} \frac{2\tan\beta}{1+\tan^2\beta} \frac{m_W^2}{M_{W_R}^2} \nonumber \\
& \ \simeq \ & - 3.5 \times 10^{-5} \times
\left( \frac{\tan\beta}{m_b/m_t} \right)
\left( \frac{g_R}{g_L} \right)
\left( \frac{M_{W_R}}{3 \, {\rm TeV}} \right)^{-2} \,, \nonumber \\ &&
\end{eqnarray}
with $\tan\beta = \kappa^\prime/\kappa$ the VEV ratio in the bidoublet sector which might also be small in light of the mass hierarchy $m_{b,\,\tau} \ll m_t$ in the SM fermion sector. Thus we neglect here also the decay $H^{\pm\pm} \to W^\pm W^\pm$. Due to the severe FCNC constraints on the heavy doublet scalars $M_{\phi_2} \gtrsim 10$ TeV~\cite{Zhang:2007da}, the decay $H^{\pm\pm} \to H^{\pm\ast} H^{\pm\ast}$ (with $H^\pm$ the singly-charged scalar from the heavy doublet) is also highly suppressed.

\begin{figure}[t!]
  \centering
  \includegraphics[width=0.4\textwidth]{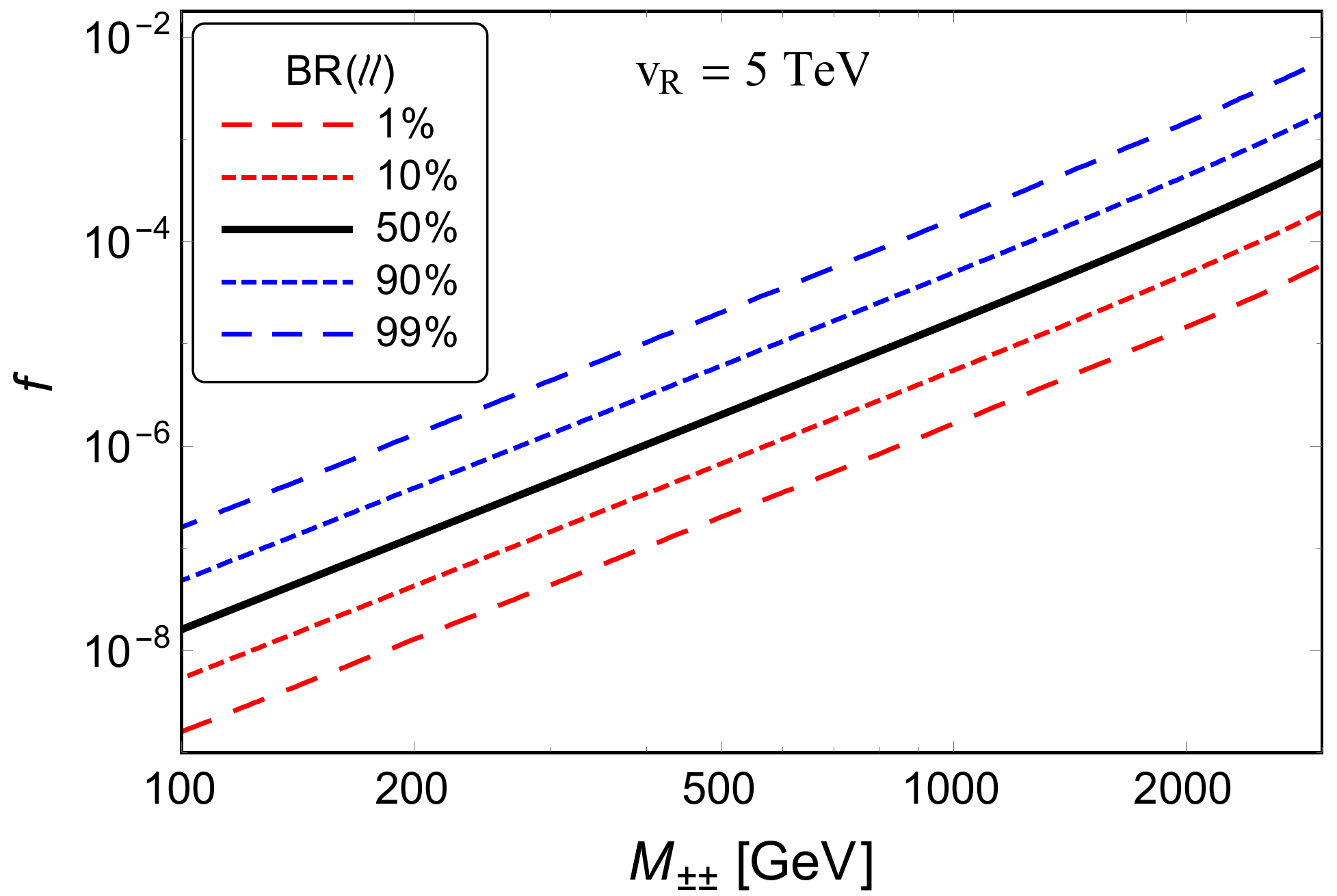}
  \caption{Contours of ${\rm BR} (H^{\pm\pm} \to \ell_\alpha^\pm \ell_\beta^\pm) = 1 - {\rm BR} (H^{\pm\pm} \to W_R^{\pm\ast} W_R^{\pm\ast}) = 1\%,\, 10\%,\, 50\%,\, 90\%,\, 99\%$ in the LRSM, as function of the doubly-charged scalar mass $M_{\pm\pm}$ and the Yukawa coupling $f$. The gauge coupling $g_R = g_L$ and the RH scale $v_R = 5$ TeV.}
  \label{fig:BR}
\end{figure}

The BRs of 1\%, 10\%, 50\%, 90\% and 99\% for the doubly-charged scalar decay into the same-sign leptons and the heavy $W_R$ boson pairs are shown in Figure~\ref{fig:BR}, where for simplicity we do not consider any of the flavor dependence in Eq.~(\ref{eqn:Hppdecay1}) and have taken $g_R = g_L$ and the RH scale $v_R = 5$ TeV in Eq.~(\ref{eqn:Hppdecay2}). For the on-shell production of doubly-charged scalar at future lepton colliders, the dilepton channel will dominate over the heavy $W_R$ channel, as long as its mass is below 1 TeV and the Yukawa coupling $f \gtrsim 10^{-4}$. For the off-shell production $e^+ e^- \to \ell_\alpha^\pm \ell_\beta^\mp$ (see Figure~\ref{fig:diagram:Hpp:3}), as long as $M_{\pm\pm} < 2 M_{W_R}$, the effects from the gauge coupling to the $W_R$ boson could be neglected; otherwise for $M_{\pm\pm} > 2 M_{W_R}$ it will contribute significantly to the total width of the doubly-charged scalar.  Then the decay branching ratios into different flavor combinations of charged leptons are simply
\begin{eqnarray}
\label{eqn:BR}
{\rm BR} (H^{\pm\pm} \to \ell_\alpha^\pm \ell_\beta^\pm) \ = \
\frac{ S_{\alpha\beta} |f_{\alpha\beta}|^2}{\sum_{\alpha,\beta}  |f_{\alpha\beta}|^2} \,.
\end{eqnarray}

\begin{figure}[t!]
  \centering
  \includegraphics[width=0.4\textwidth]{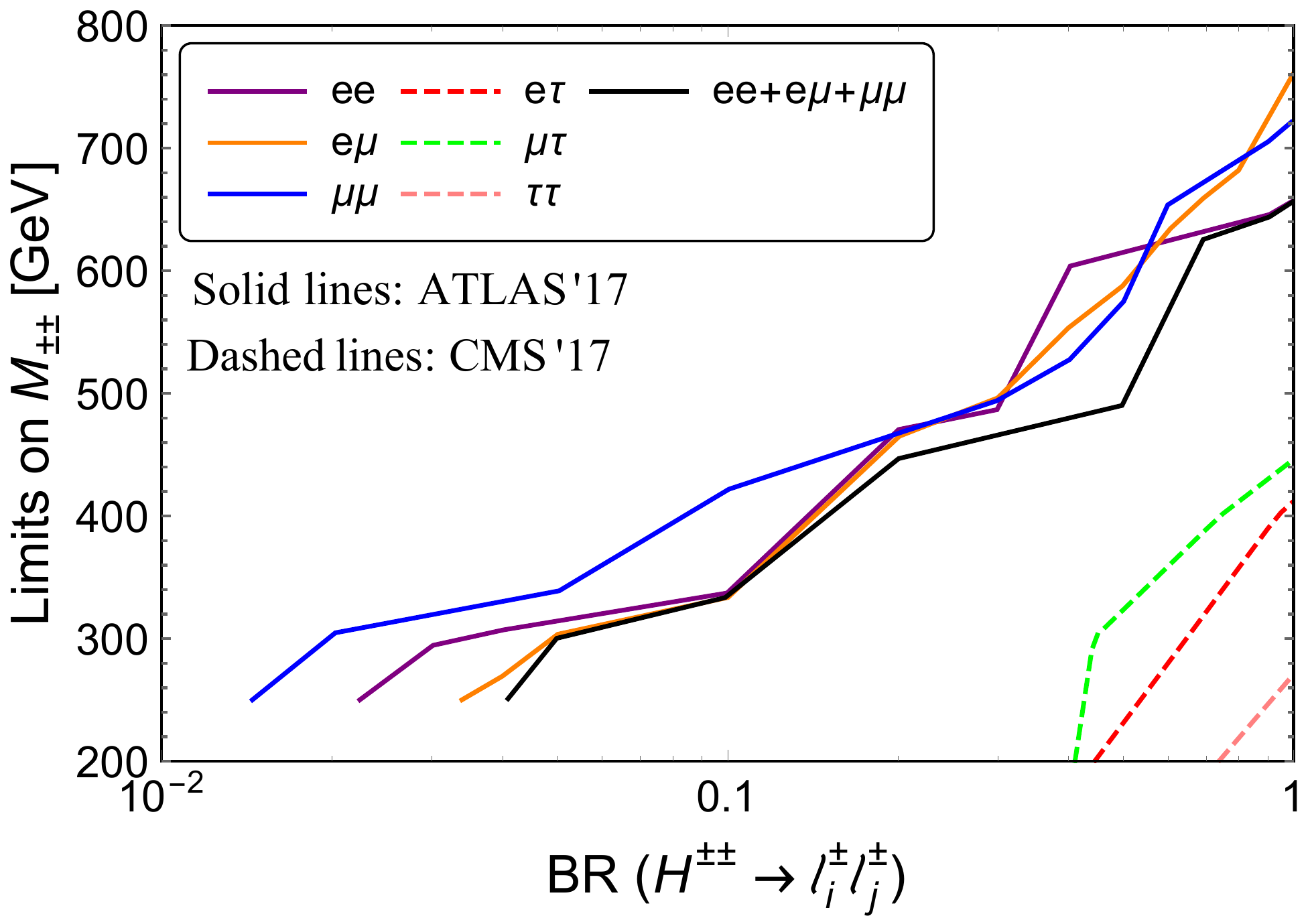}
  \caption{Same-sign dilepton limits on the mass $M_{\pm\pm}$ of RH doubly-charged scalar from ATLAS~\cite{ATLAS:2017iqw} (solid) and CMS~\cite{CMS:2017pet} (dashed), as functions of BR$(H^{\pm\pm} \to \ell_\alpha^\pm \ell_\beta^\pm)$. The black solid line combines the three decay modes of $ee$, $e\mu$ and $\mu\mu$.}
  \label{fig:limits}
\end{figure}

The latest dilepton searches of $H^{\pm\pm} \to \ell_\alpha^\pm \ell_\beta^\pm$ at the LHC can be found in Refs.~\cite{ATLAS:2017iqw, CMS:2017pet}, with the doubly-charged scalar pair produced form the Drell-Yan process $pp \to \gamma^\ast/Z^\ast \to H^{++} H^{--}$. The limits on the doubly-charged scalar mass $M_{\pm\pm}$ for all the six combinations of final states $ij = ee,\, e\mu,\, \mu\mu,\, e\tau,\, \mu\tau,\, \tau\tau$ are collected in Figure~\ref{fig:limits}, as well as the combined limit from $ee + e\mu + \mu\mu$, as functions of the corresponding BRs. For simplicity, we assume that only one of the six is open at a time and all the others are vanishing. The final states involving only the $e$ and $\mu$ flavors are the most stringent~\cite{ATLAS:2017iqw} and those with the $\tau$ flavor are much weaker~\cite{CMS:2017pet}, as a result of the poor reconstruction efficiency of the $\tau$ lepton at hadron colliders. Note that all the limits in Ref.~\cite{CMS:2017pet} are specific to the left-handed doubly-charged scalar; to interpret these limits onto the RH $H^{\pm\pm}$, we rescale down the pair production cross section $\sigma (pp \to \gamma^\ast/Z^\ast \to H^{++} H^{--})$ by a factor of 2.3, to take into account the different couplings of left- and right-handed doubly-charged scalars to the $Z$ boson.\footnote{In the LRSM, the singly-charged scalar $H^\pm$ has almost degenerate mass with the heavy neutral scalars $H$ and $A$ from the bidoublet, which is required to be beyond roughly 10 TeV by the flavor data~\cite{Zhang:2007da};  thus the single production of $H^{\pm\pm}$ in associated production with $H^\pm$ is highly suppressed, and those single production data in~\cite{CMS:2017pet} are not applicable to the LRSM, though the limits tend to be more stringent than the double production data.} All the limits in Figures 5 and 6 of~\cite{CMS:2017pet} for $H^{\pm\pm} \to e^\pm\tau^\pm,\, \mu^\pm\tau^\pm,\, \tau^\pm\tau^\pm$ have assumed a BR of 100\% into each of the final states. To obtain the constraints in Figure~\ref{fig:limits} as functions of the BRs, we rescale further down the theoretical predictions for the pair production cross sections by a factor of ${\rm BR}^2$. For a doubly-charged scalar mass of 200 GeV, the BR for the final states involving the tauon flavor could go down to 0.45, 0.41 and 0.74 for respectively $e\tau$, $\mu\tau$ and $\tau\tau$, which is much weaker than the constraints from Ref.~\cite{ATLAS:2017iqw} with only electrons and muons, as expected: If the BRs get smaller, the production cross sections are too small to be constrained by the experimental data.

The LFV couplings $f_{\alpha\beta}$ (with $\alpha \neq \beta$) could induce rare flavor violating decays and anomalous magnetic moments which are highly suppressed in the SM. The partial width for the tree level three-body decay $\ell_\alpha \to \ell_\beta \ell_\gamma \ell_\delta$ is~\cite{Akeroyd:2009nu, Dinh:2012bp}
\begin{eqnarray}
{\rm BR} (\ell_\alpha^- \to \ell_\beta^- \ell_\gamma^+ \ell_\delta^-)
p& \ \simeq \ &
\frac{|f_{\alpha\gamma}|^2 |f_{\beta\delta}|^2}{2 (1+ \delta_{\beta\delta}) G_F^2 M_{{\pm\pm}}^4} \nonumber \\
&& \times
{\rm BR} (\ell_\alpha \to e\nu\bar\nu) \,,
\end{eqnarray}
with $G_F$ the Fermi constant and $\delta_{\beta\delta} = 1 \, (0)$ for $\beta = \delta$ ($\beta\neq\delta$) the symmetry factor. As the doubly-charged scalar mass scale is much larger than the charged lepton masses, the constraints on $|f^\dagger f|/M_{\pm\pm}^2$ are almost constants, which correspond to an effective cut-off scale of $\Lambda \simeq M_{\pm\pm} / \sqrt{|f^\dagger f|}$. All the current experimental data on the rare muon and tauon decays and the corresponding upper limits on $|f^\dagger f|/M_{\pm\pm}^2$ are collected in Table~\ref{tab:limits}.

\begin{table*}[!t]
  \centering
  \caption[]{Current experimental data of the rare LFV decays $\ell_\alpha \to \ell_\beta \ell_\gamma \ell_\delta$, $\ell_\alpha \to \ell_\beta \gamma$~\cite{PDG, Amhis:2016xyh}, the electron~\cite{Hanneke:2008tm} and muon~\cite{Bennett:2006fi} $g-2$, muonium oscillation~\cite{Willmann:1998gd} and the LEP $e^+ e^- \to \ell^+ \ell^-$ data~\cite{Abdallah:2005ph}, and the resultant constraints on the couplings $|f^\dagger f|/M_{\pm\pm}^2$ for the doubly-charged scalar.}
  \label{tab:limits}
  \begin{tabular}[t]{lll}
  \hline\hline
  process & current data & constraints [${\rm GeV}^{-2}$] \\ \hline
  $\mu^- \to e^- e^+ e^-$ & $< 1.0 \times 10^{-12}$ & $|f_{ee}^\dagger f_{e\mu}|/M_{\pm\pm}^2 < 2.3 \times 10^{-11}$ \\ \hline

  $\tau^- \to e^- e^+ e^-$ & $< 1.4 \times 10^{-8}$ & $|f_{ee}^\dagger f_{e\tau}|/M_{\pm\pm}^2 < 6.6 \times 10^{-9}$ \\
  $\tau^- \to e^- \mu^+ \mu^-$ & $< 1.6 \times 10^{-8}$ & $|f_{e\mu}^\dagger f_{\mu\tau}|/M_{\pm\pm}^2 < 4.0 \times 10^{-9}$ \\
  $\tau^- \to \mu^- e^+ \mu^-$ & $< 9.8 \times 10^{-9}$ & $|f_{e\tau}^\dagger f_{\mu\mu}|/M_{\pm\pm}^2 < 5.2 \times 10^{-9}$ \\

  $\tau^- \to \mu^- e^+ e^-$ & $< 1.1 \times 10^{-8}$ & $|f_{e\mu}^\dagger f_{e\tau}|/M_{\pm\pm}^2 < 5.0 \times 10^{-9}$ \\
  $\tau^- \to e^- \mu^+ e^-$ & $< 8.4 \times 10^{-9}$ & $|f_{ee}^\dagger f_{\mu\tau}|/M_{\pm\pm}^2 < 5.5 \times 10^{-9}$ \\
  $\tau^- \to \mu^- \mu^+ \mu^-$ & $< 1.2 \times 10^{-8}$ & $|f_{\mu\mu}^\dagger f_{\mu\tau}|/M_{\pm\pm}^2 < 5.1 \times 10^{-9}$ \\ \hline

  $\mu^- \to e^- \gamma$ & $< 4.2 \times 10^{-13}$ & $|\sum_k f_{ek}^\dagger f_{\mu k}|/M_{\pm\pm}^2 < 2.7 \times 10^{-10}$ \\
  $\tau^- \to e^- \gamma$ & $< 3.3 \times 10^{-8}$ & $|\sum_k f_{ek}^\dagger f_{\tau k}|/M_{\pm\pm}^2 < 1.8 \times 10^{-7}$ \\
  $\tau^- \to \mu^- \gamma$ & $< 4.4 \times 10^{-8}$ & $|\sum_k f_{\mu k}^\dagger f_{\tau k}|/M_{\pm\pm}^2 < 2.1 \times 10^{-7}$ \\ \hline

  electron $g-2$ & $< 5.2 \times 10^{-13}$ & $\sum_k |f_{ek}|^2/M_{\pm\pm}^2 < 1.2 \times 10^{-4}$ \\
  muon $g-2$ & $< 4.0 \times 10^{-9}$ & $\sum_k |f_{\mu k}|^2/M_{\pm\pm}^2 < 1.7 \times 10^{-5}$ \\ \hline

  muonium oscillation & $<8.2 \times 10^{-11}$ &  $|f_{ee}^\dagger f_{\mu\mu}|/M_{\pm\pm}^2 < 1.2 \times 10^{-7}$ \\ \hline

  $ee \to ee$ (LEP) & $\Lambda_{\rm eff} > 5.2$ TeV & $|f_{ee}|^2/M_{\pm\pm}^2 < 1.2 \times 10^{-7}$ \\
  $ee \to \mu\mu$ (LEP) & $\Lambda_{\rm eff} > 7.0$ TeV & $|f_{e\mu}|^2/M_{\pm\pm}^2 < 6.4 \times 10^{-8}$ \\
  $ee \to \tau\tau$ (LEP) & $\Lambda_{\rm eff} > 7.6$ TeV & $|f_{e\tau}|^2/M_{\pm\pm}^2 < 5.4 \times 10^{-8}$ \\
  \hline\hline
  \end{tabular}
\end{table*}
At 1-loop level, the LFV couplings contribute to the two-body decays~\cite{Mohapatra:1992uu}
\begin{eqnarray}
{\rm BR} (\ell_\alpha \to \ell_\beta \gamma)
& \ \simeq \ &
\frac{\alpha_{\rm EM} |\sum_\delta f_{\alpha\delta}^\dagger f_{\beta\delta}|^2}{3\pi G_F^2 M_{\pm\pm}^4} \nonumber \\
&& \times p{\rm BR} (\ell_\alpha \to e\nu\bar\nu) \,,
\end{eqnarray}
where we have summed up all the diagrams involving a $\ell_\delta$ lepton running in the loop. The experimental data of $\mu \to e \gamma$, $\tau \to e\gamma$ and $\tau \to \mu \gamma$ could be used to set limits on the couplings $|\sum_\delta f_{\alpha\delta}^\dagger f_{\beta\delta}| /M_{\pm\pm}^2$, which are also presented in Table~\ref{tab:limits}. As a result of the loop factor, the constraints on the Yukawa couplings $|f^\dagger f|$ are one or two orders of magnitude weaker than those from the three-body decays $\ell_\alpha \to \ell_\beta \ell_\gamma \ell_\delta$, as shown in Table~\ref{tab:limits}.

In an analogous way, we can calculate the contributions of the doubly-charged scalar loops to the anomalous magnetic moments of electron and muon (with $\alpha=e,\,\mu$)~\cite{Leveille:1977rc, Moore:1984eg, Gunion:1989in}:
%The contribution of doubly-charged scalar loops to the electron $g-2$ is
\begin{eqnarray}
\label{eqn:g-2}
\Delta a_\alpha & \ \simeq \ &
- \frac{m_{\ell_\alpha}^2}{6\pi^2 \, M_{\pm\pm}^2} \sum_\beta |f_{\alpha\beta}|^2 \,,
\end{eqnarray}
where we have summed up the loops involving all the three flavors $\beta = e,\, \mu,\, \tau$. The current $2\sigma$ experimental uncertainty $\Delta a_e = 5.2 \times 10^{-13}$~\cite{PDG} can be used to set limits on the couplings $\sum_\beta |f_{e\beta}|^2$ as a function of the doubly-charged scalar mass. As the contributions from the doubly-charged scalar loops are always negative, the controversial theoretical and experimental discrepancy $\Delta a_\mu = (2.87 \pm 0.80) \times 10^{-9}$ can not be explained; we use instead the $5\sigma$ uncertainty of $5 \times 0.80 \times 10^{-9}$ to constrain the Yukawa couplings, as shown in Table~\ref{tab:limits}.

The muonium-antimuonium oscillation, i.e. the LFV conversion of the bound states $(\mu^+ e^-) \leftrightarrow (\mu^- e^+)$, can be induced by the effective four-fermion Lagrangian, which arises from the exchange of doubly-charged scalars in the LRSM~\cite{HM};
\begin{eqnarray}
{\cal L}_{M\overline{M}} \ = \
\frac{G_{M\overline{M}}}{\sqrt2} \Big[ \bar{\mu} \gamma_\alpha (1+\gamma_5) e \Big]
\Big[ \bar{\mu} \gamma^\alpha (1+\gamma_5) e \Big]
\end{eqnarray}
with the oscillation probability~\cite{Swartz:1989qz, Clark:2003tv}
\begin{eqnarray}
{\cal P} \ \simeq \
\frac{(\Delta M)^2}{2\Gamma_\mu^2 } \,,
\end{eqnarray}
where the mass splitting
\begin{eqnarray}
\Delta M  \ = \
2 \langle \overline{M} | {\cal L}_{M\overline{M}} | M \rangle \ = \
\frac{16 G_{M\overline{M}}}{\sqrt2 \pi a^3} \,,
\end{eqnarray}
with $a = (\alpha \mu)^{-1}$ and $\mu = m_e m_\mu / (m_e + m_\mu)$ the effective mass. By performing a Fierz transformation, the effective coefficient is related to the couplings and doubly-charged scalar mass via
\begin{eqnarray}
G_{M\overline{M}} \ = \
\frac{f_{ee} f_{\mu\mu}^\dagger}{4\sqrt2 M_{\pm\pm}^2} \,.
\end{eqnarray}
The MACS experiment~\cite{Willmann:1998gd} sets a 90\% C.L. upper bound of ${\cal P}< 8.2 \times 10^{-11}$, which requires that $|f_{ee}^\dagger f_{\mu\mu}| / M_{\pm\pm}^2 < 1.2 \times 10^{-7} \, {\rm GeV}^{-2}$, as shown in Table~\ref{tab:limits}.

There are also direct searches of doubly-charged scalars at LEP in the single~\cite{Abbiendi:2003pr} or pair~\cite{Achard:2003mv} production mode; limited by the center-of-mass energy, the constraints are very weak. An off-shell $H^{\pm\pm}$ in the $t$-channel could mediate the Bhabha scattering $e^+ e^- \to e^+ e^-$ and interfere with the SM diagrams. This alters both the total cross section and the differential distributions~\cite{Abbiendi:2003pr, Achard:2003mv}. If the Yukawa coupling $f_{ee}$ is of order one, the doubly-charged scalar $H^{\pm\pm}$ could be probed up to the TeV scale. By Fierz transformation, the doubly-charged scalar contributes to the effective contact four-fermion interaction
\begin{eqnarray}
\frac{1}{\Lambda_{\rm eff}^2} (\bar{e}_R \gamma_\mu e_R) (\bar{f}_R \gamma^\mu f_R) \,,
\label{eq:eff}
\end{eqnarray}
and is constrained by the $ee \to \ell\ell$ (with $\ell\ell = ee,\, \mu\mu,\, \tau\tau$) data in Ref.~\cite{Abdallah:2005ph}, and $\Lambda_{\rm eff} \sim M_{\pm\pm}/|f_{e\ell}|$ corresponds to the effective cutoff scale related to the doubly-charged scalar mass and the Yukawa couplings. It turns out the the LEP data in Ref.~\cite{Abdallah:2005ph} could provide more stringent limits than those in Refs.~\cite{Abbiendi:2003pr, Achard:2003mv}, so we list in Table~\ref{tab:limits} only the limits on the cutoff scale $\Lambda_{\rm eff}$ from~\cite{Abdallah:2005ph} and the consequent constraints on $|f_{e\ell}|^2/M_{\pm\pm}^2$.\footnote{For $f=e$, the effective interaction in Eq.~\eqref{eq:eff} would also induce an additional contribution to the M{\o}ller scattering and can be constrained by the upcoming MOLLER experiment~\cite{Benesch:2014bas}, which could probe the effective scale $\Lambda_{\rm eff} \simeq 5.3$ TeV, slightly stronger than the current limit from LEP data~\cite{Dev:2018sel}.}

\subsection{Production at colliders}
\label{sec:production}

\subsubsection{Pair production}

\begin{figure}[t!]
  \centering
  \includegraphics[width=0.45\textwidth]{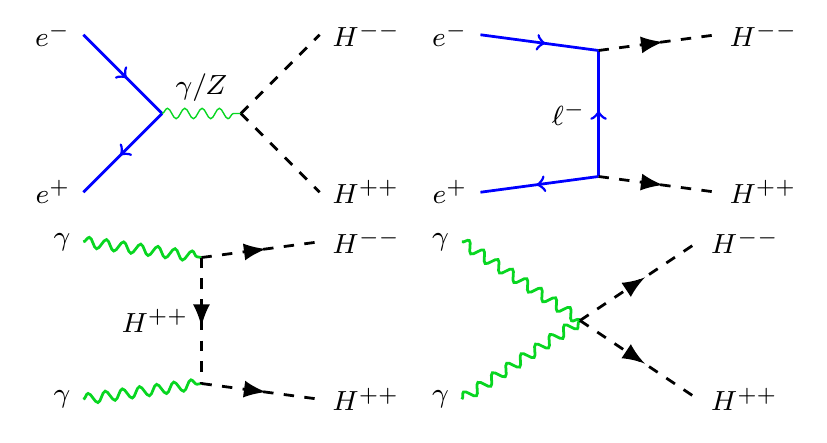} \vspace{-5pt}
  \caption{Feynman diagrams for the pair production of doubly-charged scalar $H_2^{\pm\pm}$ at lepton colliders through the Drell-Yan process (top left), the Yukawa couplings $f_{\alpha\beta}$ to the charged leptons (top right), and photon fusion (lower panels).}
  \label{fig:diagram:Hpp:1}
\end{figure}

In the LRSM, the doubly-charged scalar $H^{\pm\pm}$ can be pair-produced at lepton colliders through the Drell-Yan process and the Yukawa interaction $f_{e\ell}$ to the SM charged fermions~\cite{Kuze:2002vb}, with $\ell$ covering all the three flavors of $e$, $\mu$ and $\tau$. The Feynman diagrams are shown in the upper panels of Figure~\ref{fig:diagram:Hpp:1}. At ILC, high luminosity photon beams can be obtained by Compton backscattering of low energy, high intensity laser beam off the high energy electron beam, and then the doubly-charged scalar can be pair-produced from the ``photon fusion'' processes as shown in the lower panels of Figure~\ref{fig:diagram:Hpp:1}~\cite{Chakrabarti:1998qy}, including both the trilinear and quartic gauge-scalar couplings.\footnote{{The doubly-charged scalar contributes to the light-by-light scattering at future lepton colliders and interferes with the SM processes, see e.g. Ref.~\cite{TavaresVelasco:1999xs}.}} The effective photon luminosity distribution is given in Eq.~(\ref{eqn:photon_pdf}) above~\cite{Ginzburg:1981vm, Ginzburg:1982yr, Telnov:1989sd}.
%\begin{eqnarray}
%\label{eqn:photon_pdf}
%f_{\gamma/e} (x) \ = \
%\frac{1}{D(\xi)} \left[ (1-x) + \frac{1}{(1-x)}
%- \frac{4x}{\xi (1-x)} + \frac{4x^2}{\xi^2 (1-x)^2} \right] \,,
%\end{eqnarray}
%with
%\begin{eqnarray}
%D (\xi) \ = \
%\left( 1 - \frac{4}{\xi} - \frac{8}{\xi^2} \right) \log (1+\xi)
%+ \frac12 + \frac{8}{\xi} - \frac{1}{2(1+\xi)^2} \,,
%\end{eqnarray}
%where $x =\omega/E_e$ is the fraction of electron energy carried away by the scattered photon, with $\omega$ and $E_e$ respectively the energies of scattered photon and initial electron. The parameter $\xi = 4\omega_0 E_e / m_e^2$ depends on the energy $\omega_0$ of initial laser photon. When $\xi \gtrsim 4.8$ the photon conversion efficiency drops drastically, as a consequence of the $e^+ e^-$ pair production from the laser photons and the photon backscattering, which sets an upper bound on the energy fraction $x < x_{\rm max} = \xi / (1+\xi) \simeq 0.83$.

\begin{figure}[t!]
  \centering
  \includegraphics[width=0.4\textwidth]{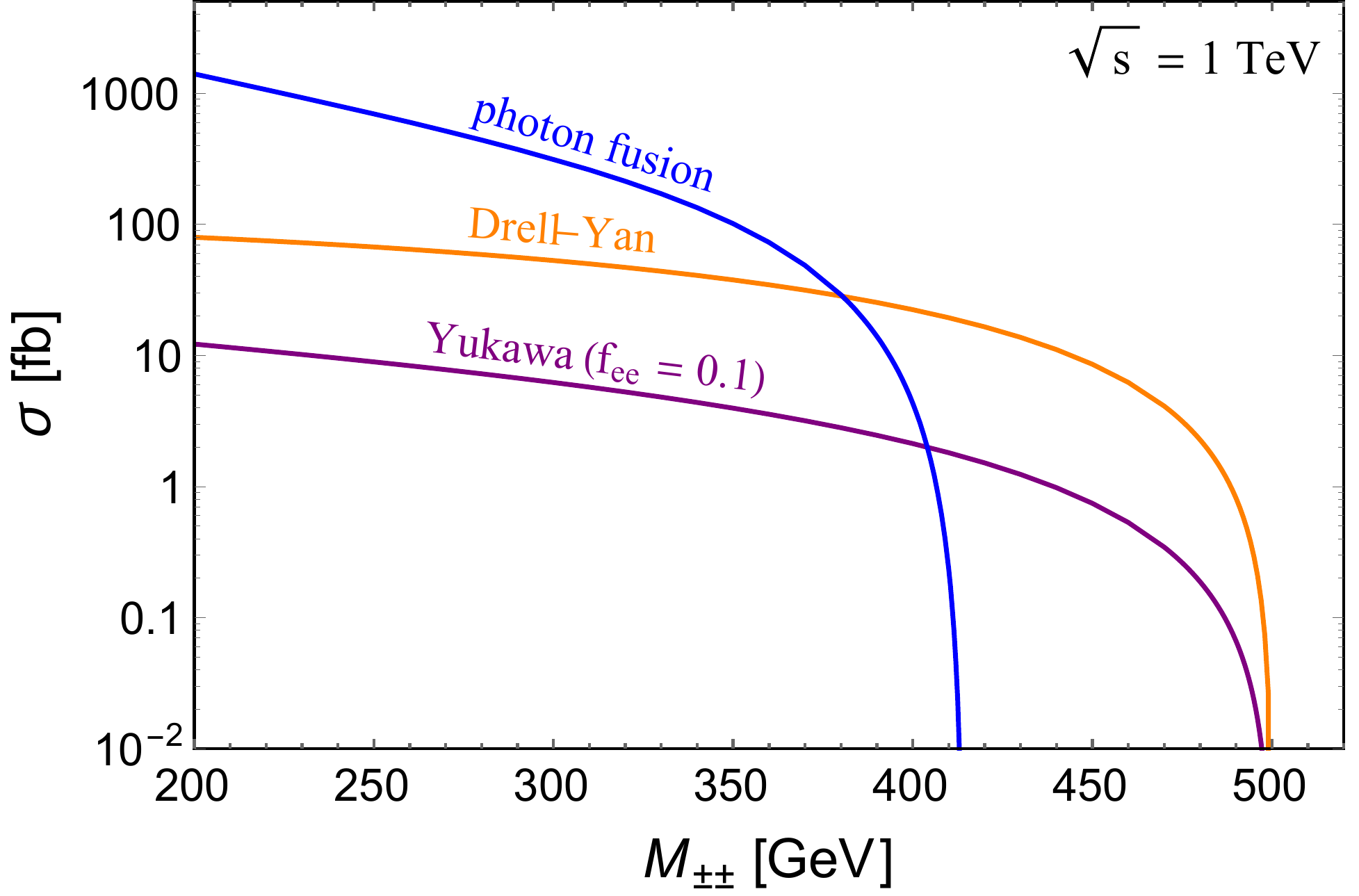}
  \caption{Pair production cross sections of the doubly-charged scalar $H^{\pm\pm}$ at ILC 1 TeV through the Drell-Yan process, the Yukawa couplings to the charged leptons (assuming $f_{ee} = 0.1$ and other couplings vanishing), and photon fusion.  {For other values of $f_{ee}$, the cross section in the Yukawa channel can be simply rescaled by a factor of $(f_{ee}/0.1)^2$, though $f_{ee} = 0.1$ has been excluded for some ranges of $M_{\pm\pm}$ (see Figure~\ref{fig:prospect:Hpp:1}).} See Figure~\ref{fig:diagram:Hpp:1} for the Feynman diagrams and text for more details.}
  \label{fig:prod:Hpp:1}
\end{figure}

The direct dilepton mass limits on the doubly-charged scalar mass $M_{\pm\pm}$ depend largely on the BRs of $H^{\pm\pm}$ decaying into different flavors of SM charged leptons; in particular, if $H^{\pm\pm}$ decays mostly into the tauon flavor, the same-sign dilepton limits are expected to be roughly below 440 GeV, as seen in Figure~\ref{fig:limits}, and $H^{\pm\pm}$ can be pair-produced on-shell at ILC running of $\sqrt{s} = 1$ TeV (for the ``photon fusion'' process, the center-of-mass energy can only go effectively up to $1 \, {\rm TeV}\times x_{\rm max} \simeq 830\, {\rm GeV}$). The total cross sections at ILC 1 TeV for the pair production of doubly-charged scalars through the Drell-Yan process, the Yukawa couplings and the photon fusion are presented in Figure~\ref{fig:prod:Hpp:1}. For concreteness, we have set $f_{ee} = 0.1$ and $f_{e\mu} = f_{e\tau} = 0$ in the Yukawa channel; for other specific values of the Yukawa couplings the cross section has to be rescaled by the factor of $|\sum_\ell f_{e\ell}|^2/0.1^4$. The electron/positron beams have been assumed to be unpolarized. For polarized beams, the cross section for the Yukawa coupling portal has to be rescaled by a factor of $(1+P_{e^-})(1-P_{e^+})$.
%For the purpose of illustration, we have also shown in Fig.~\ref{fig:prod:Hpp:1} the dilepton limits with respectively the BRs of $H^{\pm\pm}$ decaying into $\tau\tau$ of 100\%, and $\ell\ell$ of 5\%, 10\% and 20\% with $\ell = e,\, \mu$ (for the last three cases, the BRs into $\tau\tau$ are respectively 95\%, 90\% and 80\% and the tau flavor limits are much weaker those from the electrons and muons; see Fig.~\ref{fig:limits}). All of these limits are below 450 GeV, and there is large parameter space in the doubly-charged scalar sector of minimal LRSM to be probed in all the pair production channels.

\subsubsection{Single production}

\begin{figure*}[t!]
  \centering
  \includegraphics[width=0.75\textwidth]{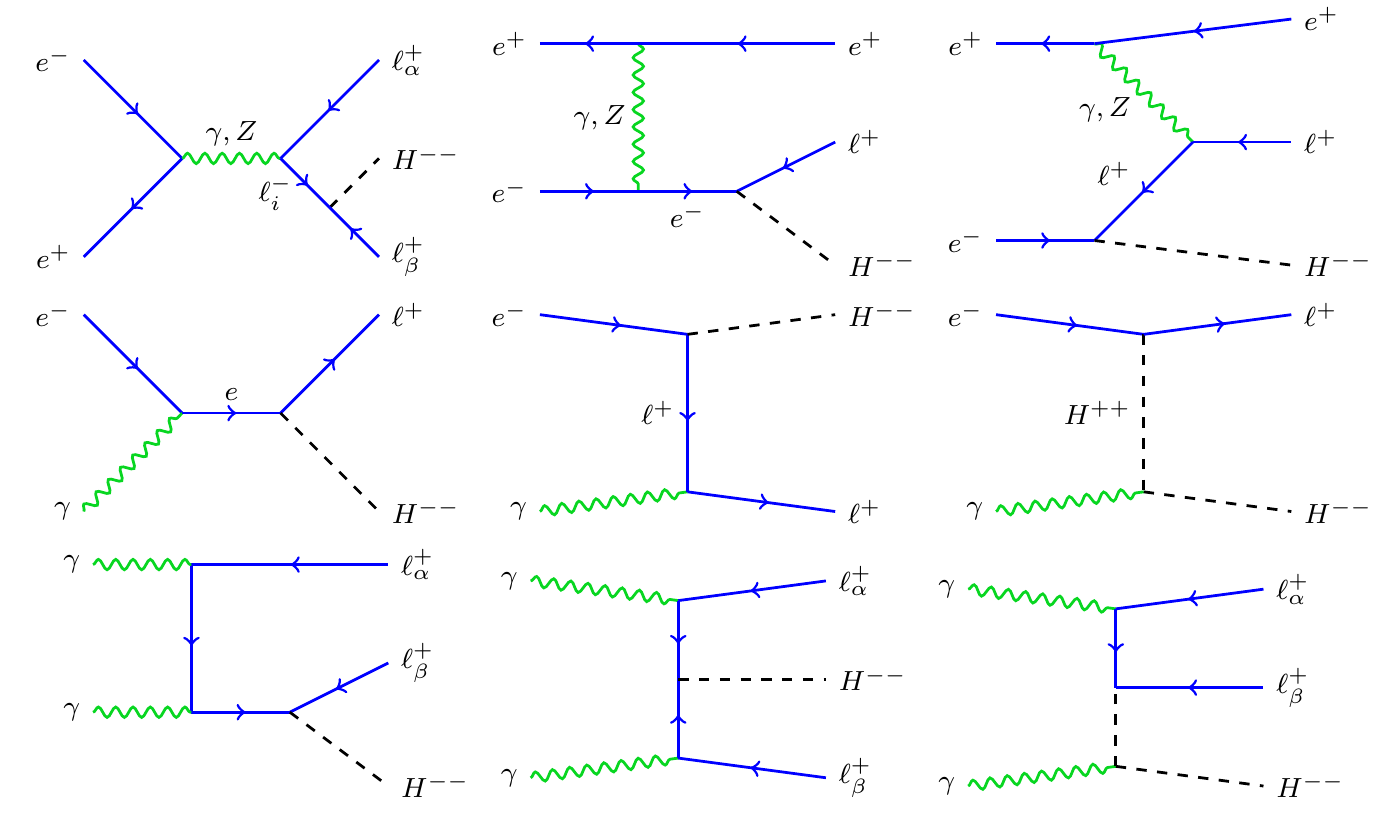}
  \caption{Representative Feynman diagrams for the single production of $H^{\pm\pm}$ at lepton colliders from the $e^+ e^-$ (upper) and $e\gamma$ (middle) and $\gamma\gamma$ (lower) processes.}
  \label{fig:diagram:Hpp:2}
\end{figure*}

When the doubly-charged scalar is too heavy to be pair-produced at lepton colliders, it could be singly produced from $e^+ e^-$~\cite{Barenboim:1996pt, Kuze:2002vb, Lusignoli:1989tr, Yue:2007kv}, $e\gamma$~\cite{Barenboim:1996pt, Godfrey:2001xb, Godfrey:2002wp, Rizzo:1981xx, Yue:2007ym} and $\gamma\gamma$ collisions, i.e.\footnote{{In the $e^+ e^-$ single production of doubly-charged scalar, we have the subprocesses $e\gamma^\ast \to \ell_\alpha H^{\pm\pm}$, with the virtual photon emitted from the initial electron/positron~\cite{Rizzo:1981xx, Lusignoli:1989tr} (see the second and third diagram in the upper row of Figure~\ref{fig:diagram:Hpp:2}), which is different from the laser photon in the $e\gamma$ collision in Eq.~(\ref{eqn:egamma}).}}
\begin{eqnarray}
\label{eqn:egamma}
e^+ e^- & \ \to \ & H^{\pm\pm} \ell_\alpha^\mp \ell_\beta^\mp \,, \nonumber \\
e^\pm \gamma & \ \to \ & H^{\pm\pm} \ell_\alpha^\mp \,, \nonumber \\
\gamma\gamma & \ \to \ & H^{\pm\pm} \ell_\alpha^\mp \ell_\beta^\mp \,,
\end{eqnarray}
with $\gamma$ the laser photon beam as in the pair production case. Some of the  representative Feynman diagrams are shown in Figure~\ref{fig:diagram:Hpp:2}. For the $e^+ e^-$ single production, it is in fact the Drell-Yan process $ee \to \ell\ell$ with one doubly-charged scalar emitted from the initial or final state and possibly changing the lepton flavor~\cite{Abbiendi:2003pr}; see the first row of diagrams in Figure~\ref{fig:diagram:Hpp:2}. The process  $e^+ e^- \to H^{++ \, \ast} H^{--}/H^{++} H^{-- \, \ast}$ should also be taken into consideration, with one of the doubly-charged scalar in the final state on-shell and the other one off-shell (cf. the first diagram in Figure~\ref{fig:diagram:Hpp:1}). Similarly, for the $\gamma\gamma$ single production, we can have the doubly-charged scalar produced by couplings to the charged leptons in $\gamma\gamma \to \ell\ell$ (the lower row in Figure~\ref{fig:diagram:Hpp:2}), or with one off-shell doubly-charged scalar in the pair production like $\gamma\gamma \to H^{++ \, \ast}H^{--}$ (cf. the last two diagrams in Figure~\ref{fig:diagram:Hpp:1}). The three diagrams in the middle row of Figure~\ref{fig:diagram:Hpp:2} are examples for the single production $e^\pm \gamma \to H^{\pm\pm} \ell_\alpha^\mp$.

\begin{figure}[t!]
  \centering
  \includegraphics[width=0.4\textwidth]{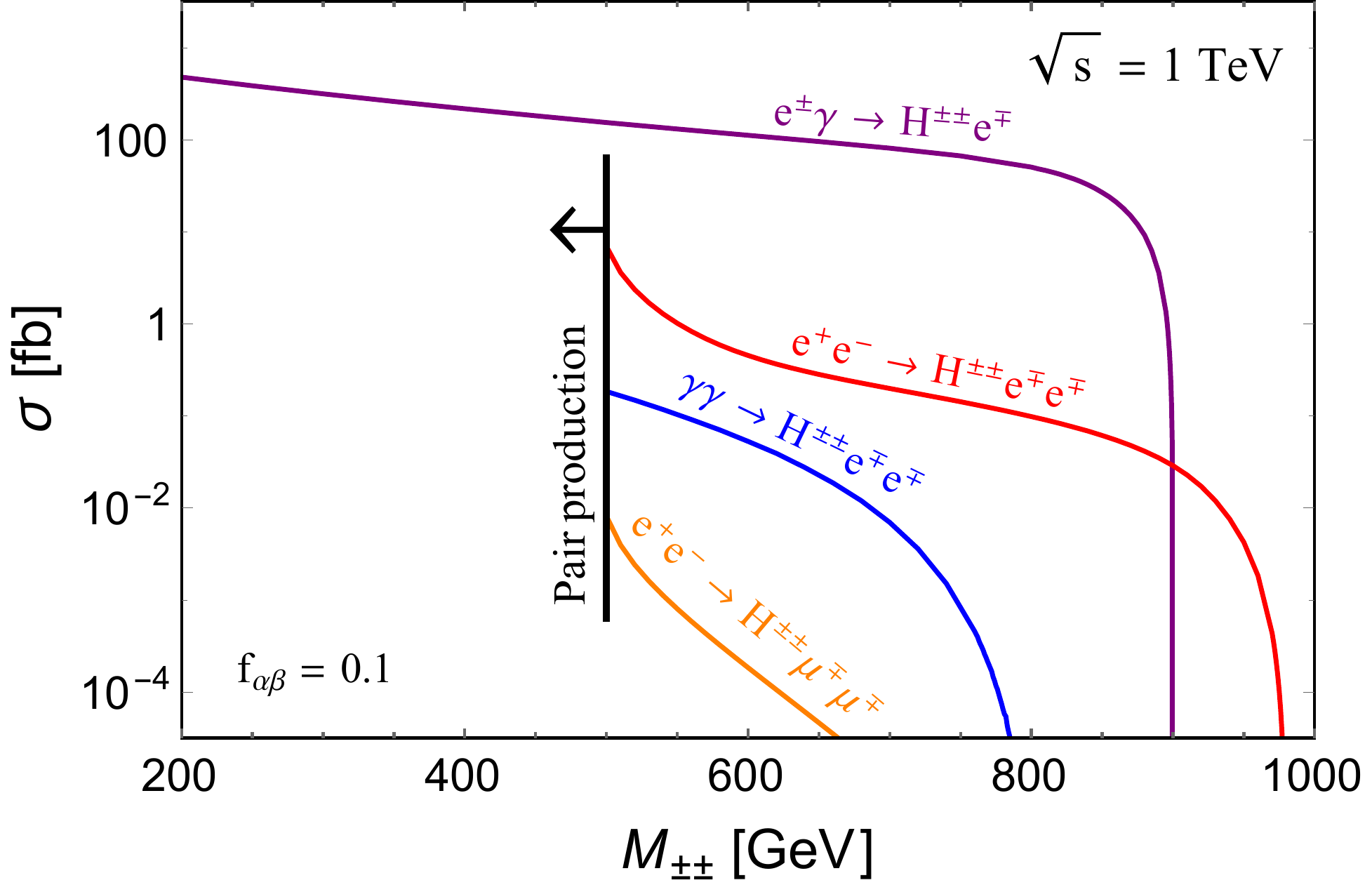}
  \caption{Single production cross sections of the doubly-charged scalar $H^{\pm\pm}$ at ILC 1 TeV through the $e^+ e^-$, $e\gamma$ and $\gamma\gamma$ processes, with all the Yukawa couplings relevant set to be $0.1$. On the left of the thick black line, the $ee$ and $\gamma\gamma$ processes are dominated by the pair production modes. {For other values of $f_{\alpha\beta}$, the cross section can be simply rescaled by a factor of $(f_{\alpha\beta}/0.1)^2$, though $f_{\alpha\beta} = 0.1$ has been excluded for some ranges of $M_{\pm\pm}$ (see Figure~\ref{fig:prospect:Hpp:1}).} See Figure~\ref{fig:diagram:Hpp:2} for the Feynman diagrams and the text for more details. }
  \label{fig:prod:Hpp:2}
\end{figure}

The cross sections for the single production of $H^{\pm\pm}$ depend largely on the production modes, the charged lepton flavors and the Yukawa couplings $f_{\alpha\beta}$ involved. Some representative production cross sections at ILC 1 TeV as functions of the doubly-charged scalar mass are presented in Figure~\ref{fig:prod:Hpp:2}, with the nominal cuts of $p_T (\ell) > 10$ GeV and $|\eta (\ell)| < 2.5$. In specific, all the Yukawa couplings involved are assumed to be 0.1.
%and the reconstruction efficiency for the tauon lepton is assumed to be 60\%~\cite{Baer:2013cma}.
When the scalar mass is roughly below $\sqrt{s}/2 = 500$ GeV, the signal of $H^{\pm\pm} \ell_\alpha^\mp \ell_\beta^\mp$ from both the $e^+ e^-$ and $\gamma\gamma$ processes would be dominated by the pair production modes in Figure~\ref{fig:diagram:Hpp:1}, and the cross sections go much larger, as indicated by the thick black line in Figure~\ref{fig:prod:Hpp:2}.

For the $e\gamma$ channel, the production cross sections are almost the same for the three flavors $\ell = e,\, \mu,\, \tau$ in the final state, up to the reconstruction efficiencies (in particular for the tauon lepton) and the Yukawa couplings, e.g. to obtain the cross section $\sigma (e^\pm \gamma \to H^{\pm\pm} \tau^\mp)$, one has only to multiply the corresponding value for $\sigma (e^\pm \gamma \to H^{\pm\pm} e^\mp)$ in Figure~\ref{fig:prod:Hpp:2} by a factor of $\epsilon_\tau \times |f_{e\tau}|^2/0.01$, with $\epsilon_\tau$ the $\tau$ efficiency factor. One should note that due to $x_{\max} \simeq 0.83 <1$ for the laser photon, the center-of-mass energy could go only up to $(500 \, {\rm GeV}) \oplus (500 \, {\rm GeV} \times x_{\max}) \simeq 911$ GeV. With the $p_T$ cut of 10 GeV on the charged leptons, in reality the doubly-charged scalar mass could only be probed up to roughly 900 GeV, as shown in Figure~\ref{fig:prod:Hpp:2}.

In the $e^+ e^- \to H^{\pm\pm} \ell_\alpha^\mp \ell_\beta^\mp$ single production of $H^{\pm\pm}$, we can have all the six distinct combinations of charged lepton flavors of $\alpha,\,\beta$ in the final state, i.e. $\ell_\alpha \ell_\beta = ee$, $e\mu$, $e\tau$, $\mu\mu$, $\mu\tau$, $\tau\tau$, which can be categorized into two groups: (a) those with an electron/positron in the final state ($ee$, $e\mu$, $e\tau$) and (b) those with only the muon or tauon flavors ($\mu\mu$, $\mu\tau$, $\tau\tau$). The production cross sections in each of the groups are roughly the same, up to the trivial symmetry factors for identical particles and the efficiency factors; two representative curves for $ee$ and $\mu\mu$ are shown in Figure~\ref{fig:prod:Hpp:2}. For instance, to obtain the cross section $\sigma (e^+ e^- \to H^{\pm\pm} e^\mp \mu^\mp)$, one needs only to multiply that for $ee$ process by a factor of  $2\epsilon_\mu \times |f_{e\mu}|^2$ with $\epsilon_\mu \simeq 1$ the muon efficiency factor. For the processes of the (b) group, only some of the diagrams in Figure~\ref{fig:diagram:Hpp:2} contribute, e.g. the first one, and the cross sections are much smaller than those of the (a) group involving an $e^\pm$, as clearly seen in Figure~\ref{fig:prod:Hpp:2} by comparing the cross sections for $ee$ and $\mu\mu$.

Regarding the production of a doubly-charged scalar in the process $\gamma \gamma \to H^{\pm\pm} \ell_\alpha^\mp \ell_\beta^\mp$, all the three lepton flavors are essentially on the same footing. In other words, all the fix different flavor combinations of $\ell_\alpha \ell_\beta$ share the same production cross sections as shown in Fig.~\ref{fig:prod:Hpp:2} (in the plot we have explicitly set $\alpha \beta = ee$ and $f_{ee} = 0.1$), up to the symmetry factors due to the identical particles $\alpha = \beta$, the Yukawa couplings $(f_{\alpha \beta}/0.1)^2$ and the efficiency factors. 
%Regarding the production of a doubly-charged scalar in the process $\gamma\gamma \to H^{\pm\pm} \ell_\alpha^+ \ell_\beta^+$, all the three lepton flavors are essentially on the same footing, i.e. all the six different flavor combinations of $\ell_\alpha \ell_\beta$ share the same production cross section as shown in Figure~\ref{fig:prod:Hpp:2} for the $ee$ case with explicitly $f_{ee} = 0.1$, up to the symmetry factors due to the identical particles $\alpha  = \beta$, the Yukawa couplings $(f_{\alpha\beta}/0.1)^2$ and the efficiency factors. 
As a result of $x_{\max} \simeq 0.83$ for the laser photons and the $p_T$ cuts, the effective center-of-mass energy could only go up to $\simeq 800$ GeV.

\subsubsection{Off-shell production}

\begin{figure}[t!]
  \centering
  \includegraphics[width=0.45\textwidth]{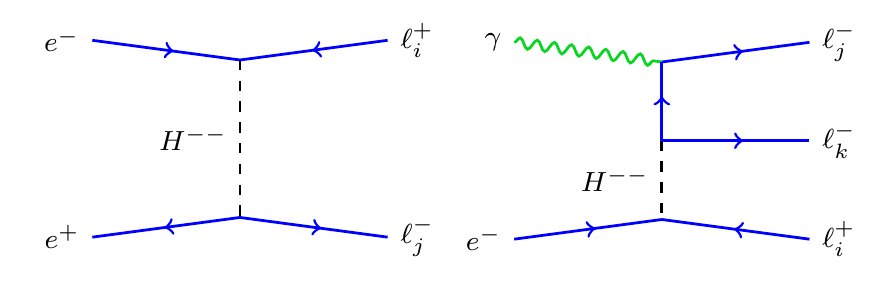}
  \caption{Feynman diagrams for the off-shell production of $H^{\pm\pm}$ at lepton colliders from the $e^+ e^-$ (left) and $e\gamma$ (right) processes.}
  \label{fig:diagram:Hpp:3}
\end{figure}

Even if the doubly-charged scalar is too heavy to be directly single/pair-produced on-shell at lepton colliders, it could still mediate the processes $e^+ e^- \to \ell_\alpha^\pm \ell_\beta^\mp$ in the $t$-channel~\cite{Kuze:2002vb} (if $\ell_\alpha \ell_\beta = e^+ e^-$, then the doubly-charged scalar contributes to the Bhabha scattering), as shown in the left panel of Figure~\ref{fig:diagram:Hpp:3} and induce LFV signals if $\alpha \neq \beta$. In an analogous way, we can have the trilepton processes $e^\pm \gamma \to \ell_\alpha^\mp \ell_\beta^\pm \ell_\gamma^\pm$~\cite{Godfrey:2001xb, Godfrey:2002wp, Rizzo:1981xx}\footnote{{It is also possible to have the off-shell production mode $\gamma\gamma \to \ell_\alpha^\pm \ell_\beta^\pm \ell_\gamma^\mp \ell_\delta^\mp$ with potentially LFV in the final state, but the cross sections are much smaller than the $ee$ and $e\gamma$ processes and are thus not considered here.}} with potential LFV signals if
\begin{eqnarray}
\alpha\beta\gamma & \ = \ &
ee\mu \,,\; ee\tau \,,\;\;
e\mu\mu \,,\;  e\mu\tau \,,\;  e\tau\tau \,, \nonumber \\
&& \mu ee \,,\; \tau ee \,,\;\; \mu e \tau \,,\; \tau e\mu \,, \;\;
\mu\mu\mu \,,\; \tau\tau\tau \,,\; \nonumber \\
&& \mu\mu\tau \,,\; \tau\tau\mu \,,\;
\mu\tau\tau \,,\; \tau\mu\mu \,,
\end{eqnarray}
as shown in the right panel of Figure~\ref{fig:diagram:Hpp:3}. As for the single production, here one has to take into account also the processes $e^\pm \gamma \to H^{\pm\pm \, \ast} \ell_\alpha^\mp \to \ell_\alpha^\mp \ell_\beta^\pm \ell_\gamma^\pm$ with the same-sign lepton pair $(\ell_\beta^\pm \ell_\gamma^\pm)$ originating from the off-shell $H^{\pm\pm}$, cf. the diagrams in the middle row of Figure~\ref{fig:diagram:Hpp:2}.

\begin{figure}[t!]
  \centering
  \includegraphics[width=0.4\textwidth]{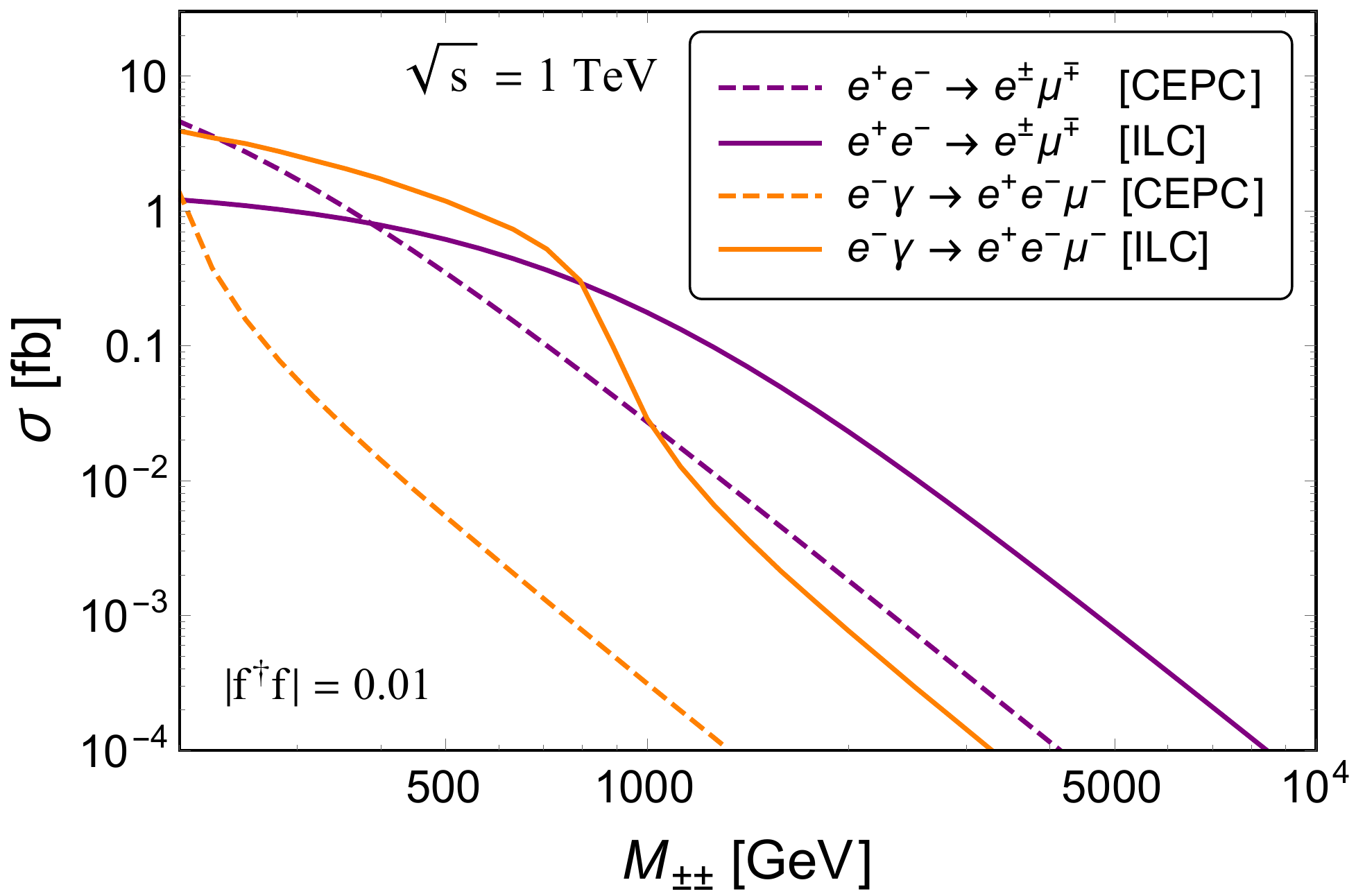}
  \caption{Single production cross sections of the doubly-charged scalar $H^{\pm\pm}$ at CEPC (dashed, $\sqrt{s} = 240$ GeV) and ILC (solid, $\sqrt{s} = 1$ TeV) through the $e^+ e^-$ (purple) and $e\gamma$ (orange) processes, with all the Yukawa couplings $|f^\dagger f|$ relevant set to be $0.01$. {For other values of $|f^\dagger f|$, the cross section can be simply rescaled by a factor of $(|f^\dagger f|/0.01)^2$, though $|f^\dagger f| = 0.01$ has been excluded for some ranges of $M_{\pm\pm}$ (see Figures~\ref{fig:prospect:Hpp:2} and \ref{fig:prospect:Hpp:3}).}  See Figure~\ref{fig:diagram:Hpp:2} for the Feynman diagrams and text for more details.}
  \label{fig:prod:Hpp:3}
\end{figure}

For illustration purpose, we show in Fig.~\ref{fig:prod:Hpp:3} the production cross sections for $e^+ e^- \to e^\pm \mu^\mp$ and $e^\pm \gamma \to e^\mp e^\pm \mu^\pm$ at both CEPC 240 GeV and ILC 1 TeV. To be concrete, the Yukawa couplings involved have set to be $|f^\dagger f| = 0.01$, for instance $|f_{ee}^\dagger f_{e\mu}| = 0.01$ in  the $e^\pm \mu^\mp$ channel (the solid and dashed purple curves in Fig.~\ref{fig:prod:Hpp:3}). 
%For the illustration purpose, we show in Figure~\ref{fig:prod:Hpp:3} the production cross sections for $e^+ e^- \to e^\pm \mu^\mp$ and $e^\pm \gamma \to e^\mp e^\pm \mu^\pm$ at both CEPC $240$ GeV and ILC $1$ TeV, with again the Yukawa couplings involved $|f^\dagger f| = 0.01$, for instance $f_{ee}^\dagger f_{e\mu} = 0.01$ for the purple curves in Figure~\ref{fig:prod:Hpp:3}. 
As stated above, all other cross sections can be trivially obtained by rescaling the curves by the proper symmetry factors, the Yukawa couplings squared, and the efficiency factors.

%With the couplings $f_{ei}$ and $f_{ej}$ we can obtain the cross sections for all the three LFV processes of $ij = e\mu,\, e\tau,\, \mu\tau$ at CEPC and ILC with respectively $\sqrt{s} = 240$ GeV and $\sqrt{s} = 1$ TeV, which are collected in Fig.~\ref{fig:prod:Hpp:3} for $f_{ei} = f_{ej} = 1$, with again the basic cuts of $p_T (\ell) > 10$ GeV, $|\eta (\ell)| < 2.5$ and an efficiency factor of 60\% for the tauon lepton at both the two colliders.

\subsection{Prospects and LFV signals}
\label{sec:prospect}

Given the cross sections in the previous subsection for all the different modes of pair production, single production and off-shell production of the RH doubly-charged scalar $H^{\pm\pm}$ in the LRSM, we estimate the prospects for the doubly-charged scalar at future colliders like CEPC and ILC, in particular with the LFV signals which are highly suppressed in the SM. All the current LFV constraints are collected in Table~\ref{tab:limits}. One should note that some of the limits in Table~\ref{tab:limits} depend on the products of two different entries in the Yukawa coupling matrix $f_{\alpha\beta}$, like the decays $\ell_\alpha \to \ell_\beta \ell_\gamma \ell_\delta$, $\ell_\alpha \to \ell_\beta \gamma$ and the muonium oscillation, while the $g-2$ and LEP $e^+e^- \to \ell^+ \ell^-$ data rely only on (combinations of) the single elements $f_{\alpha\beta}$.

Among all the on-shell and off-shell production modes in Section~\ref{sec:production}, only the Drell-Yan process $e^+ e^- \to H^{++} H^{--}$ and $\gamma\gamma \to H^{++} H^{--}$ do not depend on the Yukawa couplings $f_{\alpha\beta}$, but rather on the gauge couplings, c.f. the first, third and fourth diagrams in Figure~\ref{fig:diagram:Hpp:1}. As stated in Section~\ref{sec:constraints}, the doubly-charged scalar in the LRSM decays predominantly into the charged leptons, with the BR and flavor dependence given in Eq.~(\ref{eqn:BR}). By counting the event numbers of different lepton flavors, one could, at least in principle, fix the {\it relative} magnitudes of the elements $f_{\alpha\beta}$, e.g. the ratios $|f_{\alpha\beta}|^2 / |f_{ee}|^2$.\footnote{{In the left-right models, in the limit of small $W - W_R$ mixing and heavy $W_R$ boson, with sufficiently small Yukawa couplings $f_{\alpha\beta}$, the same-sign leptons from doubly-charged scalar decay will form displaced vertices~\cite{Dev:2018kpa}, which is expected to be a strikingly clear signal at lepton colliders. The decay length of doubly-charged scalar could, in principle, be used to determine the values of Yukawa couplings $|f_{\alpha\beta}|$ for a given doubly-charged scalar mass.}} The absolute values of $f_{\alpha\beta}$ have to been determined from other on-shell channels.

\begin{figure*}[t!]
  \centering
  \includegraphics[width=0.4\textwidth]{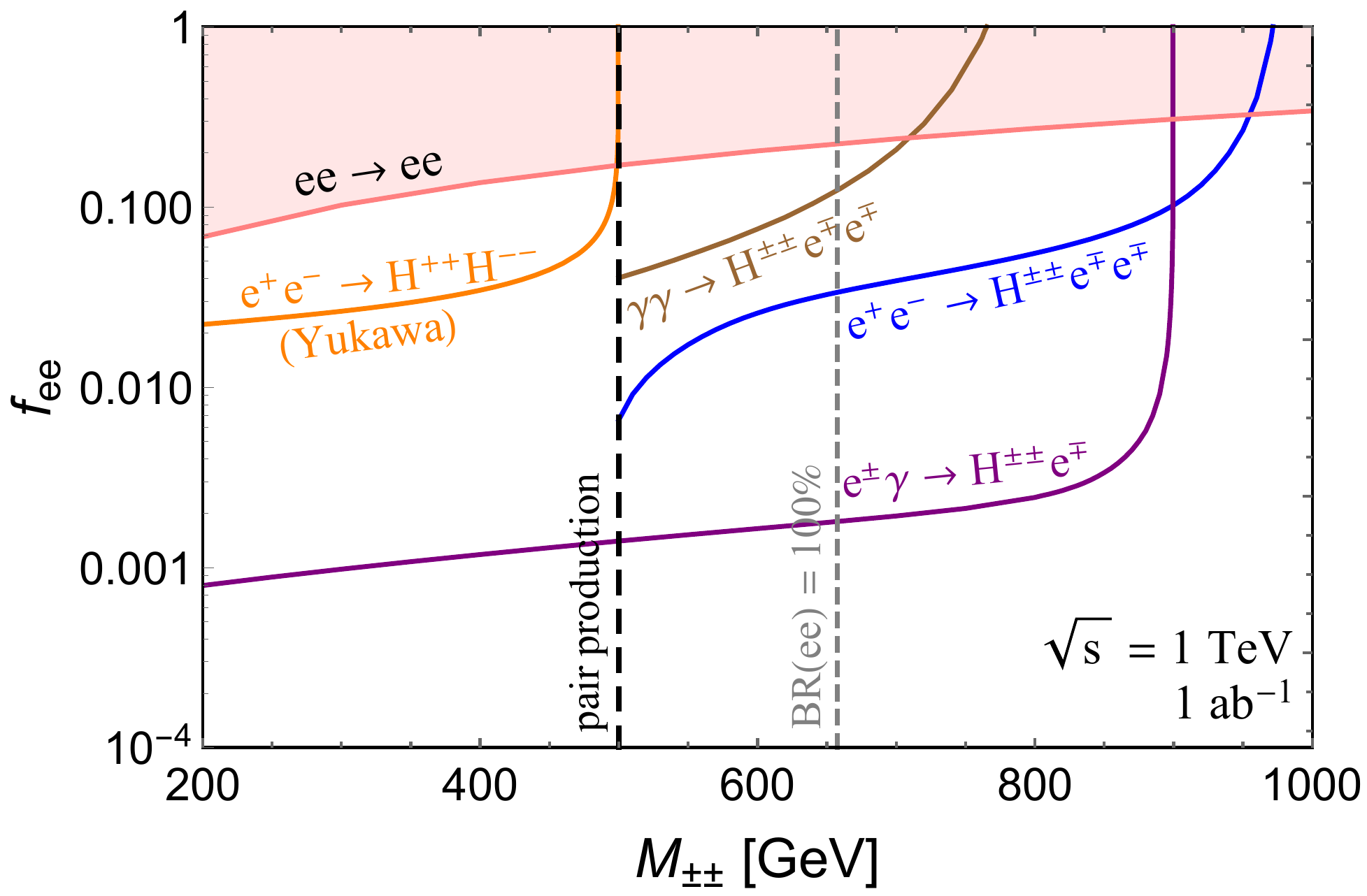}
  \includegraphics[width=0.4\textwidth]{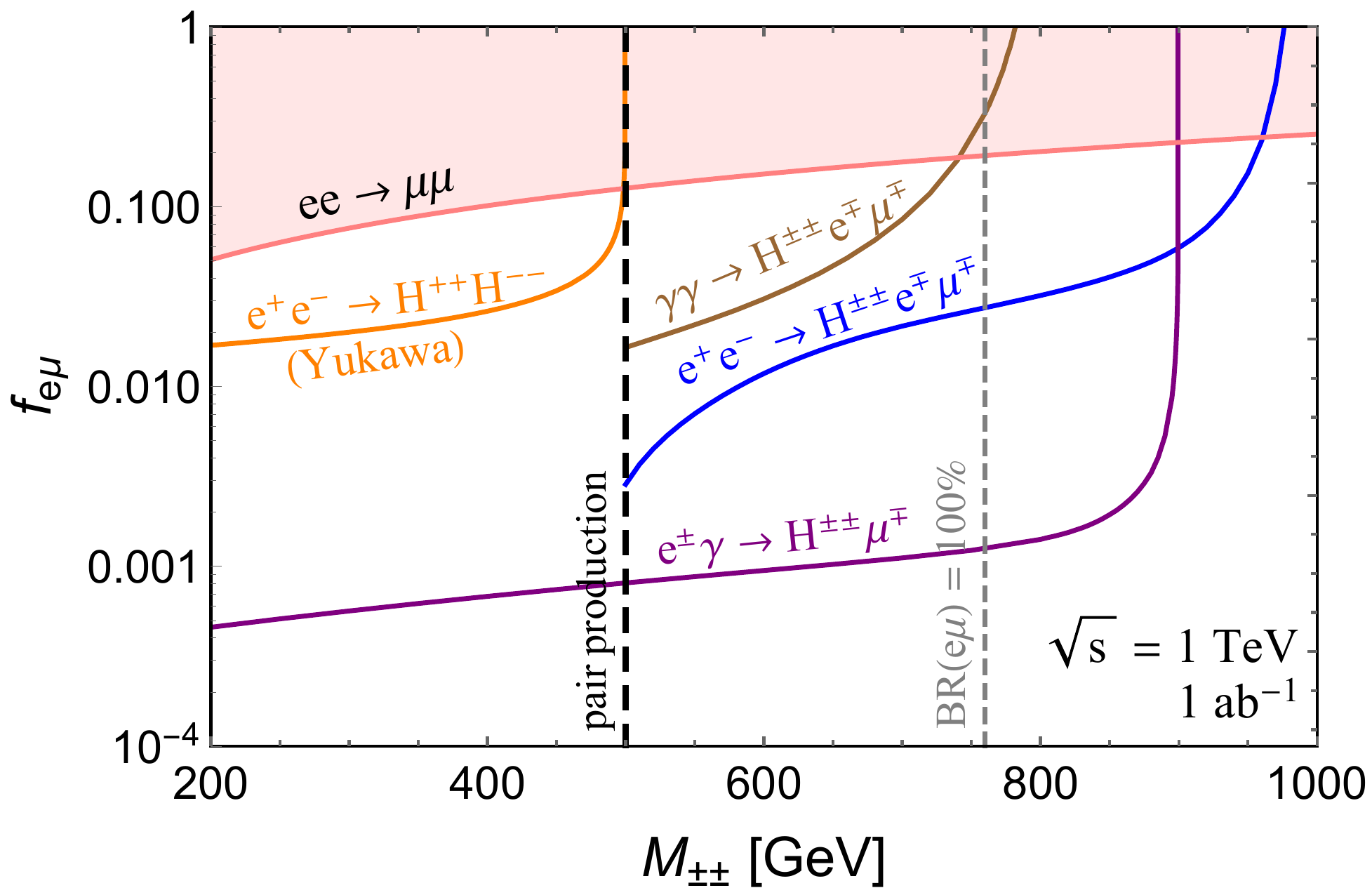} \\
  \includegraphics[width=0.4\textwidth]{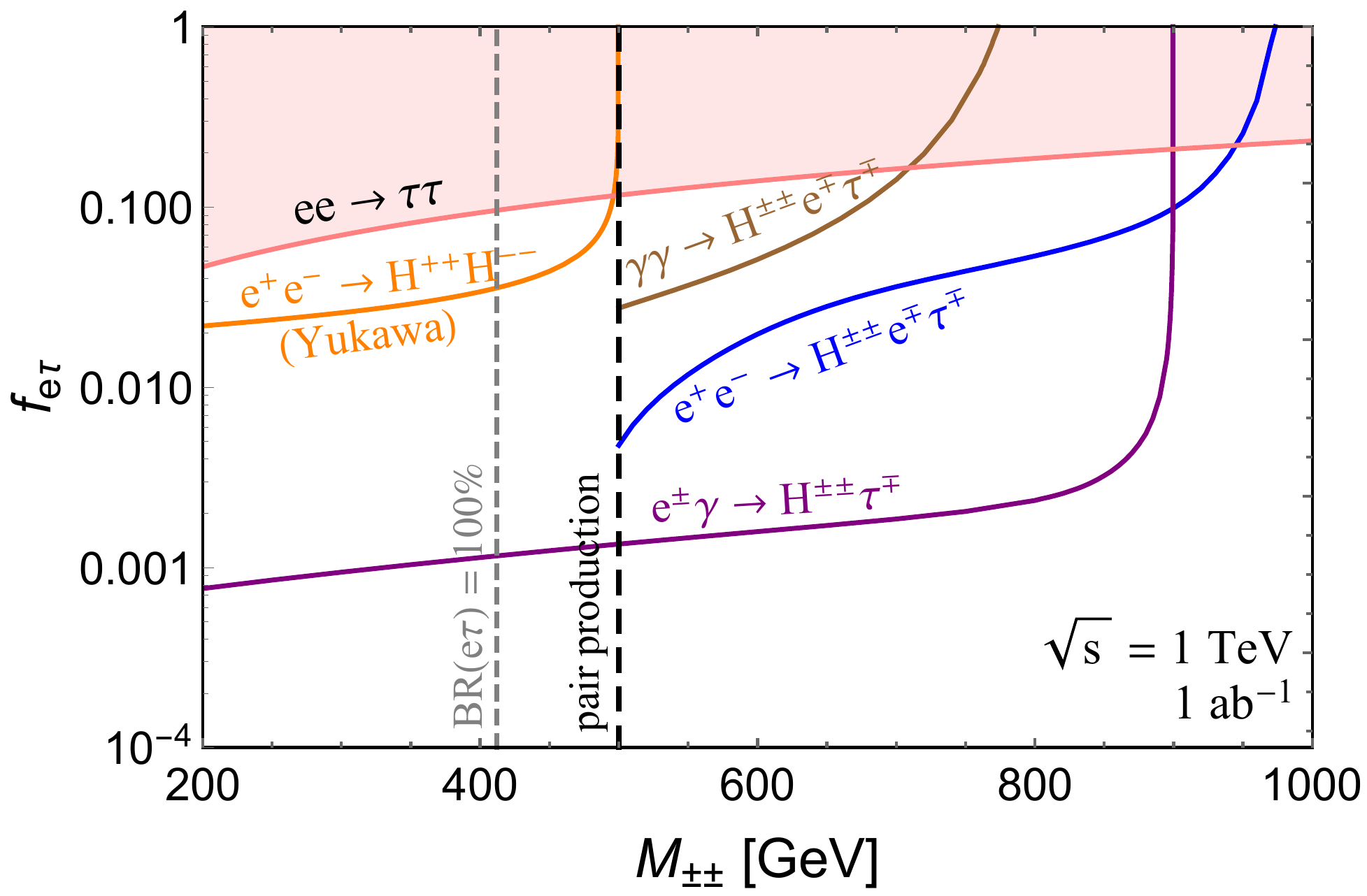}
  \includegraphics[width=0.4\textwidth]{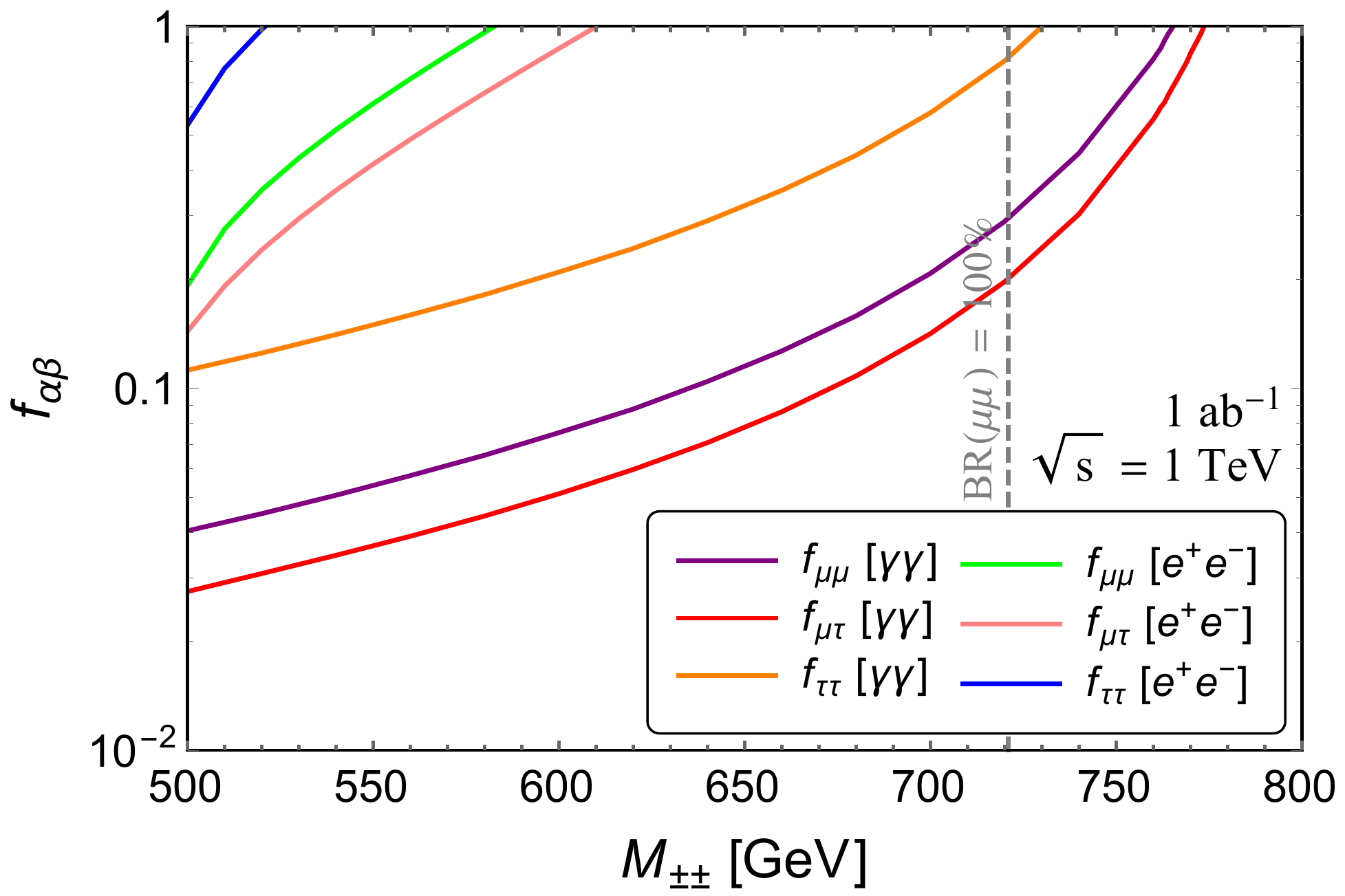}
  \caption{Prospects of the doubly-charged scalar $H^{\pm\pm}$ at ILC $1$ TeV and an integrated luminosity of 1 ab$^{-1}$.  The top left, top right and bottom left panels show both the prospects in the Yukawa pair (orange) and single production modes of the $e^+ e^-$ (blue), $e\gamma$ (purple) and $\gamma\gamma$ (brown) processes, for respectively the couplings $f_{ee}$, $f_{e\mu}$ and $f_{e\tau}$, as functions of the doubly-charged scalar mass. The pink shaded regions are excluded by the LEP $ee \to \ell\ell$ data~\cite{Abdallah:2005ph}. The prospects for the Yukawa couplings $f_{\alpha\beta}$ not involving the electron flavor are collected in the lower right panel, in both the $e^+ e^-$ and $\gamma\gamma$ processes. The vertical dashed gray lines indicate the current same-sign dilepton limits on the doubly-charged scalar mass from LHC~\cite{ATLAS:2017iqw, CMS:2017pet}, assuming a ${\rm BR} (H^{\pm\pm} \to \ell_\alpha^{\pm} \ell_\beta^{\pm}) = 100\%$. See text for more details.}
  \label{fig:prospect:Hpp:1}
\end{figure*}

At the leading order, both the amplitudes of the Yukawa pair production (the second diagram in Figure~\ref{fig:diagram:Hpp:1}) and single production channels in Figure~\ref{fig:diagram:Hpp:2} are linear functions of the couplings $f_{\alpha\beta}$. In the simplest case with only one independent element of the $f$ matrix turned on, one can predict the production cross sections as functions of the coupling $f_{\alpha\beta}$ involved, and estimate the resultant prospects for the individual couplings at future lepton colliders like CEPC, ILC and FCC-ee. To be specific, we consider only the pair and single production at ILC running of $\sqrt{s} = 1$ TeV in Figure~\ref{fig:prospect:Hpp:1}, as the planned CEPC and FCC-ee energies are lower. To be conservative, the ILC beams are assumed to be unpolarized; for the RH $H^{\pm\pm}$, polarized beams could improve to some extent the production cross sections and prospects, e.g. with $P_{e^-} = 0.8$ and $P_{e^+} = -0.3$~\cite{Baer:2013cma}. The basic cuts and efficiencies are the same as above, i.e. $p_T (\ell) = 10$ GeV, and $\epsilon_{e,\, \mu} \simeq 1$ and $\epsilon_\tau = 60\%$. The SM backgrounds for the pure lepton final states are well understood, with the $\tau$ decays causing some complications. In particular, the SM backgrounds with LFV like $(e^\pm \mu^\pm) (e^\mp \mu^\mp)$ are expected to be very low, and might originate from e.g. particle misidentification~\cite{Milstene:2006xh, Yu:2017mpx}. With the well-controlled SM background, the on-shell $H^{\pm\pm}$ is expected to give rise to sharp resonance-like peaks over the continuous kinetic distributions of the charged leptons like the invariant mass $m (\ell^\pm \ell^{\prime \, \pm})$~\cite{Dev:2017ftk}. To be concrete, we assume a total number of 30 events for the signals without apparent LFV like $H^{\pm\pm} \to e^\pm e^\pm$ and 10 for the LFV signals such as $H^{\pm\pm} \to e^\pm \mu^\pm$.

As already mentioned, only the electron and muon $g-2$ and the LEP data could be used to set unambiguous limits on the Yukawa couplings $f_{\alpha\beta}$ for the Yukawa pair and single production modes. Suppressed by the small charged lepton masses in Eq.~(\ref{eqn:g-2}), the $g-2$ limits in Table~\ref{tab:limits} are very weak and are not shown in Figure~\ref{fig:prospect:Hpp:1}. For the few hundred GeV scale $H^{\pm\pm}$, the LEP $ee \to \ell\ell$ data exclude the couplings $f_{e\ell}$ of order ${\cal O} (0.1)$ with $\ell = e,\, \mu,\, \tau$, as presented in the first three plots of Figure~\ref{fig:prospect:Hpp:1}. The same-sign dilepton limits on the doubly-charged scalar mass from the ATLAS~\cite{ATLAS:2017iqw} and CMS~\cite{CMS:2017pet} data in Figure~\ref{fig:limits} are also shown in Figure~\ref{fig:prospect:Hpp:1} as the vertical dashed gray lines, assuming a ${\rm BR} (H^{\pm\pm} \to \ell_\alpha^{\pm} \ell_\beta^{\pm}) = 100\%$ with respect to the specific lepton flavors $\ell_{\alpha,\,\beta}$. The dilepton limits for the states $\mu\tau$ and $\tau\tau$ are below 500 GeV, and are not shown in the lower right panel of Figure~\ref{fig:prospect:Hpp:1}.

When $M_{\pm\pm} \lesssim \sqrt{s}/2 = 500$ GeV at ILC, the doubly-charged scalar can both be pair or single produced, via respectively the Yukawa couplings (the second diagram in Figure~\ref{fig:diagram:Hpp:1}) and the $e\gamma$ process (the middle row in Figure~\ref{fig:diagram:Hpp:2}). In the lighter mass range, the single production like processes $e^+ e^- ,\, \gamma\gamma \to H^{\pm\pm} \ell_\alpha^\mp \ell_\beta^\mp$ is in fact dominated by the pair production modes $e^+ e^- ,\, \gamma\gamma \to H^{++} H^{--}$ in Figure~\ref{fig:diagram:Hpp:1} which does not depend on the Yukawa couplings; thus all the $ee$ and $\gamma\gamma$ single production processes in Figure~\ref{fig:prospect:Hpp:1} are truncated at 500 GeV. As the pair production amplitudes have a quadratic dependence on the couplings $f_{\alpha\beta}$ while the single $e\gamma$ process is only linear, the latter could probe a much smaller Yukawa coupling down to ${\cal O} (10^{-3})$ (the purple lines in the first three plots of Figure~\ref{fig:prospect:Hpp:1}), while the pair production only to few times $10^{-2}$ (the orange lines).

For $M_{\pm\pm} \gtrsim \sqrt{s}/2$, only the single production modes $ee$ (blue), $e\gamma$ (purple) and $\gamma\gamma$ (brown) are kinematically available. Compared to the $e\gamma$ processes, the $ee$ and $\gamma$ ones have one more lepton in the final state and thus the cross sections are comparatively smaller (see Figure~\ref{fig:prod:Hpp:2}), and the $\gamma\gamma$ modes are further suppressed by the effective photon luminosity function in Eq.~(\ref{eqn:photon_pdf}). Though the $e\gamma$ cross sections are comparatively larger, they are limited to the couplings $f_{\alpha\beta}$ involving at least one electron, cf. the middle row diagrams in Figure~\ref{fig:diagram:Hpp:2}. The Yukawa couplings in the $\mu - \tau$ sector ($f_{\mu\mu}$, $f_{\mu\tau}$ and $f_{\tau\tau}$) could only be reached via the $e^+ e^-$ and $\gamma\gamma$ collisions, with the prospects collected in the lower right panel of Figure~\ref{fig:prospect:Hpp:1}. There is currently almost no limit on these couplings, except for the muon $g-2$ on $f_{\mu\mu}$ and $f_{\mu\tau}$ which however is too weak to set any limits. Compared to the couplings involving at least one electron flavor, the production cross sections for those in the $\mu - \tau$ sector are much smaller, and the prospects are resultantly also very weaker, at most of order $10^{-2}$.  Nevertheless, all these modes are largely complementary to each other, in particular the $ee$ and $e\gamma$ processes, in producing the doubly-charged scalars, revealing the flavor structure of the Yukawa couplings $f_{\alpha\beta}$ and hunting for the LFV signals.

\begin{figure*}[t!]
  \centering
  \includegraphics[width=0.4\textwidth]{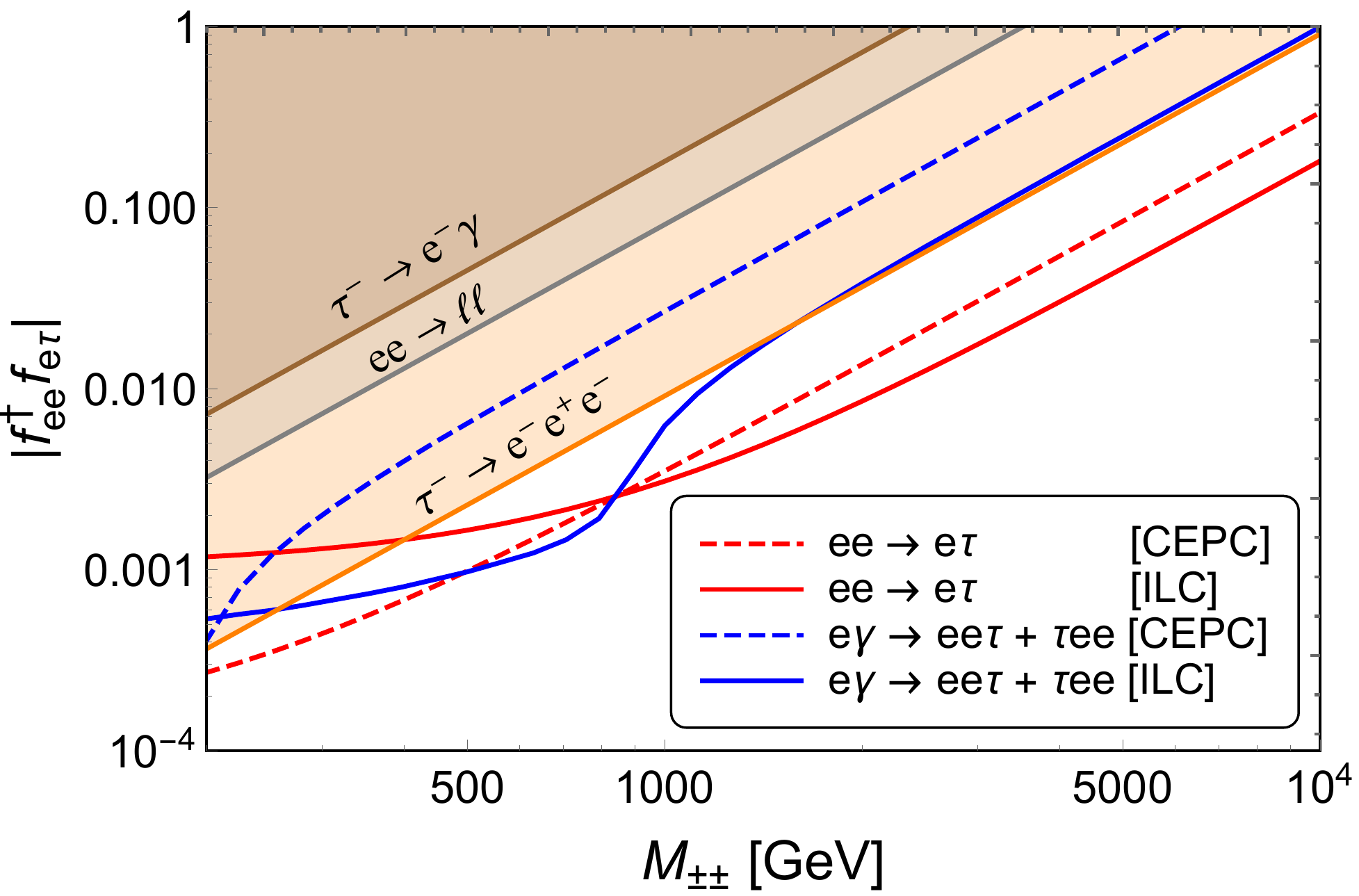}
  \includegraphics[width=0.4\textwidth]{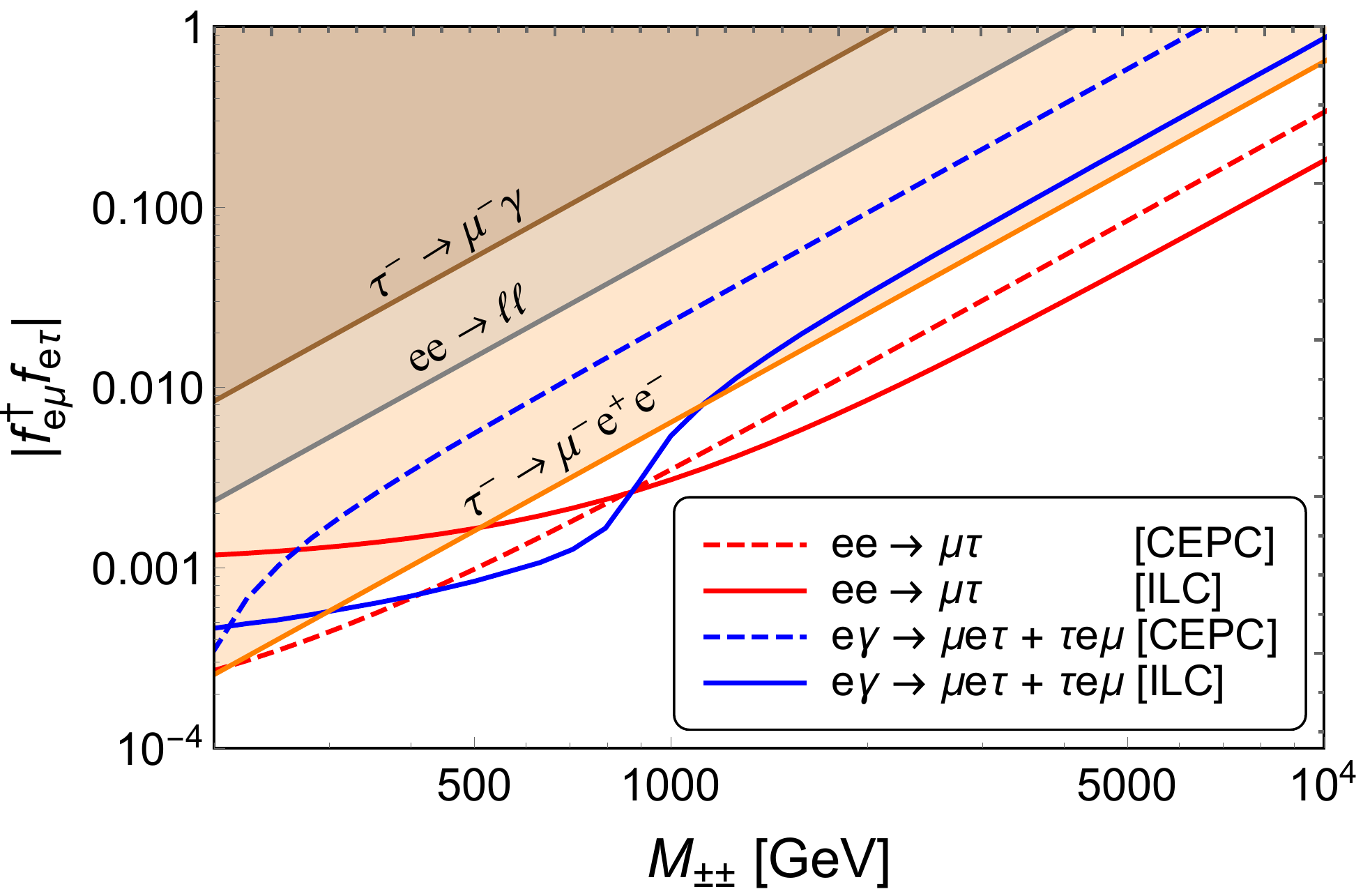}
  \caption{Prospects of the Yukawa couplings $|f_{ee}^\dagger f_{e\tau}|$ and $|f_{e\mu}^\dagger f_{e\tau}|$ for the doubly-charged scalar $H^{\pm\pm}$ production via the $ee \to \ell_\alpha \ell_\beta$ (red) and $e\gamma \to \ell_\alpha \ell_\beta \ell_\gamma$ (blue) processes, at CEPC with $\sqrt{s} = 240$ GeV and an integrated luminosity of 5 ab$^{-1}$ (dashed) and ILC with $\sqrt{s} = 1$ TeV and 1 ab$^{-1}$ (solid). The shaded regions are excluded by the rare tauon decays $\tau \to \ell_\alpha \gamma$, $\tau \to \ell_\alpha \ell_\beta \ell_\gamma$ and the LEP $ee \to \ell\ell$ data~\cite{Abdallah:2005ph}. See text for more details.}
  \label{fig:prospect:Hpp:2}
\end{figure*}

\begin{figure*}[t!]
  \centering
  \includegraphics[width=0.4\textwidth]{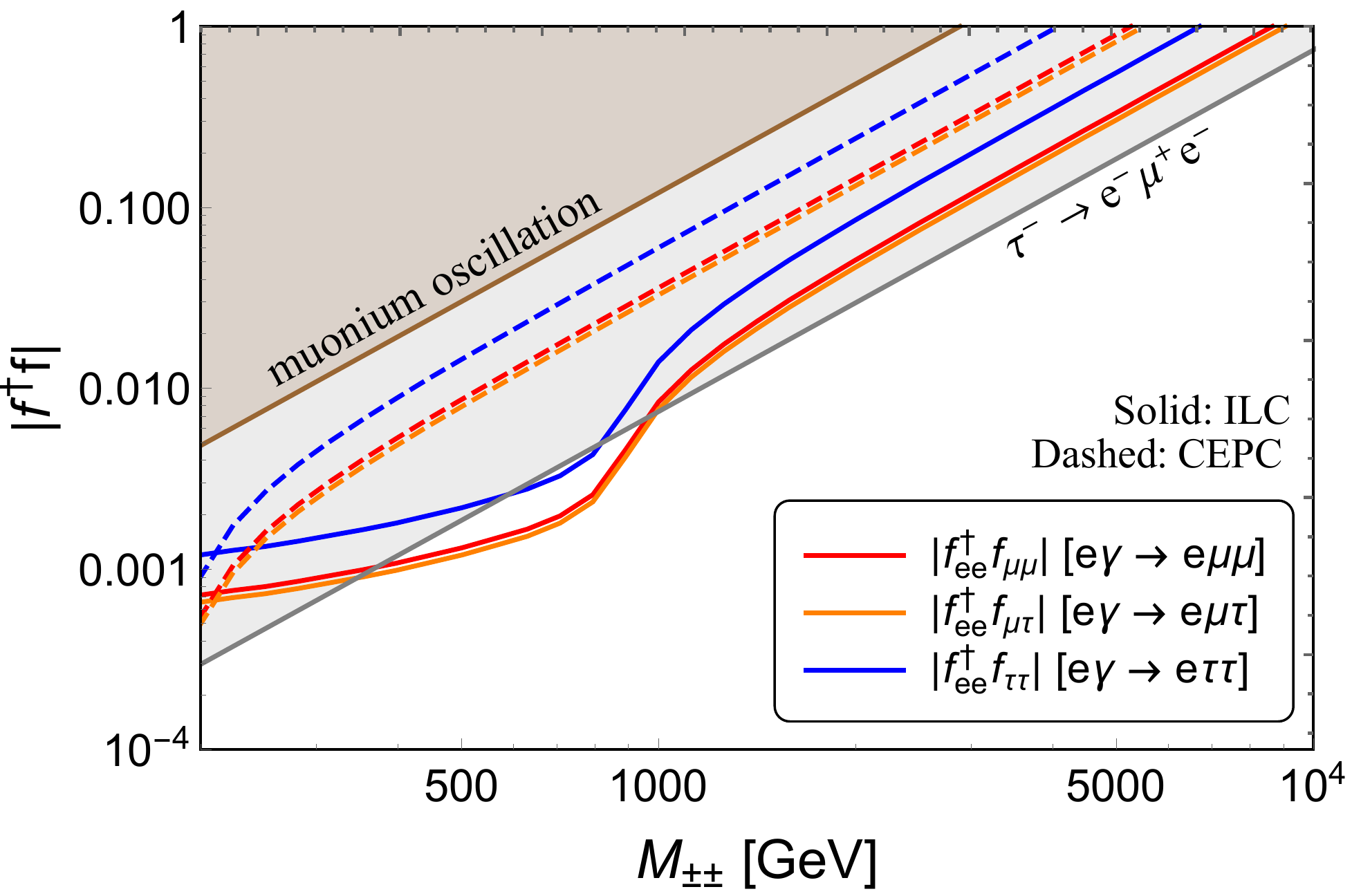} \\
  \includegraphics[width=0.4\textwidth]{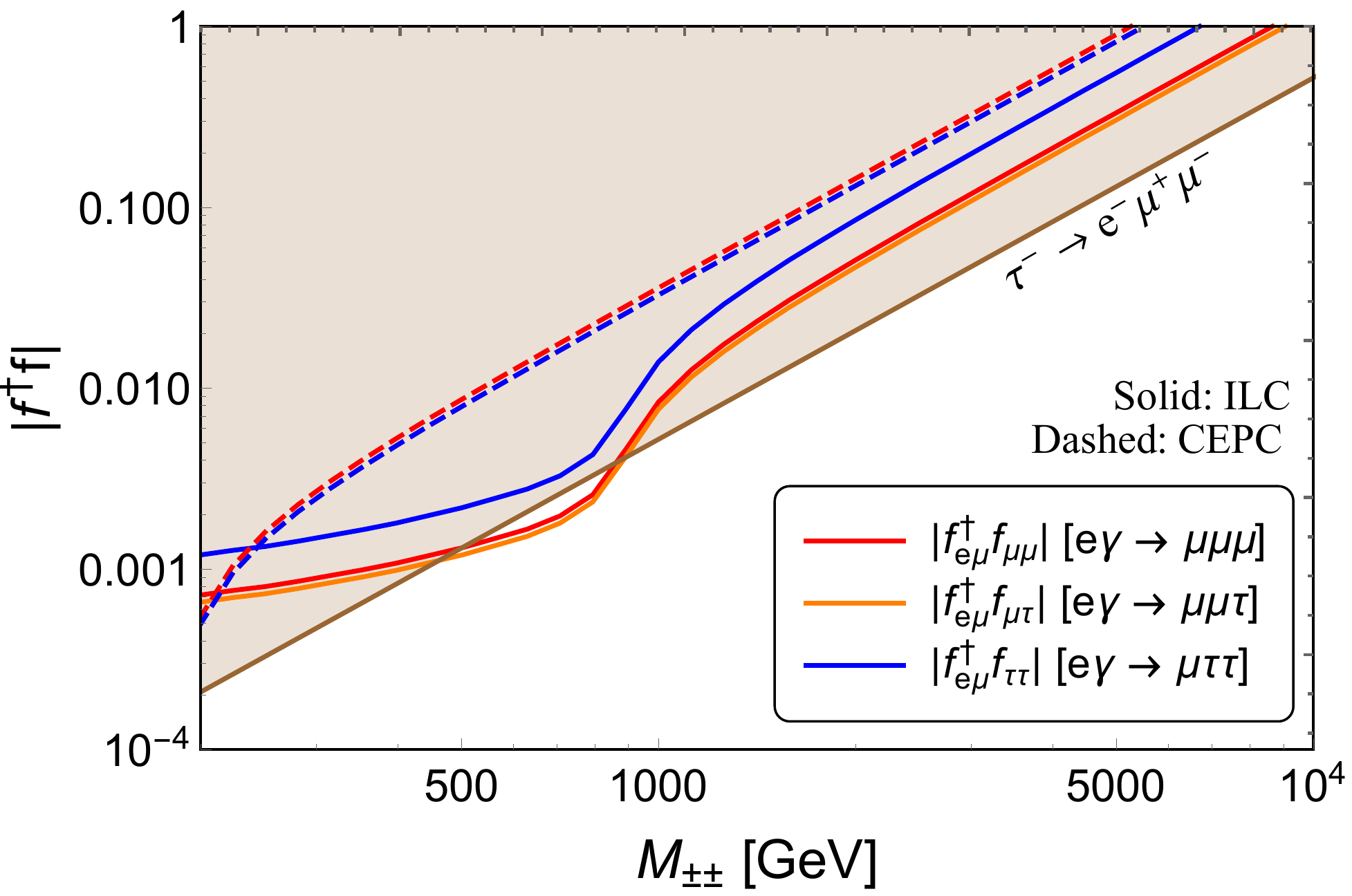}
  \includegraphics[width=0.4\textwidth]{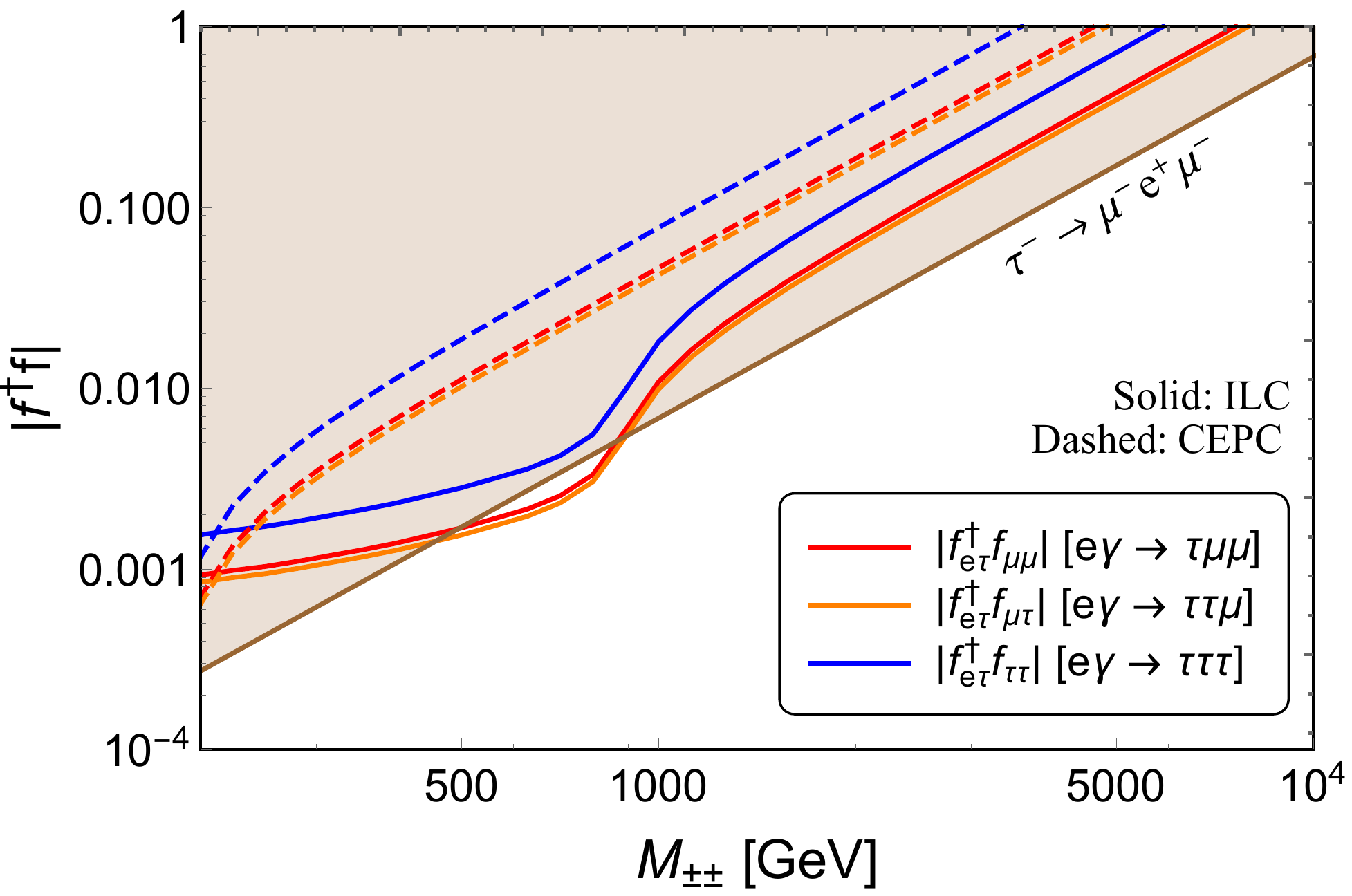}
  \caption{The same as in Figure~\ref{fig:prospect:Hpp:2}, for the couplings $|f_{ee}^\dagger f_{\mu\mu}|$ (red), $|f_{ee}^\dagger f_{\mu\tau}|$ (orange) and $|f_{ee}^\dagger f_{\tau\tau}|$ (blue) in the upper panel, $|f_{e\mu}^\dagger f_{\mu\mu}|$ (red), $|f_{e\mu}^\dagger f_{\mu\tau}|$ (orange) and $|f_{e\mu}^\dagger f_{\tau\tau}|$ (blue) in the lower left panel, and $|f_{e\tau}^\dagger f_{\mu\mu}|$ (red), $|f_{e\tau}^\dagger f_{\mu\tau}|$ (orange) and $|f_{e\tau}^\dagger f_{\tau\tau}|$ (blue) in the lower right panel, at both CEPC (dashed) and ILC (solid). The shaded brown region is excluded by muonium oscillation, which is applicable to $|f_{ee}^\dagger f_{\mu\mu}|$ (cf. Table~\ref{tab:limits}). The rare tauon decays $\tau^- \to e^- \mu^+ e^-$, $\tau^- \to e^- \mu^+ \mu^-$ and $\tau^- \to \mu^- e^+ \mu^-$ (gray) are used to constrain respectively the couplings $|f_{ee}^\dagger f_{\mu\tau}|$, $|f_{e\mu}^\dagger f_{\mu\tau}|$ and $|f_{e\tau}^\dagger f_{\mu\mu}|$ (cf. Table~\ref{tab:limits}).}
  \label{fig:prospect:Hpp:3}
\end{figure*}

For the off-shell production $e^+ e^- \to \ell_\alpha \ell_\beta$ and $e\gamma \to \ell_\alpha \ell_\beta \ell_\gamma$, the doubly-charged scalar mass could be (much) higher than the center-of-mass energy of the lepton colliders. For the sake of comparison we show in Figures~\ref{fig:prospect:Hpp:2} and \ref{fig:prospect:Hpp:3} the prospects at both CEPC with $\sqrt{s} = 240$ GeV and an integrated luminosity of 5 ab$^{-1}$ and ILC with $\sqrt{s} = 1$ TeV and a luminosity of 1 ab$^{-1}$. With a higher luminosity at 240 GeV anticipated, the FCC-ee could do better than CEPC. As the production amplitudes depend quadratically on the couplings $f_{\alpha\beta}$, the experimental constraints on the Yukawa couplings $|f^\dagger f|$ are different from those for the on-shell production in Figure~\ref{fig:prospect:Hpp:1}, as clearly seen in Table~\ref{tab:limits}. In particular, the limit from $\mu \to eee$ is so stringent that it leaves no probable space for $|f_{ee}^\dagger f_{e\mu}|$. The limits from the tauon sector are comparatively much weaker, with the prospects for $|f_{ee}^\dagger f_{e\tau}|$ and $|f_{e\mu}^\dagger f_{e\tau}|$ at CEPC and ILC via the $ee \to \ell_\alpha \ell_\beta$ and $e\gamma \to \ell_\alpha \ell_\beta \ell_\gamma$ processes collectively presented in Figure~\ref{fig:prospect:Hpp:2}. For $|f_{ee}^\dagger f_{e\tau}|$, the constraints are predominantly from the rare decays $\tau \to eee$ and $\tau \to e \gamma$, whereas the $\tau^- \to \mu^- e^+ e^-$ and $\tau \to \mu\gamma$ data can be used to set limits on $|f_{e\mu}^\dagger f_{e\tau}|$. Given the limits on $|f_{e\ell}|$ (with $\ell = e,\, \mu,\, \tau$) from the $ee \to \ell\ell$ data at LEP~\cite{Abdallah:2005ph}, these combinations $|f_{ee}^\dagger f_{e\tau}|$ and $|f_{e\mu}^\dagger f_{e\tau}|$ are also constrained by the LEP data, as shown in Figure~\ref{fig:prospect:Hpp:2}.

With only two leptons in the final state, the cross sections for the $ee \to \ell_\alpha \ell_\beta$ processes are orders of magnitude larger than those for $e\gamma \to \ell_\alpha \ell_\beta \ell_\gamma$, as seen in Figure~\ref{fig:prod:Hpp:3}; thus the former could probe smaller Yukawa couplings $|f^\dagger f|$, i.e. respectively $3.4 \times 10^{-9} \, (1.8 \times 10^{-9}) \, {\rm GeV}^{-2}$ at CEPC (ILC) for both $|f_{ee}^\dagger f_{e\tau}|$ and and $|f_{e\mu}^\dagger f_{e\tau}|$ in the heavy doubly-charged scalar limit. These are well beyond the current limits from the rare three-body tauon decays, as seen in Figure~\ref{fig:prospect:Hpp:2}. Benefiting from the larger colliding energy, the ILC running at 1 TeV outperforms the CEPC. At ILC, when the doubly-charged scalar mass $M_{\pm\pm} \lesssim 1$ TeV, the $e\gamma$ process is actually dominated by the on-shell production $e\gamma \to \ell_\alpha H^{\pm\pm}$ with the subsequent decay $H^{\pm\pm} \to \ell_\beta^\pm \ell_\gamma^\pm$ (see the middle row in Figure~\ref{fig:prod:Hpp:2} for the Feynman diagrams), and the production cross sections in Figure~\ref{fig:prod:Hpp:3} and the prospects in Figure~\ref{fig:prospect:Hpp:2} are largely enhanced, even better than the $ee$ mode.

All other combinations of the lepton flavors in $f_{\alpha\beta}^\dagger f_{\gamma\delta}$ can be collected in the form of $f_{e\alpha}^\dagger f_{\beta\gamma}$ with $\alpha = e,\, \mu,\, \tau$ and $\beta,\, \gamma = \mu,\, \tau$. These couplings could only be probed via the processes $e^\pm \gamma \to \ell_\alpha^\mp \ell_\beta^\pm \ell_\gamma^\pm$, with the corresponding prospects at CEPC and ILC presented in the plots of Figure~\ref{fig:prospect:Hpp:3}. As in Figure~\ref{fig:prospect:Hpp:2}, when the doubly-charged scalar mass is below 1 TeV, it could be produced on-shell at ILC and the prospects increase significantly. Given all the flavor constraints in Table~\ref{tab:limits}, only some of the Yukawa couplings $|f^\dagger f|$ are constrained by current experimental data: muonium oscillation can be used to set an limits on $|f_{ee}^\dagger f_{\mu\mu}|$ (red curves in the upper panel), while the rare tauon decays $\tau^- \to e^- \mu^+ e^-$, $\tau^- \to e^- \mu^+ \mu^-$ and $\tau^- \to \mu^- e^+ \mu^-$ could be used to constrain respectively the couplings $|f_{ee}^\dagger f_{\mu\tau}|$ (orange curves in the upper panel), $|f_{e\mu}^\dagger f_{\mu\tau}|$ (orange curves in the lower left panel) and $|f_{e\tau}^\dagger f_{\mu\mu}|$ (red curves in the lower right panel). It is transparent in Figure~\ref{fig:prospect:Hpp:3} that when these limits are taken into consideration, one would not see any signal at CEPC or ILC in these channels, unless the doubly-charged scalar could be produced on-shell. All the other combinations of couplings $|f^\dagger f|$ are not constrained by any data, and one could probe these couplings with the LFV signals by searching for the trileptons $e\gamma \to \ell_\alpha \ell_\beta \ell_\gamma$, even if the doubly-charged scalar mass is beyond the TeV scale.

%\newpage
%\section{Neutral scalar in the LRSM}

\subsection{Application to specific textures}

%We give three examples.

In this section, we consider three specific textures of the matrix $f_{\alpha\beta}$ in Eq.~(\ref{eqn:LYukawa}) to exemplify the flavor constraints and lepton collider prospects of the RH doubly-charged scalar in the LRSM. Typically, the property of each case starts with specific textures of the charged lepton mass matrix, the Dirac neutrino mass matrix and the heavy RHN mass matrix in the LRSM,
%that allows for sizable Dirac mass entries and corresponding $f$-texture
that gives rise to the physical charged lepton masses, the tiny neutrino masses, the lepton flavor mixing angles and the CP violating phases~\cite{Dev:2013oxa}. With regard to the doubly-charged scalar sector, only the $f$ coupling matrix is relevant, which is related to the heavy RHN masses via $M_N \simeq 2f v_R$. Three simple textures of $f$ are presented in Table~\ref{tab:examples}, i.e. the textures A, B and C, which share the same sets of parameters, with only two free parameters $f_{1,2}$ of order ${\cal O} (0.1)$, following the numerical fit of Ref.~\cite{Dev:2013oxa}. The texture zeros might originate from and be protected  by some underlying discrete symmetry groups in the lepton sector~\cite{Kersten:2007vk, Dev:2013oxa}.

\begin{table*}[!t]
  \centering
  \small
  \caption[]{Three specific sample textures A, B and C, for the coupling matrix $f$, with the fixed values of $f_{1,2}$ given in the table. We show collectively, for all the three textures, the decay BRs, the flavor limits from Table~\ref{tab:limits} (the lower bounds on $M_{\pm\pm}$ are put in the brackets with those lower than 100 GeV not shown), the LHC same-sign dilepton limits from Figure~\ref{fig:limits} (the corresponding channels are in the parentheses), the pair production mass windows at ILC 1 TeV in the $ee$ mode, the single production channels (the lepton pairs in parentheses are from on-shell $H^{\pm\pm}$ decay) in both the $ee / \gamma\gamma$ and $e\gamma$ modes, and the off-shell production channels of the $ee$ and $e\gamma$ processes. For textures B and C, the on-shell production is not possible (as denoted by $-$), because of the stringent LHC limits. See text for more details.}
  \label{tab:examples}
  \begin{tabular}[t]{c|c|c|c}
  \hline\hline
  textures & A & B & C \\ \hline
  \makecell{$f_1 = 0.1$ \\ $f_2 = 0.5$ } &
  $\left(\begin{array}{ccc} 0 & f_1 & 0\\f_1& 0 & 0\\ 0 & 0 & f_2\end{array}\right)$ &
  $\left(\begin{array}{ccc} f_2 & 0 & 0\\0 & 0 & f_1 \\ 0 & f_1 & 0 \end{array}\right)$ &
  $\left(\begin{array}{ccc} 0 & 0 & f_{1}\\0 & f_{2} & 0\\ f_{1} & 0 & 0 \end{array}\right)$ \\ \hline
  decay BRs &
  \makecell{${\rm BR} (e\mu) \ = \ 7.4\%$ \\ ${\rm BR} (\tau\tau) \ = \ 92.6\%$} &
  \makecell{${\rm BR} (\mu\tau) \ = \ 7.4\%$ \\ ${\rm BR} (ee) \ = \ 92.6\%$} &
  \makecell{${\rm BR} (e\tau) \ = \ 7.4\%$ \\ ${\rm BR} (\mu\mu) \ = \ 92.6\%$} \\ \hline
  flavor limits &
  \makecell{ \\ electron $g-2$ \\ muon $g-2$ \\ $ee \to \mu\mu$ [395 GeV]} &
  \makecell{$\tau^- \to e^- \mu^+ e^-$ [2.6 TeV] \\ electron $g-2$ \\ muon $g-2$ \\ $ee \to ee$ [1.4 TeV]} &
  \makecell{$\tau^- \to \mu^- e^+ \mu^-$ [1.7 TeV] \\ electron $g-2$ \\ muon $g-2$ \\ $ee \to \tau\tau$ [430 GeV]} \\ \hline
  %\makecell{most stringent \\ flavor limit} &
  %\makecell{395 GeV \\ (LEP $ee\to \mu\mu$)} & b & c \\ \hline
  LHC limits & 321 GeV ($e\mu$) & 648 GeV ($ee$) & 710 GeV ($\mu\mu$) \\ \hline
  \makecell{pair production \\ at ILC 1TeV} & [395, 500] GeV & $-$ & $-$ \\ \hline
  \makecell{single production \\ ($ee,\,\gamma\gamma$)} &
  \makecell{$(e^\pm \mu^\pm) \, e^\mp \mu^\mp$ \\
  $(e^\pm \mu^\pm) \, \tau^\mp \tau^\mp$ \\
  $(\tau^\pm \tau^\pm) \, e^\mp \mu^\mp$ \\
  $(\tau^\pm \tau^\pm) \, \tau^\mp \tau^\mp$} & $-$ & $-$ \\ \hline
  \makecell{single production \\ ($e\gamma$)} &
  \makecell{$(e^\pm \mu^\pm) \, \mu^\mp$ \\
  $(\tau^\pm \tau^\pm) \, \mu^\mp$} & $-$ & $-$ \\ \hline
  \makecell{off-shell production \\ ($ee$)} &
  $\mu^+ \mu^-$ & $e^+ e^-$ & $\tau^+ \tau^-$ \\ \hline
  \makecell{off-shell production \\ ($e\gamma$)} &
  \makecell{$\mu^\mp e^\pm \mu^\pm$ \\
  $\mu^\mp \tau^\pm \tau^\pm$} &
  \makecell{$e^\mp e^\pm e^\pm$ \\
  $e^\mp \mu^\pm \tau^\pm$} &
  \makecell{$\tau^\mp e^\pm \tau^\pm$ \\
  $\tau^\mp \mu^\pm \mu^\pm$} \\
  \hline\hline
  \end{tabular}
\end{table*}

%where the heavy RHN masses are related to the $f$ couplings via (the model A)
%\begin{eqnarray}
%\text{Model A: } & \ &
%M_N \ = \ v_R\left(\begin{array}{ccc} 0 & f_{e\mu} & 0\\f_{e\mu}& \epsilon & 0\\ 0 & 0 & 2f_{\tau\tau}\end{array}\right) \,,
%\end{eqnarray}
%\begin{eqnarray}
%m_D=\left(\begin{array}{ccc}m_{11} & \delta_{12} & \delta_{13}\\ m_{21} & \delta_{22} & \delta_{23}\\m_{31} & \delta_{32} & \delta_{33}\end{array}\right);~ M_N=v_R\left(\begin{array}{ccc} 0 & f_{12} & 0\\f_{12}& \epsilon & 0\\ 0 & 0 & f_{33}\end{array}\right)
%\end{eqnarray}
%We exemplify the generic form by the following example
%with $\epsilon \ll f_{\alpha\beta}$.
%For a phenomenologically favored RH scale $v_R = 5$ TeV, the Yukawa couplings $f_{e\mu,\,\tau\tau}$ are of order $0.1$ and $\epsilon$ much smaller~\cite{DLM}:
%\begin{eqnarray}
%\label{eqn:example1}
%f_{e\mu} \ = \ f_1 \ = \ 0.0813 \,, \quad
%f_{\tau\tau} \ = \ f_2 \ = \ - 0.249 \,, \quad
%\epsilon \ = \ 1.3 \times 10^{-13} \,.
%\end{eqnarray}

For the texture A, the doubly-charged scalar decays predominantly into $e^\pm \mu^\pm$ and $\tau^\pm \tau^\pm$; following Eq.~(\ref{eqn:BR}), the BR are respectively 7.4\% and 92.6\%, as collected in Table~\ref{tab:examples}. With the texture zeros, some of the flavor constraints in Table~\ref{tab:limits} are not applicable to this specific scenario; only the electron and muon $g-2$ and the LEP $ee \to \mu\mu$ data set limits on the $f_{1,\,2}$ couplings, depending on the doubly-charged scalar mass $M_{\pm\pm}$. Among them, the most stringent is from the LEP data, which requires that $M_{\pm\pm} > 395$ GeV (cf. the upper right panel of Figure~\ref{fig:prospect:Hpp:1}), and the $g-2$ limits are below 100 GeV. Regarding the ATLAS and CMS same-sign dilepton limits in Figure~\ref{fig:limits}, it is trivial to interpret the $e\mu$ and $\tau\tau$ data with respect to the corresponding BRs, and obtain the constraints on doubly-charged scalar mass, which turn out to be 321 GeV ($e\mu$) and 253 GeV ($\tau\tau$).

Combining all the limits above, there is only a narrow window left for the pair production of $H^{\pm\pm}$ at the ILC running at 1 TeV: $395 \, {\rm GeV} < M_{\pm\pm} < 500 \, {\rm GeV}$, with nevertheless strikingly clear signal in particular in the $e\mu$ channel:\footnote{For a doubly-charged scalar with mass $M_{\pm\pm} > 395$ GeV, the production $\gamma\gamma \to H^{++} H^{--}$ is only marginally allowed at ILC 1 TeV, as seen in Figure~\ref{fig:prod:Hpp:1}.}
\begin{eqnarray}
e^+ e^- & \ \to \ & \gamma^\ast/Z^\ast \to H^{++} H^{--} \nonumber \\
& \ \to \ &
(e^+ \mu^+) (e^- \mu^-), \; (e^\pm \mu^\pm) (\tau^\mp \tau^\mp),\;
(\tau^+ \tau^+) (\tau^- \tau^-) \,. \nonumber \\ &&
\end{eqnarray}
With the ${\cal O}(0.1)$ Yukawa couplings $f_{1,\,2}$, the doubly-charged scalar is hard to be missed at lepton colliders by reconstructing the invariant mass $m_{e\mu}$ and $m_{\tau\tau}$ of the same-sign dilepton pairs. One should note that the doubly-charged scalar could also be produced through the Yukawa coupling $f_{e\mu}$, cf the second diagram in Figure~\ref{fig:diagram:Hpp:1}, which is however comparatively suppressed by a factor of $(f_{1}/e)^4$ ($e$ here being the electric charge) and thus contribute only with a factor of $\sim [1+ (f_{e\mu}/e)^2]$ to the total cross section for the pair production of $H^{\pm\pm}$.

When the doubly-charged scalar is heavier than 500 GeV, it could only be single produced at ILC from the $ee$, $\gamma\gamma$ and $e\gamma$ processes, with potentially all the flavor combinations listed in the first column of Table~\ref{tab:examples}. Roughly speaking, with the initial $ee$ or $\gamma\gamma$, the doubly-charged scalar can be produced in association with the same-sign lepton pairs of $e^\pm \mu^\pm$ or $\tau^\pm \tau^\pm$; while for the $e\gamma$ process, the doubly-charged scalar can be produced with a muon, which is dictated by the $f_1$ coupling. If the doubly-charged scalar is too heavy to be directly produced on-shell, it could yet mediate the processes $e^+ e^- \to \mu^+ \mu^-$ and $e^\pm \gamma \to \mu^\mp e^\pm \mu^\pm, \, \mu^\mp \tau^\pm \tau^\pm$. The doubly-charged scalar mediated $ee \to \mu\mu$ would interfere with the SM background, altering both the total cross section and angular distributions, thus constrained by the LEP data~\cite{Abdallah:2005ph}. The doubly-charged scalar mediated $\mu^\mp e^\pm \mu^\pm$ process will also interfere with the pure SM diagrams of $e^\pm \gamma \to e^\pm \mu^\pm \mu^\mp$ and might have effects on both the cross section and differential distributions, depending on the mass $M_{\pm\pm}$. A dedicated analysis is beyond the main scope of this paper and we will not go into any further details. The $\mu^\mp \tau^\pm \tau^\pm$ final states are, on the contrary, clear signatures of LFV, even if the doubly-charged scalar goes beyond the TeV scale.

%\begin{eqnarray}
%e^+ e^- ,\, \gamma\gamma & \ \to \ &
%(e^\pm \mu^\pm) \, e^\mp \mu^\mp, \quad
%(e^\pm \mu^\pm) \, \tau^\mp \tau^\mp, \nonumber \\
%&& (\tau^\pm \tau^\pm) \, e^\mp \mu^\mp, \quad
%(\tau^\pm \tau^\pm) \, \tau^\mp \tau^\mp, \\
%e^\pm\gamma & \ \to \ &
%(e^\pm \mu^\pm) \, \mu^\mp,  \quad
%(\tau^\pm \tau^\pm) \, \mu^\mp,
%\end{eqnarray}

%Note that this texture is consistent with the rare lepton decay constraints from Table I.
%From a detailed fermion fit in  in an LR seesaw model,
%we find $f_{12}\sim 10^{-1}$ and $f_{33}\sim 1$ and are therefore clearly in the accessible range for lepton colliders.

By replacing $1\leftrightarrow 3$ and $2\leftrightarrow 3$ in the matrix $f$ of texture A, we obtain respectively the scenarios B and C. With the textures changed, the lepton flavor constraints also vary, which are all collected in the second and third column of Table~\ref{tab:examples}. For the scenario B, the most stringent limits are from the rare decay $\tau^- \to e^- \mu^+ e^-$ which requires that $M_{\pm\pm} > 2.6$ TeV for the specific values of $f_{1,2}$ (cf. the upper panel of Figure~\ref{fig:prospect:Hpp:3}). The electron and muon $g-2$ limits are below 100 GeV and are not shown in the Table. The coupling $f_{ee} = f_2$ induce also extra contribution to the Bhabha scattering and are thus tightly constrained by the LEP data with a lower bound of 1.4 TeV on the doubly-charged scalar mass. With these overwhelming limits, the $H^{\pm\pm}$ can only be produced at ILC 1 TeV in the off-shell mode, producing the LFV signal of $e^\pm \gamma \to e^\mp \mu^\pm \tau^\pm$, though the on-shell production is possible at CLIC 3 TeV.\footnote{This is currently under investigation for an upcoming Yellow report on CLIC physics potential.} In addition, the large branching fraction into $ee$ pushes the LHC dilepton limits on $M_{\pm\pm}$ much more stringent than the texture A, being 648 GeV.

The texture C is kind of similar to the texture B, with the doubly-charged scalar decaying predominantly into $\mu^\pm \mu^\pm$ with a small proportion to $e^\pm \tau^\pm$. The rare decay $\tau^- \to \mu^- e^+ \mu^-$ provides the most stringent flavor constraints of 1.7 TeV on the doubly-charged scalar mass (cf. the lower right panel of Figure~\ref{fig:prospect:Hpp:3}). The limit from the LEP $ee\to \tau\tau$ data is comparatively much weaker, roughly 430 GeV, suppressed to some extent by the coupling $f_1$. The LHC same-sign dilepton limit is somewhat stronger than in scenario B and is 710 GeV. With a beyond-TeV scale $H^{\pm\pm}$, the LFV signal in the off-shell mode is primarily $\tau^\mp \mu^\pm \mu^\pm$, in addition to the lepton-flavor-conserving-like process $e^\pm \tau^\pm \tau^\mp$, from the $e\gamma$ collision.

%\begin{eqnarray}
%\text{Model B: } & \ &
%M_N \ = \ v_R\left(\begin{array}{ccc} 2f_{ee} & 0 & 0\\0 & \epsilon & f_{\mu\tau}\\ 0 & f_{\mu\tau} & 0 \end{array}\right) \,, \\
%\text{Model C: } & \ &
%M_N \ = \ v_R\left(\begin{array}{ccc} 0 & 0 & f_{e\tau}\\0 & 2f_{\mu\mu} & 0\\ f_{e\tau} & 0 & \epsilon \end{array}\right) \,.
%\end{eqnarray}

%For the comparison reason, we take the same value of $f_{1,\,2}$ (and $\epsilon$) for the diagonal and off-diagonal elements as in Eq.~(\ref{eqn:example1}), i.e.
%\begin{eqnarray}
%\text{Model B:} & \ &
%f_{ee} \ = \ f_2 \,, \quad
%f_{\mu\tau} \ = \ f_1 \,, \\
%\text{Model C:} & \ &
%f_{e\tau} \ = \ f_1 \,, \quad
%f_{\mu\mu} \ = \ f_2 \,.
%\end{eqnarray}

Although the three textures A, B and C share the same data set of $f_{1,2}$, their implications on the searches of $H^{\pm\pm}$ at future lepton colliders are distinctly different, e.g. the flavor limits on the doubly-charged scalar mass and the on-shell and off-shell signals. Given the prospects in Section~\ref{sec:production}, these realistic models could easily be tested and distinguished at future lepton colliders, e.g. via searches of the LFV signals and direct (pair) production of doubly-charged scalars. In the LRSM with multi-TeV scale RH scale $v_R$, say 5 TeV, to accommodate TeV-scale RHNs for the seesaw mechanism, (some of) the $f$ elements are expected to be sizable. In the absence of any beyond SM signal at lepton colliders, these scenarios will be clearly ruled out. More importantly, it might imply that the RH scale $v_R$ is much heavier than the TeV scale and hence the $W_R$ mass will be pushed beyond the LHC capability by the absence of these signals, which will anyway be an interesting result in the study of  TeV scale left-right seesaw models for neutrino masses.

\section{Conclusion}
\label{sec:conclusion}

In this paper we have studied in great detail the prospects for observing the effects of triplet scalars of the left-right symmetric model in future lepton colliders, i.e. the neutral $SU(2)_R$-breaking scalar $H_3$ and its doubly-charged scalar partner $H^{\pm\pm}$.
In the original Lagrangian, the neutral scalar $H_3$ couples only to the heavy beyond SM scalar, the heavy $W_R$ and $Z_R$ gauge boson, and the heavy right-handed neutrinos, and not to any SM quarks. Thus it could easily evade all the current direct searches at the LHC. Via its mixing in the scalar sector and the neutrino sector, it could be produced in lepton colliders from fusion of the doubly-charged scalars, the radiative couplings to photon at 1-loop level, the mixing with the SM Higgs, the Yukawa couplings to the heavy neutrinos, and the mixing with the heavy neutral scalar $H_1$ from the bidoublet. From the phenomenological point of view, the most interesting channel is the last one, as the mixing with $H_1$ would lead to the LFV couplings of $H_3$, which is directly connected to the Dirac mass matrix in the seesaw formula. Given these LFV couplings, the neutral scalar $H_3$ can be produced in future lepton collider on-shell or off-shell, leading to strikingly clear LFV signals such as $e^+ e^- \to e^\pm \mu^\mp + H_3$ and $e^+ e^- \to \mu^\pm \tau^\pm$. The Feynman diagrams can be found in Figures~\ref{fig:diagram5}, \ref{fig:diagram6}, \ref{fig:diagram7} and \ref{fig:diagram8}, with the cross section depicted in Figures~\ref{fig:production4}, \ref{fig:production5}, \ref{fig:production6} and \ref{fig:production7}. With an integrated luminosity of the order of ab$^{-1}$, the couplings can be probed up to $\sim 10^{-4}$ (see Figures~\ref{fig:H3:prospect1} and \ref{fig:H3:prospect2}), depending on the center-of-mass energy, the production channel and decay products. This is well beyond the current stringent flavor constraints in the lepton sector (see Table~\ref{tab:limits:H3}) such as $\tau \to eee$ and $\mu \to e \gamma$ in a larger parameter space, and could explain the muon $g-2$ anomaly in the channel $ee \to \mu\tau + H_3$.

Regarding the right-handed doubly-charged scalar in the left-right symmetric model, its couplings to the charged leptons stem from type-II seesaw and point directly to the masses and flavor structure in the right-handed neutrino sector. We provide a full list of the production of doubly-charged scalar at future lepton colliders, through the $e^+ e^-$, $e\gamma$ and $\gamma\gamma$ collisions ($\gamma$ here refers to the real laser photon), in the channels of pair production, single production (in association with one or two additional charged leptons in the final state) and off-shell production, and with all the possible charge lepton flavor combinations. The representative Feynman diagrams can be found in Figures~\ref{fig:diagram:Hpp:1}, \ref{fig:diagram:Hpp:2} and \ref{fig:diagram:Hpp:3}, and the production cross section are collected in Figures~\ref{fig:prod:Hpp:1}, \ref{fig:prod:Hpp:2} and \ref{fig:prod:Hpp:3}. The prospects are presented in Figures~\ref{fig:prospect:Hpp:1}, \ref{fig:prospect:Hpp:2} and \ref{fig:prospect:Hpp:3}. The lepton flavor conserving and violating coupling can be probed up to $\sim 10^{-3}$ for a large range of doubly-charged scalar mass, that are well beyond the current flavor constraints listed in Table~\ref{tab:limits}. As exemplified in Table~\ref{tab:examples}, the flavor limits and lepton collider prospects depend largely on the structure of the Yukawa coupling $f$.

The neutral and doubly-charged seesaw scalars are intimately correlated to the tiny neutrino mass generation via the seesaw mechanism. The lepton collider searches of these scalars are largely complementary to the effects at current and future hadron colliders and the low-energy neutrino experiments such as those aiming at the leptonic CP violation, neutrino mass hierarchy and neutrinoless double beta decays. Due to the ``clean'' background of lepton colliders, the direct searches of the neutral and doubly-charged scalar at CEPC, ILC, FCC-ee and CLIC are very promising, and might reveal the beyond SM physics in searches of the LFV signals. These searches can even be connected to the dark matter particle in the type II  seesaw frameworks in a different class of left-right models with the lightest right-handed neutrino being the dark matter~\cite{Dev:2016qbd, Dev:2016xcp, Dev:2016qeb}.

\section*{Acknowledgments}
The work of R.N.M. is supported by the US National Science Foundation under Grant No. PHY1620074. Y.Z. would like to thank the Center for High Energy Physics, Peking University, and the Center for Future High Energy Physics, IHEP, CAS, for the hospitality and local support during the visit, where part of the work was done.

%\appendix
%\section{Fitting of the lepton data}

%Fitting of the neutrino masses and mixing angles for the three textures...

%\newpage

\end{document}